\documentclass[aps,prb,reprint,twocolumn,superscriptaddress,floatfix,longbibliography,amsmath, amssymb]{revtex4-2}
\usepackage{graphicx}
\usepackage{dcolumn}
\usepackage{bm}

\usepackage[utf8]{inputenc}
\usepackage[T1]{fontenc}
\usepackage{mathptmx}
\usepackage{etoolbox}
\usepackage{lipsum}
\usepackage[dvipsnames]{xcolor}
\usepackage{booktabs}
\usepackage{tikz}
\usepackage{comment}
\usepackage{enumitem}

\usepackage{styling}
\usepackage{tikz_commands}

\usepackage{framed}
\usepackage{empheq}
\usepackage{orcidlink}
\usepackage{diagbox}

\allowdisplaybreaks

\newcommand{\LMU}{Arnold Sommerfeld Center for Theoretical Physics, Center for NanoScience, and Munich Center for Quantum Science and Technology, Ludwig-Maximilians-Universität München, 80333 Munich, Germany}

\begin{document}

\title{Testing the parquet equations and the U(1) Ward identity for real-frequency correlation functions from the multipoint numerical renormalization group}

\author{Nepomuk Ritz\,\orcidlink{0009-0006-5173-5201}}
\affiliation{\LMU}

\author{Anxiang Ge\,\orcidlink{0009-0002-6603-4310}}
\affiliation{\LMU}

\author{Markus Frankenbach\,\orcidlink{0009-0002-8907-0031}}
\affiliation{\LMU}

\author{Mathias Pelz\,\orcidlink{0009-0001-3282-6742}}
\affiliation{\LMU}

\author{Jan von Delft\,\orcidlink{0000-0002-8655-0999}}
\affiliation{\LMU}

\author{Fabian B.\,Kugler\,\orcidlink{0000-0002-3108-6607}}
\affiliation{Institute for Theoretical Physics, University of Cologne, 50937 Cologne, Germany}
\affiliation{Center for Computational Quantum Physics, Flatiron Institute, 162 5th Avenue, New York, NY 10010, USA}

\date{\today}

\begin{abstract}
    Recently, it has become possible to compute real-frequency four-point correlation functions of quantum impurity models using a multipoint extension of the numerical renormalization group (mpNRG). In this work, we perform several numerical consistency checks of the output of mpNRG by investigating exact relations between two- and four-point functions.
    This includes the Bethe--Salpeter equations and the Schwinger--Dyson equation from the parquet formalism, which we evaluate in two formally identical but numerically nonequivalent ways. We also study the first-order U(1) Ward identity between the vertex and the self-energy for the first time in full generality in the real-frequency Keldysh formalism. We generally find good agreement of all relations, often up to a few percent, both at weak and at strong interaction.
\end{abstract}

\maketitle

\section{Introduction}

A promising route toward computing dynamical correlation functions of realistic models for correlated electronic systems lies in combining different numerical methods. One example is the idea of using the non-perturbative but local dynamical mean-field theory (DMFT) \cite{Georges1996} as a correlated starting point for subsequent diagrammatic calculations \cite{Rohringer2018}. Recent methodical advancements in the Keldysh formalism (KF) even put real-frequency dynamical correlation functions directly comparable to experiments within reach \cite{Kugler2021, Lee2021, Walter2022, Ge2024, Ritz2024}.

A suitable impurity solver for this purpose is the numerical renormalization group (NRG) \cite{Bulla2008}. In its recent multipoint extension (mpNRG) \cite{Kugler2021, Lee2021}, it can provide both the self-energy and the four-point (4p) vertex of a self-consistently determined DMFT impurity model. These may then be used as a starting point for nonlocal diagrammatic extensions \cite{Rohringer2018}, for example in the form of the dynamical vertex approximation \cite{Toschi2007, Held2008} using the parquet formalism \cite{Bickers2004} or (closely related \cite{Kugler2018a,Kugler2018b,Kugler2018c}) the functional renormalization group \cite{Metzner2012,Taranto2014}.
However, for this to be a reliable strategy, the results from mpNRG must be of sufficient quality, which a priori cannot be taken for granted due to numerical restrictions.

NRG computations converged in all numerical parameters produce numerically exact results for two-point (2p) quantities such as the self-energy in the low-energy regime. 
However, there is a danger of overbroadening at large energies due to the logarithmic bath discretization in NRG.
This may raise doubts as to how well exact relations involving integrations over all frequencies are fulfilled. Furthermore, even though the accuracy of the mpNRG 4p vertex has recently been drastically improved using the symmetric estimator technique \cite{Lihm2024}, numerical restrictions such as a relatively small number of kept states and a correspondingly large discretization parameter still hold.
It is, therefore, of interest to test to what extent the correlation functions produced by mpNRG fulfill exact relations that arise in a quantum field theory description of the many-electron problem.
In addition, the fulfillment of such relations can serve as a guide for future developments of mpNRG.

In this paper, we study a host of exact relations between real-frequency correlation functions. We perform our calculations for the single-impurity Anderson model \cite{Anderson1961}, which arises in DMFT and which NRG is tailored to solve. Along with the basics of the formalism and all employed methods, the model is introduced in Sec.\,\ref{sec:formalism}. We consider two different datasets from (mp)NRG: one at weak and one at strong interaction. In Sec.\,\ref{sec:results}, we first discuss the fulfillment of the Bethe--Salpeter equations (BSEs) and the Schwinger--Dyson equation (SDE) from the parquet formalism. Then, we consider the Ward identity (WI) arising from the local U(1) gauge invariance of the theory. For the first time, we {study} it in full generality in the KF and check its fulfillment in mpNRG. We find that both the parquet equations and the WI are fulfilled rather well, in many components up to a few percent, and comment on larger discrepancies wherever they occur.
Finally, we conclude in Sec.\,\ref{sec:conclusion} and provide details on technicalities in the Appendices \ref{app:BSE-violation}--\ref{app:calculations}.

\section{Formalism}\label{sec:formalism}

Our main objects of interest are real-frequency dynamical 2p and 4p correlation functions in the KF. Their non-trivial contributions which arise from electron-electron interactions are encapsulated in the self-energy $\Sigma$ and the 4p vertex $\Gamma$,
\begin{align}
    \Sigma =
    \tikzm{selfenergy}{
        \selfenergywithlegs{$\Sigma$}{0}{0}{1}
    }\, , \qquad
    \Gamma = 
    \tikzm{vertex}{
        \fullvertexwithlegs{$\Gamma$}{0}{0}{1}
    }
    \, .
\end{align}
The self-energy enters the Dyson equation,
\begin{align}
    G \,
    = \,
    \begin{gathered}
    \tikzm{Keldysh_formalism-Dyson_G}{
        \draw[lineWithArrowCenter] (1,0) -- (0,0);
    }
    \end{gathered}
    \, = \,
    \begin{gathered}
    \vspace{-2.5ex}
    \tikzm{Keldysh_formalism-Dyson_G0}{
        \draw[lineBareWithArrowCenter] (1,0) -- (0,0);
        \node at (0.5,-0.4) {$G_0$};
    }
    \end{gathered}
    \, + \,
    \begin{gathered}
    \vspace{-1.5ex}
    \tikzm{Keldysh_formalism-Dyson_G0SigmaG}{
        \draw[lineBareWithArrowCenter] (1,0) -- (0,0);
        \selfenergy{$\Sigma$}{1.3}{0}{1}
        \draw[lineWithArrowCenter] (2.6,0) -- (1.6,0);
        \node at (0.5,-0.4) {$G_0$};
        \node at (2.1,-0.4) {$G$};
    }
    \end{gathered}
    \, ,
\end{align}
determining the one-particle propagator $G$, where $G_0$ is the non-interacting (``bare'') propagator. From the retarded component of the propagator, the experimentally measurable spectral function is deduced as $A(\nu) = - \mathrm{Im}\, G^R(\nu)/\pi$. The vertex determines the two-particle correlation function $G^{(4)}$,
\begin{align}
    G^{(4)}
    \, =\  
    \tikzm{Keldysh_formalism-G4_dcon1}{
        \draw[lineWithArrowCenter] (0.8,-0.4) -- (0,-0.4);
        \draw[lineWithArrowCenter] (0,0.4) -- (0.8,0.4);
    }
    \ \ -\ \
    \tikzm{Keldysh_formalism-G4_dcon2}{
        \draw[lineWithArrowCenter] (0.8,-0.4) -- (0.8,0.4);
        \draw[lineWithArrowCenter] (0,0.4) -- (0,-0.4);
    }
    \ \ + \ \
    \tikzm{Keldysh_formalism-G4_con}{
        \fullvertex{$\Gamma$}{0}{0}{4./3.};
        \draw[lineWithArrowCenter] (-0.4,-0.4) -- (-1.2,-0.4);
        \draw[lineWithArrowCenter] (1.2,-0.4) -- (0.4,-0.4);
        \draw[lineWithArrowCenter] (0.4,0.4) -- (1.2,0.4);
        \draw[lineWithArrowCenter] (-1.2,0.4) -- (-0.4,0.4);
    }
    \ \ ,
\end{align}
which yields physical susceptibilities upon contracting pairs of external legs. An explicit form of the equations shown here only diagrammatically is provided in App.\,\ref{app:setup}.

\subsection{The (multipoint) numerical renormalization group}\label{sec:NRG}

The NRG is a computational technique to resolve all energy scales of quantum impurity systems in a non-perturbative fashion. Its main idea consists of logarithmically discretizing the energy spectrum of the conduction electrons and iteratively diagonalizing the resulting Hamiltonian. To this end, the discretized Hamiltonian is transformed into a semi-infinite chain with exponentially decreasing hopping amplitudes. This chain Hamiltonian is then solved iteratively by adding one energy shell at a time and diagonalizing the effective Hamiltonian at each step. By systematically keeping only the low-energy states while discarding the high-energy states from each shell, the numerical effort remains manageable. Importantly, the discarded states from each shell can be gathered into a \textsl{complete} set of approximate energy eigenstates \cite{Anders2005}.

Afterward, (multipoint) correlation functions can be computed by convolving analytically known kernel functions with a set of so-called partial spectral functions (PSFs). The latter are obtained from their respective Lehmann representations, using the eigenenergies and (discarded) eigenstates obtained from NRG.

Originally invented by Wilson to solve the Kondo problem \cite{Wilson1975}, the NRG was soon applied to the single-impurity Anderson model \cite{Krishna-murthy1980}. Later, NRG was also used as a DMFT impurity solver, first in the single-orbital context \cite{Bulla1998} and then also for multiorbital models \cite{Pruschke2005,Stadler2015,Stadler2019,Deng2019,Kugler2019,Kugler2020,Stadler2021,Kugler2022a,Kugler2024,Grundner2024,Leehand2024} and most recently in two-site cellular DMFT studies \cite{Gleis2024,Gleis2025}. The recent extension of NRG to multipoint correlation functions (mpNRG) \cite{Kugler2021, Lee2021, Lihm2024} now enables its application to vertex-based extensions of DMFT. 
This work is meant to be a preliminary step toward that goal.

Details on the NRG implementations employed in this work and the numerical parameters chosen can be found in App.\,\ref{sec:parameters}. In the following, it is self-explanatory whether the ``standard'' NRG or its multipoint extension is employed to compute 2p or higher-point functions, respectively. We will therefore not distinguish between the two in the main text.

\subsection{Single-impurity Anderson model}\label{sec:SIAM}

The Hamiltonian of the single-impurity Anderson model \cite{Anderson1961} is
\begin{align}
H &= 
\sum_{\epsilon\sigma} \epsilon c_{\epsilon\sigma}^\dagger c_{\epsilon\sigma}
+ \sum_\sigma \epsilon_d n_\sigma
+ U n_\uparrow n_\downarrow 
+
\sum_{\epsilon\sigma} (V_\epsilon d_\sigma^\dagger c_{\epsilon\sigma} + \mathrm{H.c.})
,
\label{eq:SIAM_Hamiltonian}
\end{align}
where the impurity site is described by a local $d$ level with on-site energy $\epsilon_d$. The $d$ level hybridizes with spinful conduction electrons, created by $c_{\epsilon\sigma}^\dagger$, via matrix elements $V_\epsilon$. Electrons on the impurity site, where $n_\sigma \!=\! d_\sigma^\dag d_\sigma$, interact with interaction strength $U$.
The non-interacting $c$ electrons occur quadratically in the functional integral and can be integrated out, yielding a frequency-dependent retarded hybridization function $\Delta^R(\nu)$ as an additional quadratic term for the $d$ electrons.

We choose the hybridization function as
\begin{align}
    \Delta^R(\nu) = \frac{\Delta}{\pi} \, \ln\left|\frac{\nu + D}{\nu - D}\right| - i\Delta \, \theta(D-|\nu|) ,
\end{align}
with a box-shaped imaginary part of half-bandwidth $D$ and strength $\Delta$. In the often-employed wide-band limit, its real part can be neglected and the hybridization function reduces to a constant, $\Delta^R(\nu) \overset{D\rightarrow\infty}{\longrightarrow} - i\Delta$. 

\subsection{Parquet formalism}\label{sec:parquet}

The parquet formalism \cite{Landau1954, Dominicis1964_1, Dominicis1964_2, Bickers2004} provides exact self-consistent equations for the vertex and the self-energy.
Its starting point is the parquet decomposition, which classifies all diagrammatic contributions to the vertex w.r.t.\,their two-particle reducibility,
\begin{subequations}
\begin{align}
    \Gamma &= R + \sum_{r\in\{a,p,t\}} \gamma_r \label{eq:parquet-decomp} \\
    &= \tikzm{Keldysh_formalism-Gamma0}{
        \barevertexwithlegs{0}{0}
    }
    + 
    \tikzm{mfRG-parquet-SDE0}{
            \abubblebarebarebare{0}{0}{0.85};
		\barevertexlefthalf{0}{0};
		\barevertexrighthalf{1}{0};
            \node at (0.5, 0) {$a$};
		}
    + \tfrac{1}{2}
    \tikzm{mfRG-parquet-SDE0}{
            \pbubblebarebarebare{0}{0}{0.85};
		\barevertex{0}{0};
		\barevertex{1}{0};
		\arrowslefthalfp{0}{0}{0.8};
		\arrowsrighthalfp{1}{0}{0.8};
		\node at (0,0.6){};
            \node at (0.5, 0) {$p$};
		}
    -
    \begin{gathered}
        \tikzm{mfRG-parquet-SDE0}{
            \tbubblebarebarebare{0}{0}{0.8};
		\barevertexupperhalf{0}{-1};
		\barevertexlowerhalf{0}{0};
            \node at (0, -0.5) {$t$};
		}
    \end{gathered}
    + O[(\Gamma_0)^3]\, .
\end{align}
\end{subequations}
Any diagram that contributes to {one of the $\gamma_r$} is two-particle reducible in {the corresponding} two-particle channel $a$, $p$, or $t${, sometimes also referred to as ``crossed particle-hole'' ($\overline{ph}$), ``particle-particle'' ($pp$) and ``particle-hole'' ($ph$) in the literature \cite{Rohringer2013}}. 
%\todo{Add citations to papers that use the latter notation!} 
{All other diagrams contributing to $\Gamma$ are} two-particle irreducible in all three channels and thus part of the fully two-particle irreducible vertex $R$.
The parquet formalism provides self-consistent relations for the reducible vertices $\gamma_r$ in the form of the BSEs,
\begin{align}
    \gamma_r = I_r \circ \Pi_r \circ \Gamma = \Gamma \circ \Pi_r \circ I_r\, . \label{eq:BSE}
\end{align}
Here, $I_r = \Gamma - \gamma_r$, $\Pi_r$ denotes a pair of propagators used to connect two vertices, and the symbol $\circ$ is a short-hand notation for contractions over all quantum numbers as well as frequency integrations.

In addition, self-energy and vertex are related by the SDE, 
\begin{align}
    \Sigma = -\left[\Gamma_0 + \tfrac{1}{2} \Gamma_0 \circ \Pi_r \circ \Gamma \right]\cdot G\, . \label{eq:SDEformula}
\end{align}
Here, the second term can be parametrized w.r.t.\,either of the three two-particle channels, and the symbol $\cdot$ is used to denote the contraction with a single propagator in Eq.\,\eqref{eq:SDEformula}.
Together, the BSEs and the SDE are known as the parquet equations. They are exact relations, which however require the input of the fully irreducible vertex $R$. In a purely diagrammatic treatment, approximations are employed at this stage, the most common being the parquet approximation $R \simeq \Gamma_0$, which only considers the first-order contribution to $R$ from the bare vertex $\Gamma_0$. As this neglects higher-order irreducible diagrams, which start at the fourth order in $\Gamma_0$, the parquet approximation is only justified for weak to intermediate interaction strengths. In this work, the parquet approximation is \textsl{not} employed, as the NRG provides the full vertex non-perturbatively, including its irreducible part.

\subsection{Asymptotic vertex classes}\label{sec:asymptotics}

For efficiently treating the reducible vertices $\gamma_r$, they are decomposed into asymptotic classes as introduced in Ref.\,\onlinecite{Wentzell2020}. This decomposition captures the high-frequency asymptotic behavior of the vertices by separating its diagrammatic contributions into so-called asymptotic classes in each channel $r\in\{a,p,t\}$. These asymptotic classes do or do not depend on one or both the fermionic frequencies in the natural frequency parametrization of the respective two-particle channel, $\gamma_r(\omega_r,\nu_r,\nu'_r) = K_{1,r}(\omega_r) + K_{2,r}(\omega_r,\nu_r) + K_{2',r}(\omega_r,\nu'_r) + K_{3,r}(\omega_r,\nu_r,\nu'_r)$. Diagrams that belong to classes that do not depend on a given frequency will thus give a finite contribution to the vertex in the high-frequency limit. In their remaining arguments, however, they ultimately decay. Formally, the asymptotic classes can be defined as
\begin{subequations}
\begin{align}
    K_{1,r}({\color{rot}\omega_r}) &= \lim_{{\color{blau}\nu_r} \to \infty} \ \lim_{{\color{blau}\nu_r'} \to \infty} \gamma_r({\color{rot}\omega_r},{\color{blau}\nu_r},{\color{blau}\nu'_r}) \\
    K_{2,r}({\color{rot}\omega_r},{\color{rot}\nu_r}) &=  \lim_{{\color{blau}\nu_r'} \to \infty} \gamma_r({\color{rot}\omega_r},{\color{rot}\nu_r},{\color{blau}\nu'_r}) -  K_{1,r}({\color{rot}\omega_r}) \\
    K_{2',r}({\color{rot}\omega_r},{\color{rot}\nu'_r}) &=  \lim_{{\color{blau}\nu_r} \to \infty} \gamma_r({\color{rot}\omega_r},{\color{blau}\nu_r},{\color{rot}\nu'_r}) -  K_{1,r}({\color{rot}\omega_r}) \\
    K_{3,r}({\color{rot}\omega_r},{\color{rot}\nu_r},{\color{rot}\nu'_r}) &= \gamma_r({\color{rot}\omega_r},{\color{rot}\nu_r},{\color{rot}\nu'_r}) - K_{1,r}({\color{rot}\omega_r}) \nonumber \\
    &\quad - K_{2,r}({\color{rot}\omega_r},{\color{rot}\nu_r}) - K_{2',r}({\color{rot}\omega_r},{\color{rot}\nu'_r})\, ,
\end{align}
\end{subequations}
in each channel $r$. They can be visualized graphically as 
\begin{align}
    &\gamma_a (\omega_a, \nu_a, \nu'_a) \
	=
	\tikzm{Keldysh_vertex_parametrization-diagclass-gamma_a}{
		\fullvertexwithlegs{$\gamma_a$}{0}{0}{1}
		\node[left] at (-0.6,0.7) {\small$\nu_a \! + \! \tfrac{\omega_a}{2}$};
		\node[right] at (0.6,0.7) {\small$\nu_a' \! + \! \tfrac{\omega_a}{2}$};
		\node[left] at (-0.6,-0.7) {\small$\nu_a \! - \! \tfrac{\omega_a}{2}$};
		\node[right] at (0.6,-0.7) {\small$\nu_a' \! - \! \tfrac{\omega_a}{2}$};
	} 
	\nonumber \\
	&=
	\tikzm{Keldysh_vertex_parametrization-diagclass-K1a}{
		\arrowslefthalf{0}{0}
		\Konea{$K_{1,a}$}{0}{0}{1}
		\arrowsrighthalf{1.4}{0}
		\draw[lineBareWithArrowEnd,blau] (-0.6,0.3) to [out=315, in=45] (-0.6,-0.3) -- (-0.6006,-0.301);
		\node at (-0.8,0) {{\color{blau} \small$\nu_a$}};
		\draw[lineBareWithArrowEnd,blau] (2,-0.3) to [out=135, in=225] (2,0.3) -- (2.0006,0.301);
		\node at (2.2,0) {{\color{blau} \small$\nu_a'$}};
		\draw[lineBareWithArrowEnd,rot] (-0.15,0.6) to [out=340, in=200] (1.55,0.6);
		\node at (0.7,0.7) {{\color{rot} \small$\omega_a$}};
	}
	 +
	\tikzm{Keldysh_vertex_parametrization-diagclass-K2a}{
		\arrowslefthalffull{0.3}{0}{1}
		\Ktwoa{$K_{2,a}$}{0}{0}{1}
		\arrowsrighthalf{1.2}{0}
		\draw[lineBareWithArrowEnd,rot] (-0.45,0.4) to [out=315, in=45] (-0.45,-0.4) -- (-0.4507,-0.401);
		\node at (-0.65,0) {{\color{rot} \small$\nu_a$}};
		\draw[lineBareWithArrowEnd,blau] (1.8,-0.3) to [out=135, in=225] (1.8,0.3) -- (1.8006,0.301);
		\node at (2,0) {{\color{blau} \small$\nu_a'$}};
		\draw[lineBareWithArrowEnd,rot] (-0.15,0.7) to [out=330, in=190] (1.35,0.5);
		\node at (0.6,0.7) {{\color{rot} \small$\omega_a$}};
	}
	\nonumber \\
	&\quad +
	\tikzm{Keldysh_vertex_parametrization-diagclass-Kb2a}{
		\arrowslefthalf{0}{0}
		\Ktwoab{$K_{2',a}$}{0}{0}{1}
		\arrowsrighthalffull{0.9}{0}{1}
		\draw[lineBareWithArrowEnd,blau] (-0.6,0.3) to [out=315, in=45] (-0.6,-0.3) -- (-0.6006,-0.301);
		\node at (-0.8,0) {{\color{blau} \small$\nu_a$}};
		\draw[lineBareWithArrowEnd,rot] (1.65,-0.4) to [out=135, in=225] (1.65,0.4) -- (1.6507,0.401);
		\node at (1.85,0) {{\color{rot} \small$\nu_a'$}};
		\draw[lineBareWithArrowEnd,rot] (-0.15,0.5) to [out=350, in=210] (1.35,0.7);
		\node at (0.6,0.7) {{\color{rot} \small$\omega_a$}};
	}
	+
	\tikzm{Keldysh_vertex_parametrization-diagclass-K3a}{
		\fullvertexwithlegs{$K_{3,a}$}{0}{0}{1}
		\draw[lineBareWithArrowEnd,rot] (-0.75,0.4) to [out=315, in=45] (-0.75,-0.4) -- (-0.7507,-0.401);
		\node at (-0.95,0) {{\color{rot} \small$\nu_a$}};
		\draw[lineBareWithArrowEnd,rot] (0.75,-0.4) to [out=135, in=225] (0.75,0.4) -- (0.7507,0.401);
		\node at (0.95,0) {{\color{rot} \small$\nu_a'$}};
		\draw[lineBareWithArrowEnd,rot] (-0.4,0.7) to [out=315, in=225] (0.4,0.7) -- (0.401,0.7007);
		\node at (0,0.9) {{\color{rot} \small$\omega_a$}};
	}
\end{align}
for the $a$ channel, and correspondingly for the $p$- and $t$ channels. Note that the symmetric estimator technique in NRG provides the $K_{1,r}$, $K_{2,r}$ and $K_{2',r}$ classes in each channel $r$ separately, but not so for $K_{3}$. Instead, it only gives the sum of the irreducible vertex $R$ and all three $K_{3,r}$ classes, obtained by subtracting the asymptotic contributions from the full vertex. That object is used to define the ``vertex core'',
\begin{subequations}
\begin{align}
    \Gamma_{\mathrm{core}} &= \Gamma - [\Gamma_0 + \sum_{r\in\{a,p,t\}} (K_{1,r} + K_{2,r} + K_{2',r})] \\
    &= R - \Gamma_0 + \sum_{r\in\{a,p,t\}} K_{3,r}\, ,
\end{align}
\end{subequations}
which thus contains all diagrams that genuinely depend on three frequencies and decay in every direction.

As a side note, the vertex can alternatively be parametrized using the single-boson exchange (SBE) decomposition \cite{Krien2019, Krien2019b, Krien2019c, Krien2020, Krien2020b, Krien2021}, which classifies the diagrams according to their interaction reducibility instead of their two-particle reducibility. This formalism naturally exploits the asymptotic behavior of individual classes of diagrams as well. In fact, asymptotic classes can be related to SBE objects and vice-versa \cite{Gievers2022}. We will not employ the SBE decomposition in this work.

\subsection{Keldysh formalism}\label{sec:KF}

The KF \cite{Keldysh1965, Schwinger1961, Kadanoff1962} is an alternative to the widespread Matsubara formalism and is applicable both in and out of thermal equilibrium. In this work, we use it exclusively for equilibrium computations. There, its main advantage over the {Matsubara formalism} lies in the fact that it enables computing  dynamical correlation functions in \textsl{real} time or frequency, whereas the {Matsubara formalism} gives an imaginary frequency description. Obtaining experimentally measurable observables such as the spectral function in the {Matsubara formalism} thus requires analytical continuation, a numerically ill-conditioned problem. 

Working in the KF entails significant practical complications. For instance, since each operator has a contour index, the 4p vertex has $2^4=16$ Keldysh components.
Next, unlike in the finite-temperature {Matsubara formalism}, KF objects have a continuous frequency dependence, which must be discretized in numerical treatments.
This is especially challenging for 4p functions that depend on three independent frequencies (in equilibrium) and limits the accuracy of the computations if done naively. Also, in bubble or loop contractions of diagrams, integrations over frequencies have to be performed instead of simple summations over Matsubara frequencies, which again is numerically much more demanding.

For computing correlation functions with mpNRG, most of these complications become relevant only in later stages of the calculations.
As explained in detail in Ref.\,\onlinecite{Kugler2021}, the actual NRG algorithm is agnostic of the formalism. From the approximate eigenenergies and eigenstates of the impurity model with a discretized bath, one obtains a set of PSFs. These, in turn, can be used to compute correlation functions in any formalism, be it the {Matsubara formalism}, the zero-temperature formalism, or the KF.
To this end, the PSFs are convoluted with a set of formalism-dependent kernel functions; for the KF, the Keldysh index structure enters via the Keldysh kernels. Only at this step, namely the convolutions, does it become necessary to specify a discretized frequency grid, which can be chosen arbitrarily in principle. Numerically, the convolution of the PSFs with the kernel functions is easy to perform. Since the PSFs consist of delta peaks, the frequency integrals reduce to simple sums. Beyond that point, no frequency integrations are necessary within mpNRG. In this work, they enter at a later stage, when the output of mpNRG is used to evaluate the parquet equations or the U(1) WI. Moreover, the vertex is related to the 4p correlator through the amputation of external legs. Naively, this requires divisions of the 4p correlator by 2p propagators, which can become numerically unstable. Using the recently developed symmetric improved estimators \cite{Lihm2024}, the vertex can be computed using only element-wise multiplications and additions.

\section{Results}\label{sec:results}

We consider two separate parameter sets throughout{, both times studying the particle-hole symmetric case}. One is at weak interaction, $u=U/(\pi\Delta)=0.5$, and $T/U=0.01$ ($T/\Delta \approx 0.16$) in the wide-band limit, where $\Delta^R(\nu) \rightarrow -i\Delta$, corresponding to one of the datasets studied in Ref.\,\onlinecite{Ge2024}. For this parameter set, the {parquet approximation} is justified and a self-consistent solution of the parquet equations can easily be obtained. The other one is at strong interaction, at a much lower temperature and for a finite bandwidth, $U/\Delta=5$ ($u \approx 1.59$), $T/\Delta = 0.0025$ and $D/\Delta = 25$, corresponding to one of the datasets studied in Ref.\,\onlinecite{Lihm2024}. Here, the {parquet approximation} is not justified anymore and a self-consistent solution of the parquet equations in the {parquet approximation} with the methods employed in Ref.\,\onlinecite{Ge2024} is out of reach in the KF. Using the standard formula \cite{Tsvelick1983} 
for the Kondo temperature at particle-hole symmetry,
$    T_K  \simeq \sqrt{\tfrac{U\Delta}{2}} \, \exp\left(-\tfrac{\pi U}{8\Delta} + \tfrac{\pi \Delta}{2U}\right)$,
one obtains $T_\mathrm{K}/\Delta \approx 1.3$ and $T_\mathrm{K}/\Delta \approx 0.30$ for the weak- and strong-interaction parameter sets, respectively.

For each parameter set, we performed a ``standard'' NRG calculation for the self-energy and a multipoint NRG calculation for the vertex. The NRG parameters of these calculations are summarized in App.\,\ref{sec:parameters}. The self-energy and vertex obtained this way were then used to evaluate all equations of interest here, utilizing the \textsc{KeldyshQFT} codebase \cite{Ritz_KeldyshQFT}.

All NRG vertex data was generated on a logarithmic frequency grid, $\nu/\Delta \in \{-10^2, \ldots, -10^{-2}, 0, 10^{-2},\ldots, 10^2\}$, with $30$ points per decade, i.e.\,$241$ points per frequency axis. The vertex was computed in the $t$ channel parametrization according to the conventions of Ref.\,\onlinecite{Lihm2024}. This data had to be transferred to the conventions of Ref.\,\onlinecite{Walter2022} (see App.\,\ref{app:conventions} for details), during which not only the frequency parametrizations were adapted, but also the data was interpolated onto the non-linear grids introduced in \cite{Walter2022, Ge2024} and implemented in \cite{Ritz2024, Ritz_KeldyshQFT}. The transferred data was subsequently used to evaluate all equations relevant to this work.

All one-dimensional functions of interest here are either symmetric or antisymmetric in frequency. We hence restrict their plots to positive frequencies using semi-logarithmic axes.
For comparing two dynamical quantities $a(\nu)$ and $b(\nu)$, we use their maximal relative difference, which we define as 
\begin{align}
    \delta^\mathrm{max}_\mathrm{rel}(a,b) \equiv \max_\nu |a(\nu) - b(\nu)| / \max_\nu |b(\nu)|\, .
\end{align}
We normalize w.r.t.\,the maximal absolute value of $b$ across the whole real-frequency axis to avoid an overemphasis on deviations in regions where the functions $a$ and $b$ are small. {Throughout this work, we always assign the output of \mbox{mpNRG} to the quantity $b(\nu)$, while $a(\nu)$ denotes the results obtained from evaluating the BSEs, the SDE, or the WI.}

The results in the main text are shown for a single Keldysh component, since the other components follow from (generalized) fluctuation-dissipation relations (FDRs) in thermal equilibrium. In the case of 2p functions, we focus on the retarded component. For the self-energy, the other non-trivial ``Keldysh'' component obeys the standard fermionic FDR,
\begin{align}
    \Sigma^K(\nu) = 2i \, \tanh(\tfrac{\nu}{2T})\, \mathrm{Im}\, \Sigma^R(\nu)\, . \label{eq:FDT_Sigma}
\end{align}
As explained in detail in Ref.\,\onlinecite{Walter2022}, for the $K_{1,r}$ classes (corresponding to bosonic 2p functions), symmetries and causality reduce the number of naively 16 Keldysh components to only two. These are related via the standard bosonic FDR,
\begin{align}
    K_{1,r}^K(\omega_r) = 2i\, \coth\left(\tfrac{\omega_r}{2T}\right) \, \mathrm{Im}\, K_{1,r}^R(\omega_r)\, , \label{eq:FDT_K1}
\end{align}
where the ``retarded'' $R$ component refers to the $11|21$ component, and the ``Keldysh'' $K$ component refers to the $11|22$ component for the $t$ channel or the $12|12$ component for the $a$ and $p$ channel, respectively.

For 2p functions, we compute only the retarded components with NRG and deduce the Keldysh component, if needed, directly from the FDRs.
Generalized FDRs that relate different Keldysh components of the full three-dimensional vertex in thermal equilibrium have been derived in \cite{Heinz2009, Ge2020, Halbinger2024}. On the 4p level, these were already studied in Ref.\,\onlinecite{Lihm2024} (see Fig.\,19 therein), so we refrain from repeating such an analysis here. We only comment on the generalized FDR for one special Keldysh component of the $K_2$ class in App.\,\ref{app:BSE-violation}, for which the BSE studied in Sec.\,\ref{sec:BSE} is violated comparatively strongly.

\subsection{Bethe--Salpeter equations}\label{sec:BSE}

We begin by testing the fulfillment of the BSEs, considered separately for $K_1$ and $K_2$. Since NRG does not provide the individual $K_3$ classes but only the vertex core, the BSEs for the $K_3$ classes cannot be verified explicitly. Indeed, while a full parquet decomposition of the vertex in the {Matsubara formalism} proceeds by (matrix-) inversion of the BSEs, this has not yet been done in the KF, where the frequency dependence of all functions is continuous. Therefore, it is not possible at this point to study the BSEs for the full $\gamma_r$s in the KF.

\begin{figure*}
    \centering
    \includegraphics[height=.25\textwidth]{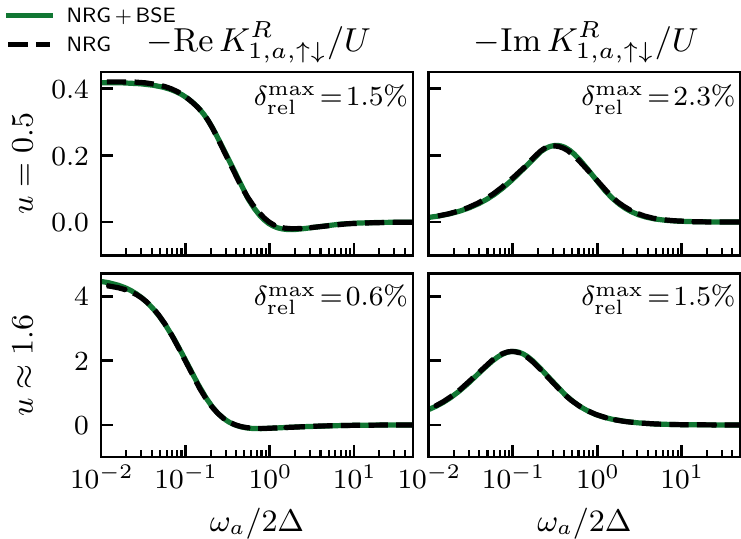}
    \includegraphics[height=.238\textwidth]{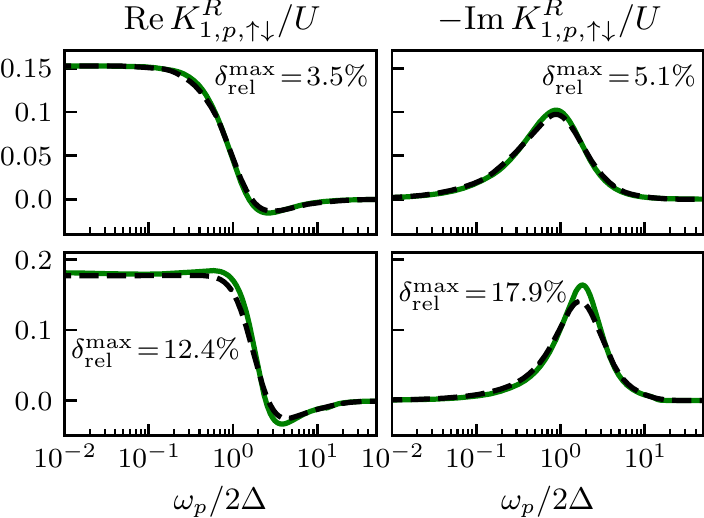}
    \includegraphics[height=.238\textwidth]{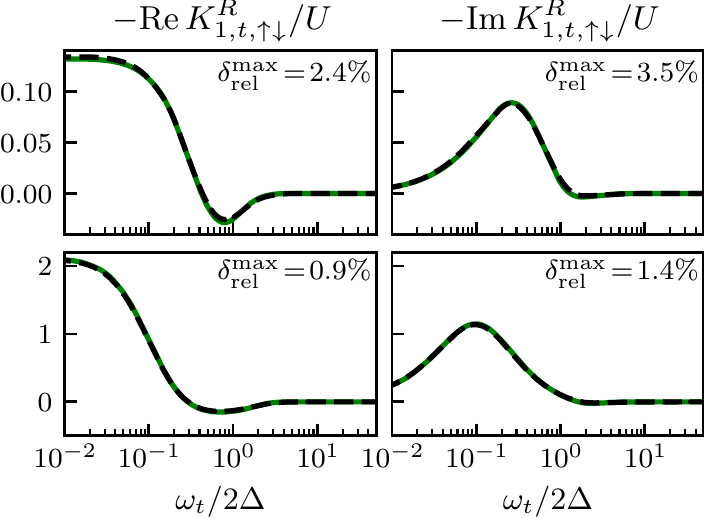}
    \caption{
    Retarded Keldysh component and $\uparrow\downarrow$ spin component of the $K_{1,r}$ vertex classes in the three two-particle channels $r\in\{a,p,t\}$. We compare the result of NRG (black dashed lines) and the result after one evaluation of the BSEs according to Eq.\,\eqref{eq:BSE_K1} (green lines), using $K_{1,r}$ and $K_{2,r}$ as well as the self-energy inside $\Pi_r$ from NRG, using symmetric estimators for all quantities \cite{Kugler2022, Lihm2024}. In this and all following plots, results for two separate data sets are shown: the top panels are at ``weak interaction'', for which $u=U/(\pi\Delta)=0.5$ and $T/U=0.01$ ($T/\Delta \approx 0.16$) in the wide-band limit $D\rightarrow\infty$, the bottom panels at ``strong interaction'' $U/\Delta=5$,$T/\Delta = 0.0025$, $D/\Delta = 25$. All quantities shown here are symmetric or anti-symmetric in frequency, thus the plots are restricted to positive frequencies. Real- and imaginary parts are related by Kramers--Kronig relations, which are enforced in NRG but not in the implementation of the BSEs. Other spin components follow from crossing symmetry. The only other non-trivial Keldysh component is related via the fluctuation-dissipation theorem, Eq.\,\eqref{eq:FDT_K1}. We observe excellent agreement up to a few percent, particularly for the dominating $a$ channel. Especially at strong interaction, the slight deviations in the $p$ channel are negligible, as $K_{1,p}$ is smaller by about one order of magnitude compared to the other channels.
    }
    \label{fig:K1}
\end{figure*}

The BSEs for $K_1$ follow from the limit $\nu_r,\nu'_r \rightarrow \infty$ of Eq.\,\eqref{eq:BSE},
\begin{subequations}
\begin{align}
    K_{1,r} &= \Gamma_0 \circ \Pi_r \circ (\Gamma_0 + K_{1,r} + K_{2,r}) \label{eq:BSE_K1} \\ \nonumber \\
    &=  (\Gamma_0 + K_{1,r} + K_{2',r}) \circ \Pi_r \circ \Gamma_0 \, ,
\end{align}
\end{subequations}
or, diagrammatically,
\begin{align}
    \tikzm{K1a}{
        \arrowslefthalf{0}{0}
        \Konea{$K_{1,a}$}{0}{0}{1.2}
        \arrowsrighthalf{1.68}{0}
	}
 &=
    \tikzm{BSE_K1a}{
        \barevertexlefthalf{0}{0}
        \abubblebarefull{0}{0}{0.85}{1.2}
        \Ktwoawide{\hspace{10pt}$\Gamma_0 + K_{1,a} + K_{2,a}$}{1}{0}{1.2}{2.7}
        \arrowsrighthalf{3.7}{0}
    } \nonumber \\
&=
    \tikzm{BSE_K1a_via_K2p}{
        \arrowslefthalf{0}{0}
        \abubblefullbare{2.7}{0}{0.85}{1.2}
        \Ktwoabwide{\hspace{-5pt}$\Gamma_0 + K_{1,a} + K_{2',a}$}{0}{0}{1.2}{2.7}
        \barevertexrighthalf{3.7}{0}
    } \, 
\end{align}
in the $a$ channel and likewise in the $p$ and $t$ channels. We verified that it makes no difference numerically if $K_{2,r}$ or $K_{2',r}$ is used in the BSEs. Figure\,\ref{fig:K1} shows the fulfillment of the BSEs for $K^R_{1,r,\uparrow\downarrow}$, the retarded Keldysh component of the $\uparrow \downarrow$ spin component in all three two-particle channels. All other spin components are related via crossing and SU(2) spin symmetry. We show both the real and imaginary parts even though, for these retarded functions, they are connected by Kramers--Kronig relations. Indeed, NRG exploits the Kramers--Kronig relations, fulfilling them by construction. However, the implementation of the BSEs does not enforce them explicitly but evaluates real and imaginary parts separately. 

For both parameter sets we observe excellent fulfillment of the BSEs for $K_1$ up to a few percent. The agreement is particularly good for the $a$ channel that dominates already at weak interaction and more so at strong interaction. Since the $a$ channel is related to the $t$ channel by crossing symmetry, it is no surprise that the agreement of the BSE in the $t$ channel is excellent as well. Only in the $p$ channel do the deviations reach about 18\% for the strong-interaction dataset. In particular, the peak in the imaginary part, which lies at larger frequencies compared to the other channels, is not reproduced perfectly. This is to be expected as NRG becomes less accurate at larger frequencies. Still, since $K_{1,p}$ is smaller compared to the other two channels by about one order of magnitude at strong interaction, these deviations are arguably negligible.

\begin{figure*}
    \centering
    \includegraphics[height=.24\textwidth]{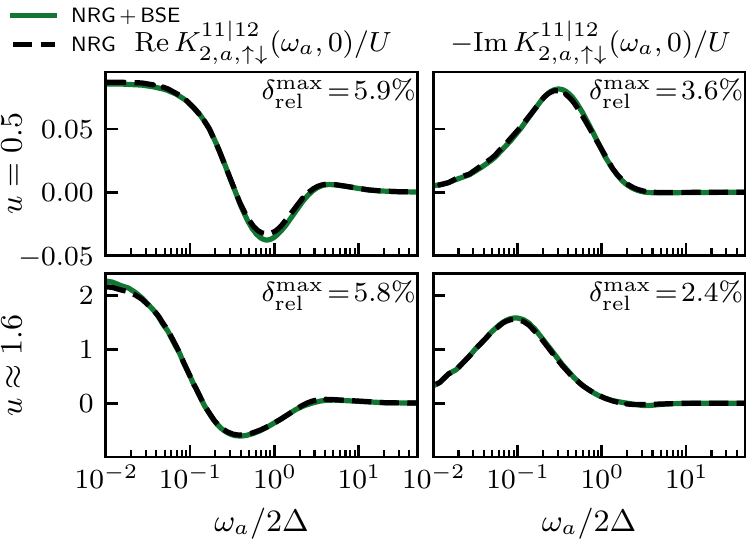}
    \includegraphics[height=.23\textwidth]{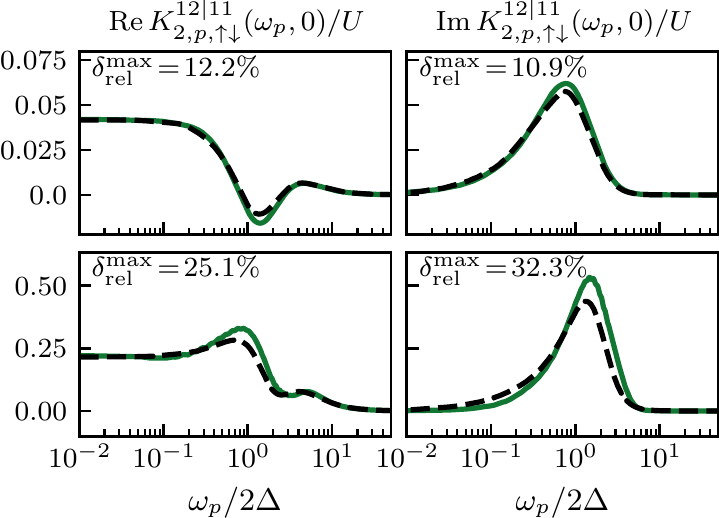}
    \includegraphics[height=.23\textwidth]{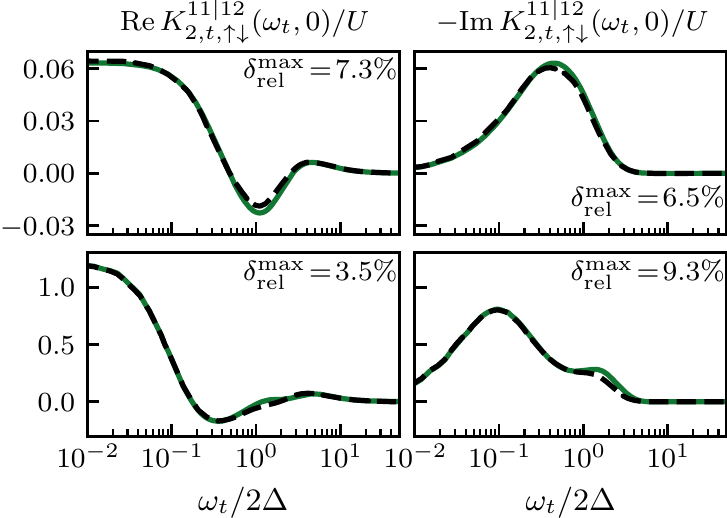}
    \caption{
    Fulfillment of the BSE for the $\uparrow\downarrow$ spin component of $K_{2,r}$ in all channels $r$ at zero fermionic frequency, for the $11|12$ Keldysh component in channels $a,t$ and the $12|11$ Keldysh component in channel $p$. There generally is good agreement up to a few percent. The strongest violations occur at the peak in the imaginary part of the $p$ channel. As for $K_1$, the peaks in the $p$ channel lie at larger frequencies than in the $a$ and $t$ channels. The slight violation of the BSE at those peaks reflects the fact that, due to the logarithmic bath discretization, NRG is less accurate at large frequencies than at small frequencies.
    }
    \label{fig:K2_nu0}
\end{figure*}

\begin{figure*}
    \centering
    \includegraphics[height=.245\textwidth]{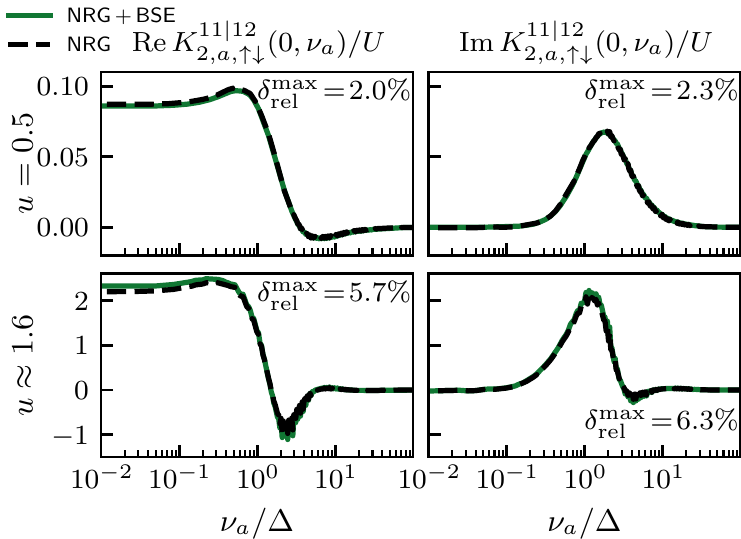}
    \includegraphics[height=.235\textwidth]{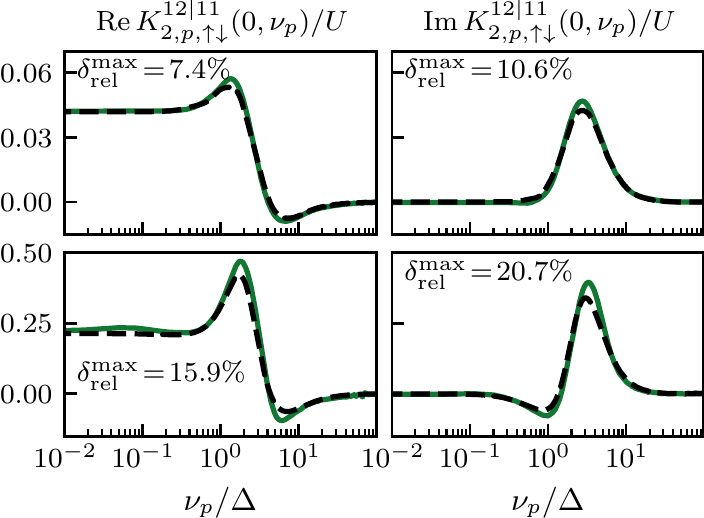}
    \includegraphics[height=.235\textwidth]{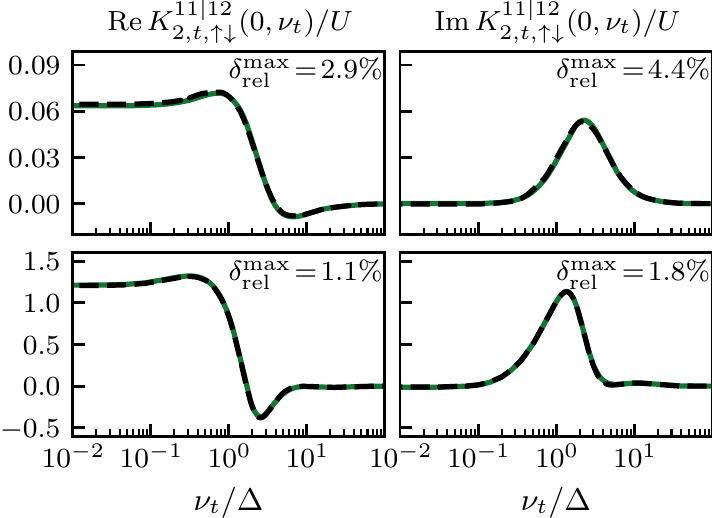}
    \caption{Fulfillment of the BSE for $K_{2,r}$ as in Fig.\,\ref{fig:K2_nu0}, but at zero bosonic frequency. We observe that it is satisfied to a similar degree. 
    The rugged structures at small frequencies, particularly visible in the $a$ channel at strong interaction, can be attributed to minor interpolation errors due to different frequency parametrizations used in NRG and {quantum field theory}. 
    }
    \label{fig:K2_omega0}
\end{figure*}

Taking the limit $\nu_{r'}\rightarrow\infty$ of Eq.\,\eqref{eq:BSE} gives the BSEs for the sum of $K_1$ and $K_2$ in channel $r$,
\begin{align}
    K_{1,r} + K_{2,r} &= \lim_{\nu'_r \rightarrow \infty} \Gamma \circ \Pi_r \circ I_r = \Gamma \circ \Pi_r \circ \Gamma_0\, .
\end{align}
To obtain $K_2$, $K_1$ hence has to be subtracted, which diagrammatically gives
\begin{align}
    \tikzm{K2}{
        \arrowslefthalffull{0.3}{0}{1}
        \Ktwoa{\hspace{0pt}$K_{2,a}$}{0}{0}{1}
        \arrowsrighthalf{1.2}{0}
    }
    &=
    \tikzm{K1plusK2_BSE}{
        \arrowslefthalffull{0}{0}{1}
        \fullvertex{$\Gamma$}{0}{0}{1}
        \abubblefullbare{0.3}{0}{1}{1}{1}
        \barevertexrighthalf{1.5}{0}
    }
    \
    -
    \
    \tikzm{K1a}{
        \arrowslefthalf{0}{0}
        \Konea{$K_{1,a}$}{0}{0}{1.2}
        \arrowsrighthalf{1.68}{0}
	} 
    \label{eq:BSE_K2}
\end{align}
in the $a$ channel, and likewise in the $p$ and $t$ channels. Similarly, taking the limit $\nu_r \rightarrow \infty$ yields the BSEs for $K_{2'}$. As $K_{2'}$ and $K_2$ are related by crossing symmetry, we found equivalent results in both cases up to numerical errors. In Fig.\,\ref{fig:K2_nu0}, we show a one-dimensional slice of the fulfillment of the BSEs for $K_{2,r}$ w.r.t.\,$\omega_r$ at $\nu_r=0$ for the $11|12$ Keldysh component in channels $a,t$ and 
$12|11$ in the $p$ channel. We chose these 
Keldysh components to avoid situations where the data vanish identically. 
Of course, $K_2$ depends on two frequencies independently, and we show another one-dimensional slice of the BSEs w.r.t.\,$\nu_r$ at $\omega_r=0$ in Fig.\,\ref{fig:K2_omega0}. The $K_2$ classes have five Keldysh components that are not related by causality and symmetries, which in thermal equilibrium, however, are again related via (generalized) FDRs. We show the full two-dimensional frequency dependence of all of them in Figs.\,\ref{fig:K2_a}, \ref{fig:K2_p} and \ref{fig:K2_t} in App.\,\ref{app:K2}. 

For the one-dimensional cut through $K_2$ at $\nu_r=0$ in Fig.\,\ref{fig:K2_nu0}, we observe a generally good fulfillment of the BSEs, again up to a few percent in the $a$ and $t$ channels. As for $K_1$ discussed previously, the strongest violations occur in the $p$ channel. Especially in the imaginary parts, the peaks become slightly broader and higher after one evaluation of the BSE. As for $K_1$, these peaks lie at larger frequencies than for the $a$ and $t$ channels. Since NRG is less accurate at large frequencies due to the logarithmic bath discretization, such a discrepancy is, therefore, unsurprising. Improving the NRG computations in this regard requires a convergence analysis in the bath discretization parameter while retaining a sufficient number of kept states. At present, this is one of the main bottlenecks and out of reach for multipoint calculations.

The other one-dimensional cut through $K_2$ in Fig.\,\ref{fig:K2_omega0} for $\omega_r=0$, shows a similar result. However, for strong interaction in the $a$ channel, the data is not entirely smooth. Still, the slightly rugged structures can be argued to be negligible in practice. They can be attributed to the conversions between different frequency parametrizations, see App.\,\ref{app:conventions}.

\begin{figure*}
    \centering
    \includegraphics[height=.245\textwidth]{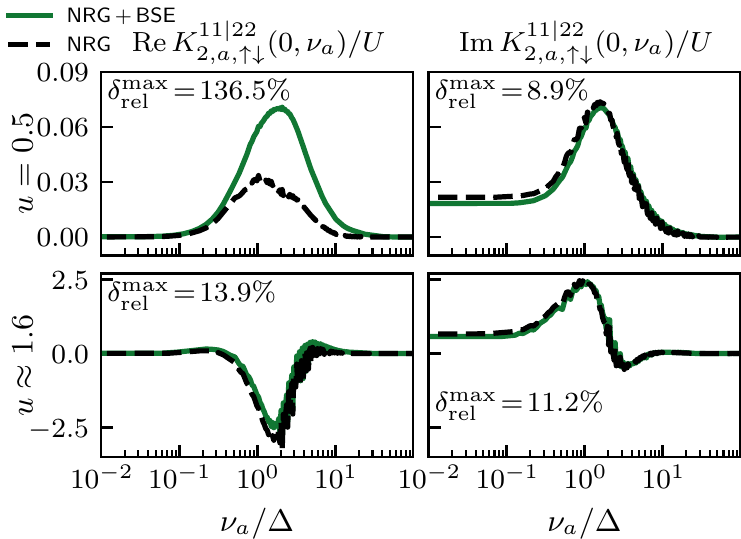}
    \includegraphics[height=.235\textwidth]{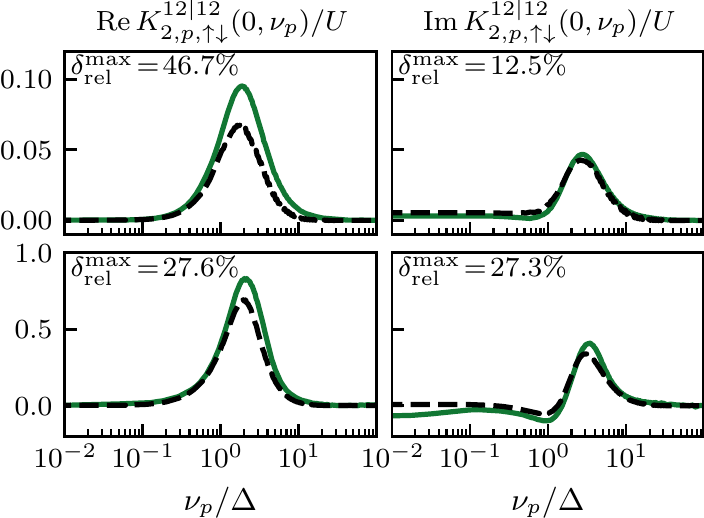}
    \includegraphics[height=.235\textwidth]{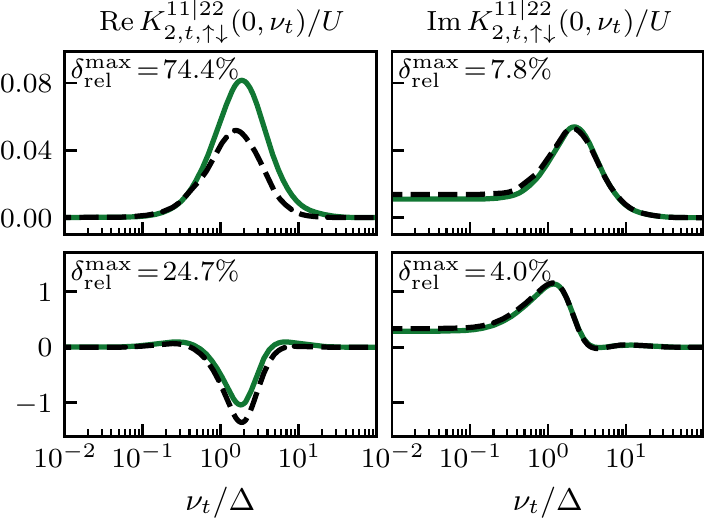}
    \caption{Fulfillment of the BSE as in Fig.\,\ref{fig:K2_omega0}, but showing the Keldysh components $11|22$ in the $a$ and $t$ channels and $12|12$ in the $p$ channel. We observe significant mismatches in the real parts in all three channels, especially at weak interaction.}
    \label{fig:BSE_K2p_Keldysh}
\end{figure*}

Looking closely at the two-dimensional plots for the $K_2$ classes in App.\,\ref{app:K2}, one notices that some Keldysh components fulfill the BSE less accurately than others.  To highlight this fact, we plot another one-dimensional slice of $K_2$ at zero bosonic frequency in Fig.\,\ref{fig:BSE_K2p_Keldysh}, this time for the $11|22$ component in the $a$ and $t$ channels and the $12|12$ component
in the $p$ channel. We observe significant mismatches, especially in the real parts, in all three channels. The different Keldysh components of $K_2$ are related by generalized FDRs. Since the BSE is very well fulfilled for some components but less for others, one could suspect that generalized FDRs are violated by NRG. However, in App.\,\ref{app:BSE-violation} we exemplarily study the generalized FDR for the component shown in Fig.\,\ref{fig:BSE_K2p_Keldysh} and find that it is very well fulfilled. 
In App.~\ref{app:BSE-violation}, we also discuss a symmetry relating $K_{2,p}$ and $K_{2,t}$, observing that it is very well fulfilled, too. We leave it for future work to identify the origin of the discrepancy in the BSE for some Keldysh components of $K_2$. Problems with overbroadening of PSFs at very small bosonic frequencies have previously been observed in mpNRG \cite{lihm_notitle_2025}, which might also account for the current inconsistencies. 

We finally note that, for the strong-interaction dataset, the magnitude of $K_2$ is comparable to $K_1$ shown before, whereas at weaker interaction $K_2$ is much smaller. This shows that the strong-interaction parameters correspond to a regime in which low-order perturbation theory cannot be applied anymore, and evaluating the BSEs thus constitutes a highly non-trivial consistency check of the quality of the NRG data.

\subsection{Schwinger--Dyson equation}\label{sec:SDE}

\begin{figure}
    \centering
    \includegraphics[width=.48\textwidth]{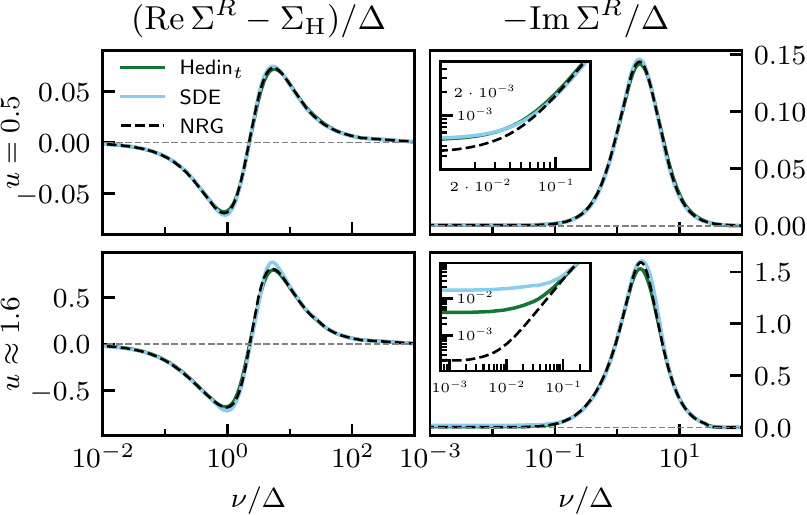}
    \caption{
Real (left) and imaginary (right) parts of the dynamical part of the retarded self-energy. Results for weak and strong interaction are shown in the top and bottom panels, respectively. We compare the results from (i) the ``Hedin$_t$'' version, Eq.\,\eqref{eq:SDE-Hedin}, where a loop is closed directly over the sum of $K_{1,t}$ and $K_{2,t}$, (ii) the ``SDE'' version, Eq.\,\eqref{eq:SDE}, where $\Gamma$ is decomposed into $K_1$, $K_2$, and $K_{2'}$, contracted with $\Gamma_0$ in the respective channel, and the contribution from $\Gamma_\mathrm{core}$ is contracted in the $t$ channel, and (iii) a ``standard'' 2p NRG calculation.
The ``Hedin'' version better captures the peaks at finite frequencies and is more accurate in the limit $\nu\rightarrow 0$ than the ``SDE'' version.
}
    \label{fig:SigmaR}
\end{figure}

\begin{figure}
    \centering
    \includegraphics[width=.48\textwidth]{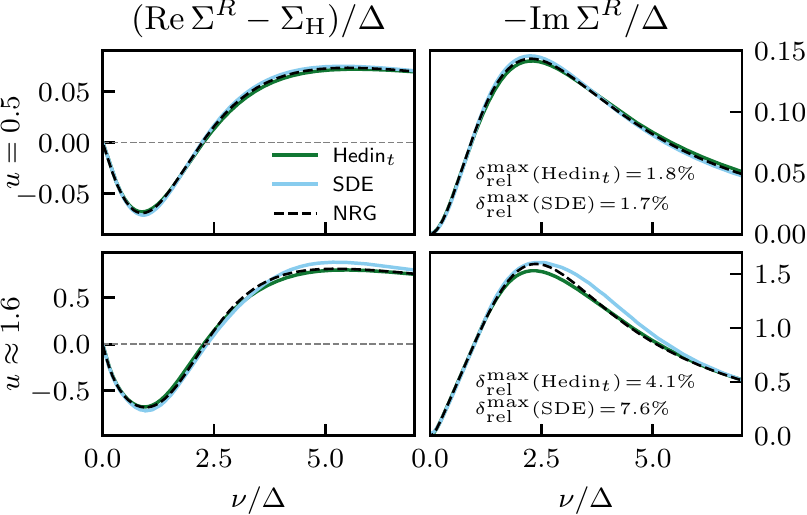}
    \caption{Same as in Fig.\,\ref{fig:SigmaR}, but on a linear frequency scale to make deviations in the high-energy peaks more apparent. Although not entirely perfect, the ``Hedin'' strategy yields more accurate results at strong interaction. This is easily understood since the ``SDE'' version requires $\Gamma_{\mathrm{core}}$, which is more difficult to resolve numerically than $K_1$ and $K_2$ used in the ``Hedin'' version and becomes increasingly more important at large interactions.}
    \label{fig:SigmaR_linear}
\end{figure}

The first term of the SDE \eqref{eq:SDEformula} for the self-energy is {the constant Hartree term{, which here reads} $\Sigma_\mathrm{H} = U/2$}. The second term can be evaluated in multiple ways, and we discuss three formally identical methods in the following. First, one can view the full vertex as a single entity and contract it with the bare vertex in any channel $r$, followed by a loop contraction with $G$. Diagrammatically, this can be visualized as 
\begin{align}\label{eq:SDE}
    \tikzm{selfenergy}{
        \selfenergywithlegs{$\Sigma$}{0}{0}{1}
    }
    =
    -
    \tikzm{mfRG-parquet-SDE0}{
			\barevertex{0}{0}
			\loopbarevertex{lineWithArrowCenter}{0}{0}
			\arrowslowerhalf{0}{0}
			\node at (0,0.45) {};
		}
    -
    \frac{1}{2} \
    \tikzm{mfRG-parquet-SDE_Gamma}{
			\barevertex{0}{0}
			\abubblebarefull{0}{0}{1}{1}
			\fullvertex{$\Gamma$}{1.5}{0}{1}
			\draw[lineWithArrowCenterEnd] (0,0) -- (-0.3,-0.3);
			\draw[lineWithArrowCenterEnd] (2.1,-0.6) -- (1.8,-0.3);
			\draw[lineWithArrowCenter] (1.8,0.3) .. controls ++(45:0.5) and ++(0:0.5) .. (0.9,0.8) .. controls ++(180:0.5) and ++(135:0.5) .. (0,0);
			\node at (0,0.85) {};
		}
    . 
\end{align}
Numerically, this is the least favorable way to evaluate the SDE, as interpolations of $K_{1,r'\neq r}$, $K_{2^{(')},r'\neq r}$ vertex components and $\Gamma_{\mathrm{core}}$ are required to compute the bubble contraction in channel $r$, due to the different native frequency parametrizations in the three channels. Inaccuracies from channel transformations can be reduced by applying the parquet decomposition to the vertex and contracting each reducible vertex $\gamma_r$ with the bare vertex in its native frequency parametrization, closing the missing loop subsequently. We call this strategy simply ``SDE'', and it is {explained and depicted explicitly} in {Eq.\,(D1) and} Fig.\,16 of Ref.\,\onlinecite{Ge2024}{, where it is called ``SDE1''}. Using a vertex from NRG, this method can only be applied to $K_1$, $K_2$, and $K_{2'}$ in each channel, since the vertex core (including $K_3$) is treated as a single entity (which is here parametrized in the $t$ channel, as the original NRG vertex). 

The third way to evaluate the SDE utilizes the BSEs. Contracting the full vertex with the bare vertex in channel $r$ yields $K_{1,r} + K_{2,r}$ in that channel (see, e.g., Eq.\,\eqref{eq:BSE_K2}). 
Assuming fulfillment of the BSEs, one can thus evaluate the SDE by closing a loop over $K_{1,r} + K_{2,r}$ \textsl{directly}, without a prior bubble contraction with the bare vertex. Since $K_{1,r} + K_{2,r}$ is a three-point object, we call this the ``$\mathrm{Hedin}_r$'' strategy \cite{Hedin1965}, depending on the channel $r$ used. Diagrammatically, it can be depicted as
\begin{align}
\Sigma_{\mathrm{Hedin}_a} &=
\tikzm{K1plusK2}{
        \Ktwoawide{\hspace{7pt}$K_{1,a}+K_{2,a}$}{0}{0}{1.2}{1.8}
        \draw[lineWithArrowCenter] (1.8,0) .. controls ++(45:0.5) and ++(0:0.5) .. (0.9,0.8) .. controls ++(180:0.5) and ++(135:0.5) .. (0,0.35);
        \draw[lineWithArrowCenterEnd] (0,-0.35) -- (-0.3,-0.65);
        \draw[lineWithArrowCenterEnd] (2.1,-0.3) -- (1.8,0);
    } \label{eq:SDE-Hedin}
\end{align}
in the $a$ channel, and similarly in the other channels. Here, we will use the ``Hedin$_t$'' version to minimize numerical interpolation errors, as the NRG vertex is paragrametrized in the $t$ channel. Note that it makes no difference whether one uses the sum of $K_{1,r}$ and $K_{2,r}$ or $K_{2',r}$, as both versions are related by crossing symmetry. For a numerically exact result that fulfills the BSEs exactly, all ways of evaluating the SDE should give identical results. However, as seen previously, the NRG vertices satisfy the BSEs only up to a few percent. Furthermore, the vertex core only enters the ``SDE'' version, which, being the only genuinely three-dimensional object, is more difficult to resolve numerically than $K_1$ and $K_2$ used in the ``Hedin'' version. Lastly, the ``SDE'' version requires one evaluation of the BSEs for $K_1 + K_2$, as a contraction with the bare vertex to be computed in the first step. This brings about additional interpolation and integration errors. 

Indeed, while both methods yield almost identical results at weak interaction, we see at strong interaction in Figs.\,\ref{fig:SigmaR} and \ref{fig:SigmaR_linear} that the ``Hedin'' way of evaluating the SDE reproduces the NRG self-energy more accurately than the ``SDE'' strategy:
Although not entirely perfect, its deviations from the 2p computation at the peaks of the real and imaginary parts of the retarded component, most clearly shown in Fig.\,\ref{fig:SigmaR_linear}, are smaller. Furthermore, both methods deviate slightly from the 2p result at very small frequencies. As shown in the insets of Fig.\,\ref{fig:SigmaR}, the asymptotic value of $\mathrm{Im}\, \Sigma^R$ in the limit $\nu\rightarrow 0$ is more accurate for the ``Hedin'' result, at least for the strong-interaction dataset. We conclude that the NRG vertex fulfills the SDE well, especially if evaluated with the ``Hedin'' version. Whether this observation carries over to lattice problems, where the self-energy has an additional momentum dependence and cannot be computed with NRG alone, remains to be studied. Indeed, the two strategies for evaluating the SDE might require different ways of treating the momentum dependence; see, e.g., a recent study using the SBE formalism and a truncated-unity approach for the momenta \cite{Patricolo2025}.

\subsection{U(1) Ward identity}\label{sec:WI}

Finally, we discuss the first-order U(1) Ward identity (WI), which is an exact relation between the self-energy and the vertex. It arises from a local U(1) gauge invariance of the action and all correlation functions. This implies a local continuity equation for the density operator \cite{Krien2018}.
For electronic models such as the Anderson or Hubbard models, the U(1) WI has been extensively studied in the {Matsubara formalism} \cite{Katanin2004, Kopietz2010, Krien2017, Krien2018, Chalupa2022}. In the KF, however, so far only its dependence on a single frequency argument for the special case of vanishing transfer frequency was investigated {explicitly} \cite{Heyder2017, Walter2022, Oguri2022}. Here, we study a more general ``two-dimensional'' version (depending on two independent frequencies) of this WI in the KF. Using frequency conservation, spin conservation, and spin-flip symmetry (the latter two following from SU(2) spin symmetry), it reads 
\begin{align}
 &\Sigma^{\overline{\alpha}_{1'}|\alpha_{1}}(\nu_+) - \Sigma^{\alpha_{1'}|\overline{\alpha}_{1}}(\nu_-) \nonumber \\
&\overset{!}{=}\hspace{-2ex}  \sum_{\substack{ \alpha_{2'} \alpha_{2} \alpha_{\tilde{2}}}} \!\int\! \frac{d\tilde{\nu}}{2\pi i}\! \bigg\{ \!\omega \, G^{\alpha_{\tilde{2}}|\alpha_{2'}}(\tilde{\nu}_+) \Gamma_{D}^{{\alpha}_{2'}\alpha_{1'}|{\alpha}_{2}\alpha_{1}}(\omega, \nu, \tilde{\nu}) G^{\alpha_{2}|\alpha_{\tilde{2}}}(\tilde{\nu}_-) \nonumber \\
&+ \sum_{\alpha_{\tilde{1}}}  \Big[ \Delta^{\overline{\alpha}_{\tilde{2}}|{\alpha}_{\tilde{1}}}(\tilde{\nu}_+) G^{\alpha_{\tilde{1}}|\alpha_{2'}}(\tilde{\nu}_+) \Gamma_{D}^{{\alpha}_{2'}\alpha_{1'}|{\alpha}_{2}\alpha_{1}}(\omega, \nu, \tilde{\nu}) G^{\alpha_{2}|\alpha_{\tilde{2}}}(\tilde{\nu}_-)  \nonumber \\
&-  G^{\alpha_{\tilde{2}}|\alpha_{2'}}(\tilde{\nu}_+)  \Gamma_{D}^{{\alpha}_{2'}\alpha_{1'}|{\alpha}_{2}\alpha_{1}}(\omega, \nu, \tilde{\nu}) G^{\alpha_{2}|\alpha_{\tilde{1}}}(\tilde{\nu}_-) \Delta^{{\alpha}_{\tilde{1}}|\overline{\alpha}_{\tilde{2}}}(\tilde{\nu}_-) \Big] \bigg\} \, , \label{eq:WI_main}
\end{align}
where $\Gamma_{D} = \Gamma_{t,\uparrow\uparrow}+\Gamma_{t,\uparrow\downarrow}$ and we defined the short-hand notation $\nu_{\pm} = \nu \pm\frac{\omega}{2}$ (and, likewise, for $\tilde{\nu}$).
A bar over a Keldysh index means that this index is flipped ($\bar{1}=2;\ \bar{2}=1$). We provide a detailed derivation of Eq.\,\eqref{eq:WI_main} in App.\,\ref{app:WI} and the appendices referenced therein. Let us note that Eq.\,\eqref{eq:WI_main} is not restricted to thermal equilibrium but holds in the non-equilibrium steady-state as well. For explicitly time-dependent problems, the more general form, Eq.\,\eqref{eq:WI_sigma_gamma}, also derived in App.\,\ref{sec:WI}, should be used. Let us also note that there is no contribution to Eq.\,\eqref{eq:WI_main} to first order in the bare interaction $\Gamma_0$: For the self-energies on the LHS, the first-order contribution comes simply from the constant Hartree term and vanishes upon taking the difference. Consequently, the first-order contribution to the RHS must vanish, too. This is easily verified by replacing $\Gamma\rightarrow\Gamma_0$ and $G\rightarrow G_0$ and performing the integral (which can be done analytically). Therefore, the WI provides a non-trivial consistency check for the higher-order dynamical parts of $\Gamma$.

Note that another WI follows from SU(2) spin symmetry. It is almost identical to Eq.\,\eqref{eq:WI_main}, the only difference being that, instead of $\Gamma_{D}$, the $\Gamma_{M} = \Gamma_{t,\uparrow\uparrow}-\Gamma_{t,\uparrow\downarrow}$ component is required on the RHS. For more details 
on the SU(2) WI, see App.~\ref{app:spin}.

{Finally, let us point out that the general two-dimensional form of the U(1) WI, Eq.\,\eqref{eq:WI_main}, connecting $\Sigma$ and $\Gamma$, can also be obtained by combining a Ward identity for the three-point vertex with its relation to the four-point vertex $\Gamma$, as derived in Eqs.\,(104) and (109) of Ref.\,\onlinecite{Oguri2022}. That 
work did not study the two-dimensional form of the U(1) WI further, however, but instead went on to investigate the one-dimensional limit $\omega=0$, similarly to Refs.\,\onlinecite{Heyder2017} and \onlinecite{Walter2022}.}

We now restrict ourselves to  $\alpha_{1'} = \alpha_{1} = 2$ and consider two one-dimensional limits: First, as shown in App.\,\ref{app:Heyder}, in the wide-band limit and for $\omega = 0$, one recovers the special form of the WI studied in Refs.\,\onlinecite{Walter2022} and \onlinecite{Heyder2017},
\begin{align}
    &-2\, \mathrm{Im} \Sigma^{R}(\nu) = \frac{\Delta}{i\pi} \int \mathrm{d}{\tilde{\nu}}\, G^R(\tilde{\nu}) G^A(\tilde{\nu}) \Big\{ \Gamma^{12|21}_{\overline{\uparrow\downarrow} + \uparrow\uparrow}(\tilde{\nu},\nu|\nu,\tilde{\nu}) \nonumber \\
    &\quad - 
    \tanh(\tfrac{\tilde{\nu}}{2T})
    \left[ \Gamma^{12|22}_{\overline{\uparrow\downarrow} + \uparrow\uparrow}(\tilde{\nu},\nu|\nu, \tilde{\nu}) - \Gamma^{22|21}_{\overline{\uparrow\downarrow} + \uparrow\uparrow}(\tilde{\nu},\nu|\nu,\tilde{\nu})\right] \Big\}\, . 
    \label{eq:WI_1D}
\end{align} 
Note that $\Sigma$ could generally retain an additional anomalous contribution coming from the RHS of Eq.\,~\eqref{eq:WI_main} in the limit $\omega\rightarrow 0$ if the vertex behaves like $1/\omega$. Since the vertex of the Anderson impurity model is continuous and non-singular, we neglect this part here. 

Second, for the case of particle-hole symmetry, one obtains another equation for the imaginary part of $\Sigma^R$ from the other one-dimensional limit $\nu=0$: Using $\Sigma^R(\nu) - \Sigma_{\mathrm{H}} = -[\Sigma^A(-\nu) - \Sigma_{\mathrm{H}}]$ at particle-hole symmetry, its LHS becomes
\begin{align}
    \Sigma^R(\tfrac{\omega}{2}) - \Sigma^A(-\tfrac{\omega}{2}) = 2 [\Sigma^R(\tfrac{\omega}{2}) - \Sigma_{\mathrm{H}}] 
\, .
 \label{eq:WI_LHS}
\end{align}

For completeness, we list all four Keldysh components of the LHS of Eq.\,\eqref{eq:WI_main} for the special cases $\omega=0$ in the top part of Tab.\,\ref{tab:WI_LHS} and $\nu=0$ together with particle-hole symmetry in the bottom part of Tab.\,\ref{tab:WI_LHS}. Since all components are related either via complex conjugation or via the FDR, Eq.\,\eqref{eq:FDT_Sigma}, we focus on only one component, $\alpha_{1'} = \alpha_{1} = 2$.

\begin{table}
    \centering
    \renewcommand\arraystretch{1.25}
    \begin{tabular}{c|>{\centering\arraybackslash}p{2.2 cm} >{\centering\arraybackslash}p{2.2 cm}}
        \diagbox[width=\dimexpr \textwidth/24+2\tabcolsep\relax, height=0.75cm]{$\alpha_{1'}$}{$\alpha_1$} & 1 & 2 \\ \hline
         1 & $-2i \, \mathrm{Im}\, \Sigma^R(\nu)$ & $-\Sigma^K(\nu)$ \\
         2 & $\Sigma^K(\nu)$ & $2i\, \mathrm{Im}\, \Sigma^R(\nu)$ 
    \end{tabular}
    \begin{tabular}{c|>{\centering\arraybackslash}p{2.2 cm} >{\centering\arraybackslash}p{2.2 cm}}
        \diagbox[width=\dimexpr \textwidth/24+2\tabcolsep\relax, height=0.75cm]{$\alpha_{1'}$}{$\alpha_1$} & 1 & 2 \\ \hline
         1 & $2 [\Sigma^A(\tfrac{\omega}{2}) - \Sigma_\mathrm{H}]$ & $\Sigma^K(\tfrac{\omega}{2})$ \\
         2 & $\Sigma^K(\tfrac{\omega}{2})$ & $2 [\Sigma^R(\tfrac{\omega}{2}) - \Sigma_\mathrm{H}]$ 
    \end{tabular}
    \caption{Top: LHS of Eq.\,\eqref{eq:WI_main} for $\omega=0$. Bottom: LHS of Eq.\,\eqref{eq:WI_main} for $\nu=0$ and particle-hole symmetry.}
    \label{tab:WI_LHS}
\end{table}

\begin{figure}
    \centering
    \includegraphics[width=\linewidth]{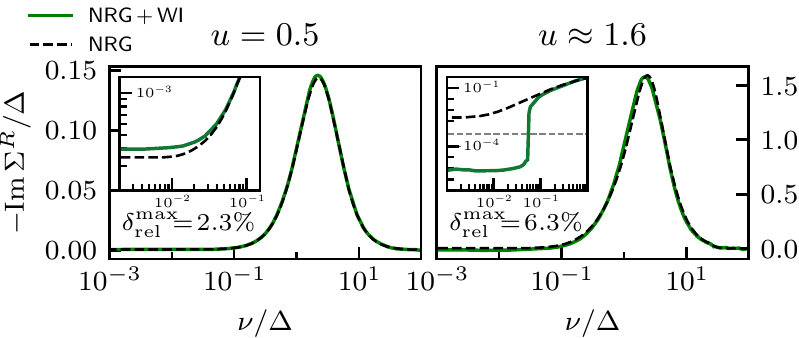}
    \caption{
    Fulfillment of the U(1) WI \eqref{eq:WI_1D} for $\omega=0$ and $\alpha_{1'} = \alpha_{1} = 2$. We observe excellent fulfillment, especially at weak interaction. Only at strong interaction and very small frequencies, $-\mathrm{Im}\, \Sigma^R$ reaches unphysical negative values.}
    \label{fig:1D_WI_v}
\end{figure}

\begin{figure}
    \centering
    \includegraphics[width=\linewidth]{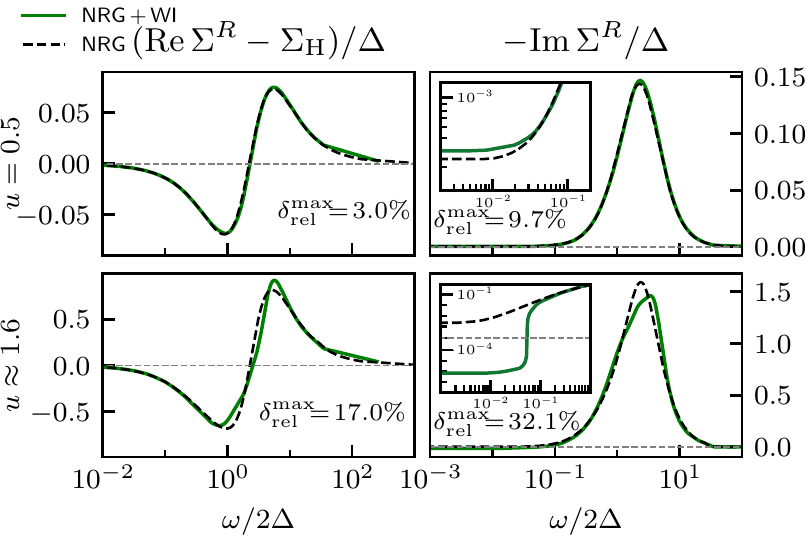}
    \caption{
    Fulfillment of the U(1) WI \eqref{eq:WI_LHS} valid at particle-hole symmetry at $\nu=0$ for $\alpha_{1'} = \alpha_{1} = 2$. The real and imaginary parts are shown on the left and right, weak and strong interactions at the top and bottom, respectively. We observe good fulfillment of the WI, especially at weak interaction. At strong interaction, the peaks of both $\mathrm{Re}\, \Sigma^R$ and $\mathrm{Im}\, \Sigma^R$ are not reproduced very accurately by the WI, and $-\mathrm{Im}\, \Sigma^R$ again shows unphysical negative values at small frequencies.}
    \label{fig:1D_WI}
\end{figure}

We first test the WI for $\omega=0$, which yields the imaginary part of $\Sigma$, see Eq.\,\eqref{eq:WI_1D} and Tab.\,\ref{tab:WI_LHS}. In Fig.\,\ref{fig:1D_WI_v}, we observe excellent fulfillment of the WI, especially at weak interaction. Only at strong interaction, $-\mathrm{Im}\, \Sigma^R$ reaches unphysical negative values at small frequencies, albeit of rather small magnitude. In NRG, the correct sign of $\mathrm{Im}\, \Sigma^R$ is enforced by the symmetric improved estimator \cite{Kugler2022}.

Next, we investigate the $\nu=0$ limit of the WI, which gives both real and imaginary parts of $\Sigma$. Figure\,\ref{fig:1D_WI} shows good fulfillment of the WI throughout, especially at weak interaction. At strong interaction, the peaks in both $\mathrm{Re}\, \Sigma^R$ and $\mathrm{Im}\, \Sigma^R$ are captured less accurately and $-\mathrm{Im}\, \Sigma^R$ again becomes negative at small frequencies. The inaccuracies in the peaks probably stem from the first term on the RHS of Eq.\,\eqref{eq:WI_main}, involving a factor $\omega$ which might exacerbate the numerical inaccuracies of the NRG vertex at large $\omega$. By contrast, in the other one-dimensional limit $\omega=0$, this term is zero.

\begin{figure*}
    \centering
    \includegraphics[height=.35\linewidth]{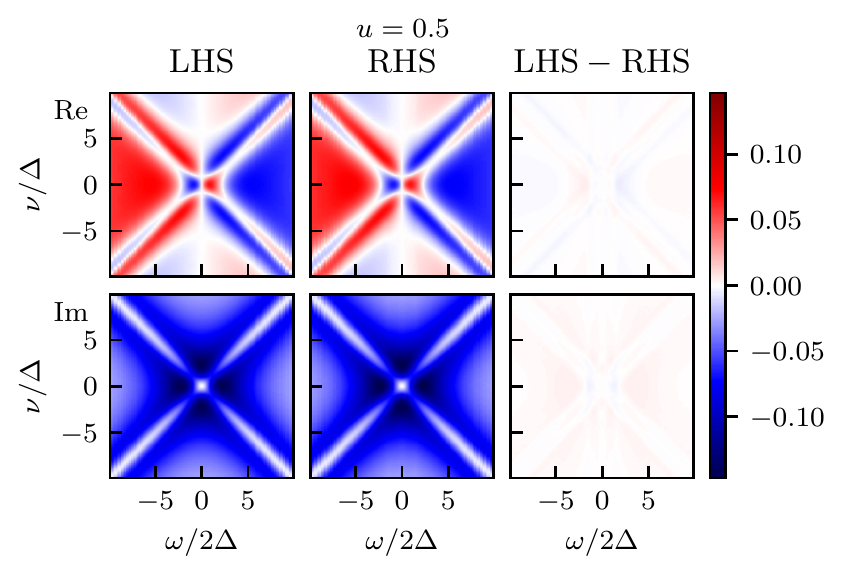}\vline
    \includegraphics[height=.35\linewidth]{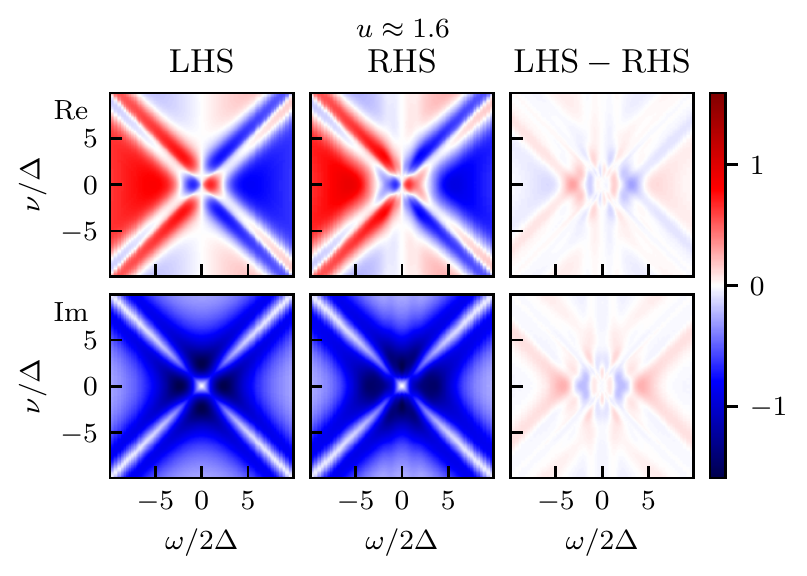}
    \caption{Generalized U(1) WI \eqref{eq:WI_main} for $\alpha_{1'}=\alpha_1=2$, across its full two-dimensional real-frequency dependence for both weak and strong interaction. The qualitative fulfillment of the WI is excellent throughout.}
    \label{fig:2D_WI_weak}
\end{figure*}

The full two-dimensional frequency dependence of the generalized WI, one of the main results of this work, is plotted in Fig.\,\ref{fig:2D_WI_weak}. There, we see once more that the qualitative fulfillment of the WI is excellent throughout. Quantitatively, the largest deviations occur along the one-dimensional cuts at $\nu=0$, shown already in Fig.\,\ref{fig:1D_WI}.

\section{Conclusion}\label{sec:conclusion}

In this paper, we scrutinized the real-frequency 4p vertex of the single-impurity Anderson model as computed by NRG. 
We performed numerical consistency checks for the 2p self-energy and the 4p vertex based on the parquet equations and the generalized U(1) WI. The latter was derived {and}, for the first time, {studied} in full generality in the KF. We investigated two data sets: One at weak interaction, where perturbative approaches like the parquet approximation are justified, and one in a non-perturbative regime at strong interaction. We generally found good agreement throughout, often up to a few percent. Only in a small number of cases did major discrepancies, worth addressing in the future, appear. Some underestimated peaks in a few Keldysh components of $K_2$ suggest that the multipoint NRG calculations might not have been converged in all numerical parameters.

We tested two numerically nonequivalent ways of evaluating the SDE for the self-energy and found that it is fulfilled well both times, but especially using the ``Hedin'' strategy, where the $K_1$ and $K_2$ classes of the vertex are used directly. This is because the more naive evaluation of the SDE includes the vertex core and requires an intermediate contraction with a bare vertex, which introduces additional numerical errors. In the final part of the paper, we observed that the generalized WI is fulfilled well for both datasets. Only at strong interaction, minor deviations appeared, particularly in the imaginary part at small frequencies. 

The very good fulfillment of the {quantum field theory} equations studied in this work in our view encourages the use of the NRG vertex and self-energy as a starting point for a non-local diagrammatic extension of DMFT for lattice problems. To this end, several further steps need to be taken. First, the computation of correlation functions such as the vertex from PSFs should be significantly accelerated: Using quantics tensor cross interpolation (QTCI) \cite{Shinaoka2022, Fernandez2022, Ritter2023, Fernandez2024}, an exponentially fine resolution for the vertex can be afforded at linear cost, provided the vertex is compressible. Indeed, in a recent proof-of-principle study in the {Matsubara formalism}, the parquet equations for the single-impurity Anderson model were solved entirely in the QTCI framework \cite{Rohshap2024}. First numerical experiments indicate that the vertex is compressible even in the KF, at least up to the percent level. Furthermore, the computation of the vertex from PSFs can be formulated and carried out entirely in the QTCI language, thereby significantly reducing the required numerical costs. An efficient implementation of this procedure is underway \cite{Frankenbach2025}. Second, including additional momentum dependencies of correlation functions in the KF has so far not been feasible due to the additional numerical cost and, especially, the memory demand. Again, the QTCI framework promises a solution to that problem, as it can be generalized to functions that depend on arbitrarily many multidimensional variables. 

Third, to enable calculations for experimentally studied correlated materials, the formalism and numerical codes must be generalized to multi-orbital models. Here, NRG quickly encounters a fundamental barrier, as the numerical effort of NRG computations for multi-orbital models increases exponentially in the number of orbitals. At the time of this writing, standard NRG calculations are limited to four orbitals and multipoint NRG is limited to at most two orbitals. One could try using a different method than NRG for computing the local self-energy and vertex. A promising candidate currently being developed is a ``tangent-space Krylov solver'' \cite{Kovalska2025}, a tensor-network technique that iteratively generates dynamical contributions on top of a ground state produced by the density matrix renormalization group \cite{White1992}. First numerical experiments show that this approach can be straightforwardly applied to multi-orbital models. Furthermore, it promises to be 
more accurate than NRG at large frequencies, since it does not rely on logarithmic discretization \cite{Picoli2025}.
However, this approach has not yet been generalized to finite temperature and, most importantly, to multipoint functions. 

Regarding the WI, for future perturbative diagrammatic calculations which employ, e.g., the parquet approximation, one might think of replacing the SDE of the parquet formalism with the WI. For instance, the one-dimensional special case, Eq.\,\eqref{eq:WI_1D}, could be used to compute the imaginary part of the retarded self-energy from the vertex. Using the Kramers--Kronig relation and the FDR, all components of $\Sigma$ follow from that result. At the cost of possibly violating the SDE, the U(1) local gauge invariance implying fulfillment of the local continuity equation for the density operator would then be granted on the 2p and 4p level, which is not given in the standard parquet approximation with the SDE. Especially in the context of non-equilibrium calculations in the KF, where charge conservation is essential, this might prove useful.

\section*{Data and Code availability}

NRG computations were performed with the MuNRG package \cite{Lee2016, Lee2017,Lee2021} based on the QSpace tensor library \cite{Weichselbaum2012, Weichselbaum2012b, Weichselbaum2020, Weichselbaum2024}. The latest version of QSpace is available \cite{Weichselbaum2024b}, and a public release of MuNRG is intended. The code used for the evaluation of the parquet equations and the WI is an extension of the \textsc{KeldyshQFT} package and can be found on GitHub, see Ref.\,\onlinecite{Ritz_KeldyshQFT}. The raw data, data analysis, and plotting scripts can be found in Ref.\,\onlinecite{data}.

\section*{Acknowledgments}

We thank Jeongmin Shim, Jae-Mo Lihm and Seung-Sup Lee for generating and sharing some of the PSF data used for the vertex computations and Jae-Mo Lihm and Seung-Sup Lee for insightful comments on the manuscript.
The authors gratefully acknowledge the computational resources given by grant INST 86/1885-1 FUGG of the German Research Foundation (DFG) and the Gauss Centre for Supercomputing e.V. for funding this project by providing computing time on the GCS Supercomputer SuperMUC-NG at the Leibniz Supercomputing Centre in Munich.
NR acknowledges funding from a graduate scholarship from the German Academic Scholarship Foundation (``Studienstiftung des deutschen Volkes'') and additional support from the ``Marianne-Plehn-Programm'' of the state of Bavaria. NR, AG, MF, MP, and JvD were supported by the Deutsche Forschungsgemeinschaft under Germany’s Excellence Strategy EXC-2111 (Project No.~390814868), and through the project LE3883/2-2, and by the Munich Quantum Valley, supported by the Bavarian state government with funds from the Hightech Agenda Bayern Plus. 
NR, AG, MF, and MP acknowledge support from the International Max Planck Research School for Quantum Science and Technology (IMPRS-QST). FBK acknowledges funding from the Ministry of Culture and Science of the German State of North Rhine-Westphalia (NRW R\"uckkehrprogramm). The Flatiron Institute is a division of the Simons Foundation.

\appendix

\section*{APPENDICES}

In App.\,\ref{app:BSE-violation}, we comment on the surprisingly large violation of the BSE for $K_{2,p}$, observed in Fig.\,\ref{fig:K2_omega0} at weak interaction.
The following appendices provide details on many technical aspects:
In App.\,\ref{sec:parameters}, we specify the numerical parameters chosen for the self-energy and vertex computations with NRG. In App.\,\ref{app:conventions}, we summarize the differences between the conventions used in the mpNRG and \textsc{KeldyshQFT} codes and explain how to convert the vertex from one convention to the other. In App.~\ref{app:BSE_asymp}, we take a closer look at the BSE at very large frequencies and show that inaccuracies due to the finite size of the frequency grid are minimal. In App.\,\ref{app:K2}, we show the full frequency dependence of all non-trivial Keldysh components of $K_2$ and their BSEs, which were omitted in the main text. In App.\,\ref{app:WI}, we derive the generalized WI in the KF studied in the main text. Finally, the subsections of App.\,\ref{app:calculations} detail several, in part lengthy calculations required for the preceding sections.

\section{Comment on the violation of the BSE for $K_2$}\label{app:BSE-violation}

\begin{figure}
    \centering
    \includegraphics[width=\linewidth]{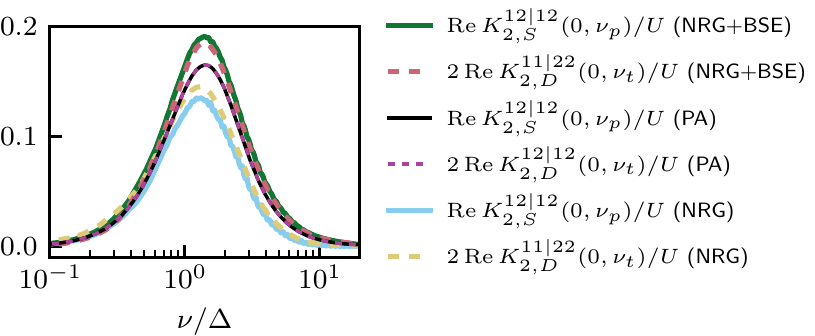}
    \caption{
    Both sides of Eq.\,\eqref{eq:symmetry_K2} for the NRG vertex and for the result after a single evaluation of the BSE at weak interaction, $u=0.5$. Like in Fig.\,\ref{fig:K2_omega0}, the BSE is clearly violated. However, the symmetry relation~\eqref{eq:symmetry_K2} is fulfilled very well by NRG and, by extension, also for the NRG+BSE result. We conclude that the discrepancy is not due to an inconsistency of the NRG vertex on the level of the symmetry, Eq.\,\eqref{eq:symmetry_K2}. Instead, it seems that some Keldysh components of the $K_2$ vertex suffer from inaccuracies in NRG, given the numerical settings summarized in App.\,\ref{sec:parameters}.
    This suspicion is supported by the deviation of NRG from an independently obtained solution of the parquet equations in the {parquet approximation}, shown as a black dotted line, which are unexpectedly large at the weak interaction $u=0.5$.
    }
    \label{fig:K2-symmetry}
\end{figure}

In Fig.\,\ref{fig:BSE_K2p_Keldysh}, we observed a surprisingly large mismatch in the height of the peak of the real part in all channels at weak interaction. Here, we take a closer look at this discrepancy and perform two more consistency checks: First, as derived in Ref.\,\onlinecite{Rohringer2013}, particle-hole symmetry and SU(2) spin symmetry can be exploited to relate certain spin components of the 2PR vertices in the $p$ and $t$ channels. Using the notation 
$\Gamma_{\uparrow\downarrow|\uparrow\downarrow} \equiv \Gamma_{\uparrow\downarrow}$, $\Gamma_{\uparrow\downarrow|\downarrow\uparrow} \equiv \Gamma_{\overline{\uparrow\downarrow}}$, $\Gamma_{\uparrow\uparrow|\uparrow\uparrow} \equiv \Gamma_{\uparrow\uparrow}$, we define the commonly used ``singlet'', ``triplet'', ``magnetic'' and ``density'' spin components as
\begin{subequations}\label{eq:spin-components}
\begin{align}
    S/T &=\, \uparrow\downarrow \mp\, \overline{\uparrow\downarrow}\\
    M/D &=\, \uparrow \uparrow \mp\, \uparrow\downarrow
\end{align}
\end{subequations}
By particle-hole, SU(2) spin and crossing symmetry, the full vertex $\Gamma$  fulfills the relation (see Eq.~(2.135) in \cite{Rohringer2013})
\begin{align}
    \Gamma^{\uparrow\downarrow}_{1'2'|12}
    =
    \Gamma^{\uparrow\uparrow}_{21'|12'}
    +
    \Gamma^{\uparrow\downarrow}_{11'|22'}\, , \label{eq:rohringer}
\end{align}
where the multi-indices comprise all vertex arguments except spin. Combining Eqs.\,\eqref{eq:spin-components} and Eq.\,\eqref{eq:rohringer}, one obtains 
\begin{align}
    \Gamma^{S/T}_{1'2'|12} = \Gamma^{D/M}_{21'|12'} \pm \Gamma^{D/M}_{11'|22'}\, ,\label{eq:A3}
\end{align}
where crossing symmetry was employed once. From Eq.\,\eqref{eq:A3}, we can derive corresponding equations for the asymptotic classes. Focusing on $K_{2}$, where the large discrepancy occurs in Fig.~\ref{fig:BSE_K2p_Keldysh}, we insert the native frequency parametrization in the $p$ channel \cite{Walter2022},
\begin{align}
    (\nu_{1'}, \nu_{2'}|\nu_1, \nu_2)_p = (\tfrac{\omega}{2} + \nu, \tfrac{\omega}{2} - \nu| \tfrac{\omega}{2} + \nu', \tfrac{\omega}{2} - \nu')\, ,
\end{align}
which gives
\begin{align}
    &\Gamma^{S/T}_{1'2'|12}(\tfrac{\omega}{2} + \nu, \tfrac{\omega}{2} - \nu| \tfrac{\omega}{2} + \nu', \tfrac{\omega}{2} - \nu') \nonumber\\
    &= \Gamma^{D/M}_{21'|12'}(\nu'-\tfrac{\omega}{2}, \tfrac{\omega}{2} + \nu | \tfrac{\omega}{2}+\nu', \nu - \tfrac{\omega}{2}) \nonumber \\
    &\quad \pm  \Gamma^{D/M}_{11'|22'}(-\nu' - \tfrac{\omega}{2}, \tfrac{\omega}{2}+\nu | \tfrac{\omega}{2} - \nu', \nu - \tfrac{\omega}{2})\, . \label{eq:rohringer-frequencies}
\end{align}
Exchanging external legs from in- to outgoing or vice versa leads to a sign flip in the corresponding frequency arguments. This is due to our convention used for Fourier transforms, see also App.\,\ref{app:fourier} below, and has been accounted for in Eq.\,\eqref{eq:rohringer-frequencies}.
The remaining indices now only label Keldysh components. Comparing to the native parametrization in the $t$ channel \cite{Walter2022},
\begin{align}
    (\nu_{1'}, \nu_{2'}|\nu_1, \nu_2)_t = (\nu'+\tfrac{\omega}{2}, \nu-\tfrac{\omega}{2}|\nu'-\tfrac{\omega}{2}, \nu+\tfrac{\omega}{2})\, , 
\end{align}
we can write Eq.\,\eqref{eq:rohringer-frequencies} as 
\begin{align}
    \Gamma^{S/T}_{1'2'|12;\, p}(\omega, \nu, \nu')&= \Gamma^{D/M}_{21'|12';\, t}(-\omega, \nu, \nu') \nonumber \\
    &\quad \pm  \Gamma^{D/M}_{11'|22';\, t}(-\omega, \nu, -\nu')\, ,
\end{align}
where the additional subscript labels the native frequency parametrization used. Taking the limit $\nu'\rightarrow \infty$ results in an equation for $K_2$. Focusing on the $S$ spin component and the $12|12$ Keldysh component (see Fig.\,\ref{fig:BSE_K2p_Keldysh}) gives 
\begin{align}
    K_{2,S}^{12|12}(\omega_p,\nu_p) &= K_{2,D}^{21|12}(-\omega_t,\nu_t) + K_{2,D}^{11|22}(-\omega_t,\nu_t) \nonumber\\
    &= 2K_{2,D}^{11|22}(-\omega_t,\nu_t)\, , \label{eq:symmetry_K2}
\end{align}
where we used that the $21|12$ and $11|22$ Keldysh components of $K_{2,t}$ are identical, since they are connected by parity, see Eq.\,(4.48b) in Ref.\,\onlinecite{Walter2022}.

Setting $\omega_r=0$, we plot in Fig.\,\ref{fig:K2-symmetry} both sides of Eq.\,\eqref{eq:symmetry_K2} for the NRG vertex and for the result after a single evaluation of the BSE at weak interaction. As was the case in Fig.\,\ref{fig:BSE_K2p_Keldysh}, there is a significant mismatch between the two results. However, Eq.\,\eqref{eq:symmetry_K2} is fulfilled very well for the NRG vertex. Since the BSEs are symmetric by construction, the NRG+BSE result is then symmetric as well, which is indeed confirmed in Fig.\,\ref{fig:K2-symmetry}. We conclude that the violation of the BSE is not inherent to the $p$ channel alone but that the NRG vertex is consistent on the level of Eq.\,\eqref{eq:symmetry_K2}. For comparison, in Fig.\,\ref{fig:K2-symmetry}, we also plot the result from a solution of the parquet equations in the {parquet approximation}, independently obtained with our \textsc{KeldyshQFT} code \cite{Ritz_KeldyshQFT}. Due to the small influence of the higher-order contributions to the 2PI vertex $R$, neglected in the {parquet approximation}, at weak interaction $u=0.5$, one would expect good agreement between the {parquet approximation} and NRG. However, while this is the case at small frequencies, the peak for $\nu/\Delta\gtrsim 1$ does not match. It is hence no surprise that the NRG result does not fulfill the BSE, since it deviates from the {parquet approximation} result (which fulfills the BSE by construction). Improving the NRG result requires numerically more challenging parameter settings: Increasing the number of frequency bins per decade or reducing the discretization parameter would presumably give more accurate results (see also App.\,\ref{sec:parameters} for details on the NRG parameters). Finding suitably optimized parameter settings for NRG would go beyond the scope of this paper and is left for future work.

\begin{figure*}
    \centering
    \includegraphics[height=.325\linewidth]{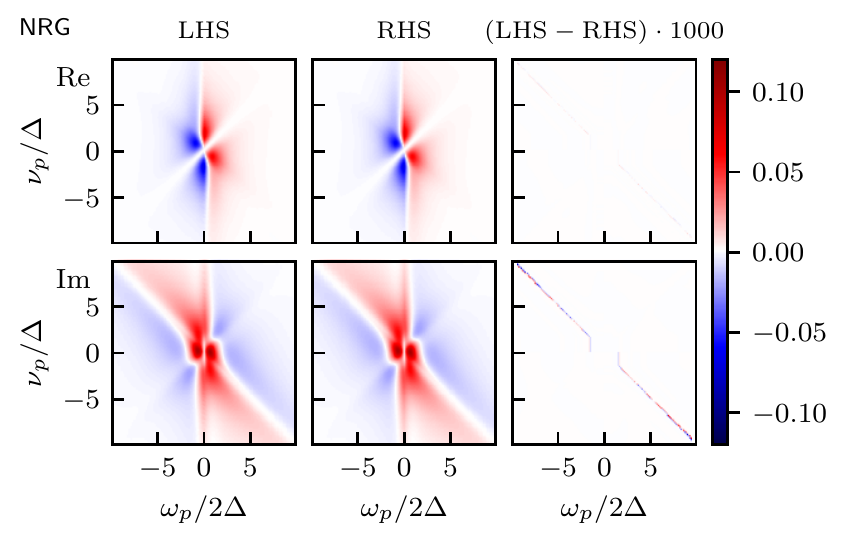}\vline
    \includegraphics[height=.325\linewidth]{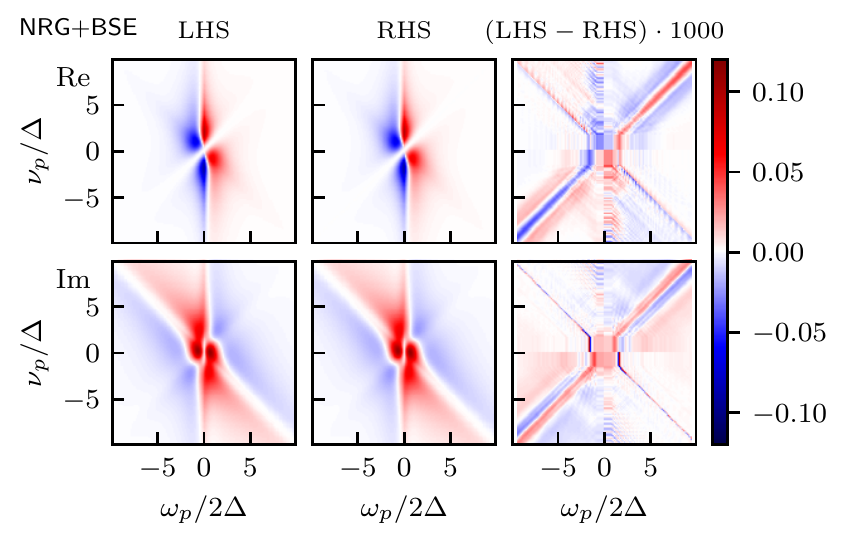}
    \caption{Fulfillment of the generalized FDR~\eqref{eq:FDT_K2p} for $K_{2, S}^{12|12}(\omega_p,\nu_p)$ at weak interaction from the NRG vertex (left) and after one evaluation of the BSE (right). Both times, the FDR is fulfilled exceptionally well. The slight discrepancies across the anti-diagonal $\nu_p + \omega_p/2=0$ are negligible interpolation errors where the second term in Eq.\,\eqref{eq:FDT_K2p} vanishes. In addition, very minor additional inaccuracies appear in the NRG+BSE result. We attribute these to the finite numerical accuracy of the integrations required for evaluating the BSE.}
    \label{fig:FDT_K2}
\end{figure*}

\begin{figure}
    \centering
    \includegraphics[width=\linewidth]{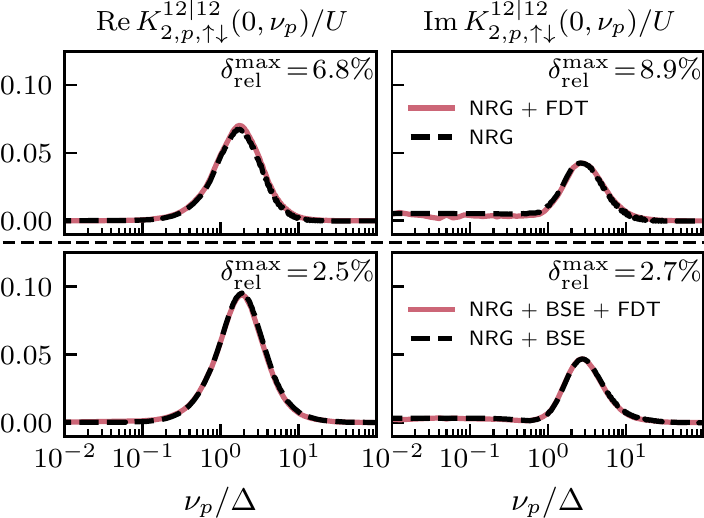}
    \caption{\
    Same as Fig.\,\ref{fig:FDT_K2} for $\omega_p\rightarrow 0$, evaluated as explained in Eq.\,\eqref{eq:FDT-finite-diff}. Also in this case, the generalized FDR is fulfilled very well. The minor wiggles in the top right panel at small frequencies probably stem from the finite difference used in Eq.\,\eqref{eq:FDT-finite-diff}.}
    \label{fig:1D_FDT}
\end{figure}

Second, we exemplarily study the generalized FDR for $K_{2, S}^{12|12}$. It can be derived from Eq.\,(84) in Ref.\,\cite{Halbinger2024} and reads
\begin{align}
    &K_{2, S}^{12|12}(\omega_p,\nu_p) = \nonumber \\
    &\quad \coth\left(\tfrac{-\omega_p}{2T}\right) \Big[ [K_{2, S}^{21|11}(\omega_p,\nu_p)]^* - K_{2, S}^{12|11}(\omega_p,\nu_p) \Big]
    \nonumber \\
    &\quad +
    \tanh\left(\tfrac{\nu_p+\omega_p/2}{2T}\right) \Big[ [K_{2, S}^{21|11}(\omega_p,\nu_p)]^* - K_{2, S}^{22|12}(\omega_p,\nu_p) \Big]\, . \label{eq:FDT_K2p}
\end{align}
Evaluating and comparing the LHS and the RHS of Eq.\,\eqref{eq:FDT_K2p} for the NRG vertex and the NRG vertex after one evaluation of the BSE at weak interaction yields Fig.\,\ref{fig:FDT_K2}.  We see that the FDR is fulfilled exceptionally well both times. Very minor inaccuracies occur in the NRG+BSE result, which can be attributed to the finite numerical accuracy of the integrations required for evaluating the BSE. Strictly speaking, the one-dimensional cut at $\omega_p=0$ had to be excluded in Fig.\,\ref{fig:FDT_K2}, due to the diverging $\coth$ term on the RHS of Eq.\,\eqref{eq:FDT_K2p}. Therefore, we show an additional one-dimensional plot at $\omega_p=0$ in Fig.\,\ref{fig:1D_FDT}, taking the limit properly: Using the short-hand notation $\Big[ [K_{2, S}^{21|11}(\omega_p,\nu_p)]^* - K_{2, S}^{12|11}(\omega_p,\nu_p) \Big]\equiv \tilde{K}_2(\omega_p)$, we employ L'Hôpital's rule to approximate the first term on the RHS of Eq.\,\eqref{eq:FDT_K2p} as 
\begin{align}
    &\lim_{\omega_p\rightarrow 0} \coth\left(\tfrac{-\omega_p}{2T}\right) \tilde{K}_2(\omega_p) \!\approx\!  -2T\,  \tfrac{\tilde{K}_2(0 + \Delta\omega_p) - \tilde{K}_2(0 - \Delta\omega_p)}{2\Delta\omega_p}\, , \label{eq:FDT-finite-diff}
\end{align}
where we approximated the derivative by a finite difference ($\Delta\omega_p$ is the step size of the frequency grid around $\omega_p=0$). This way, we obtain Fig.\,\ref{fig:1D_FDT}, where, for $\omega_p=0$ too, the generalized FDR is fulfilled very well. We conclude that the NRG vertex is consistent on the level of the symmetries and the generalized FDR exemplarily checked in this section.

\section{NRG computations}\label{sec:parameters}

\begin{table}[]
    \setlength{\tabcolsep}{5pt}
    \centering
    \begin{tabular}{l| c c c c c c c c}
    \toprule
    & $\Lambda$ & $n_z$ & $N_{\mathrm{keep}}$ & $E_{\mathrm{step}}$ & $\sigma_\mathrm{LG}$ & $\gamma_\mathrm{L}$ & $\alpha$ & $\gamma$ \\ \midrule
    $\Sigma$ & 2 & 6 & 5000 & $-$ & $-$ & $-$ & 2 & 4 \\
    $K_1, K_2$ & 4 & 4 & 300/200 & 16 & 0.4 & $T$ & $-$ & $-$ \\
    $\Gamma_\mathrm{core}$ & 4 & 4 & 300/200 & 8/16 & 0.4 & $T$ & $-$ & $-$ \\
    \bottomrule
    \end{tabular}
    \caption{NRG parameters for the self-energy and vertex calculations. If two values are specified, the first (second) one corresponds to the setting for weak (strong) interaction.}
    \label{tab:NRG_params}
\end{table}

The NRG computations performed for this work are based on the QSpace tensor library \cite{Weichselbaum2012, Weichselbaum2012b, Weichselbaum2020, Weichselbaum2024, Weichselbaum2024b}. We employ the full density-matrix NRG \cite{Peters2006, Weichselbaum2007}, using adaptive broadening \cite{Lee2016, Lee2017} for obtaining 2p dynamical correlators. The 4p vertex was computed using the recent generalization of the NRG method to multipoint functions \cite{Kugler2021, Lee2021}. Symmetric improved estimators were used both for the self-energy \cite{Kugler2022} and the vertex \cite{Lihm2024}.
To compute the vertex, the PSF produced by NRG had to be convoluted with the appropriate kernel functions. In order to do so on logarithmic grids with reasonable computational effort, we employed the following strategy (described in more detail in Ref.~\cite{Frankenbach2025}): The broadened Keldysh frequency kernels were first precomputed on extremely fine, equidistant, one-dimensional grids with a grid spacing of $100/2^{15}\approx0.003$ in units of the hybridization parameter $\Delta$. The resulting kernel functions were brought into matrix form and compressed using SVDs with a tolerance of $10^{-6}$. To obtain the vertex, these compressed kernel matrices were contracted with the PSFs, using trilinear interpolation from points on a cuboid surrounding the respective frequency points of the logarithmic grid.  

We state the numerical parameters chosen for the NRG calculations in Tab.\,\ref{tab:NRG_params}.
$\Lambda$ is the Wilson parameter used to logarithmically discretize the non-interacting bath. (The limit $\Lambda\searrow 1$ would correspond to the original continuous bath.) Spectral data are averaged over $n_z$ shifted versions of the logarithmic discretization grid, following Žitko's discretization scheme \cite{Zitko2009, Zitko2009b}. $N_{\mathrm{keep}}$ specifies the maximal number of kept SU(2) multiplets in each shell during the iterative diagonalization. In principle, a convergence analysis in both $n_z$ and $N_{\mathrm{keep}}$ would be required to produce optimal results. While $N_{\mathrm{keep}}=5000$ from experience is large enough to compute the self-energy accurately, this is unfeasible numerically for the multi-point vertex computations at this point. 

$E_\mathrm{step}$ specifies the number of frequency bins per decade on the logarithmic grid for the PSFs of the vertex. $\sigma_\mathrm{LG}$ and $\gamma_\mathrm{L}$ are broadening parameters used for the log-Gaussian broadening of the PSFs, see, e.g., see App.\,E.2 in Ref.\,\onlinecite{Lihm2024}. In contrast to Ref.\,\onlinecite{Lihm2024}, where $\sigma_\mathrm{LG}=0.3$ and $\gamma_\mathrm{L}=0.5 T$ were used for the vertex at strong interaction, we chose the slightly larger broadening employed already for weak interaction. The reason is that we observed slight under-broadening of $K_{2,r}$ at small frequencies with the broadening parameters of Ref.\,\onlinecite{Lihm2024}. $\alpha$ and $\gamma$ are similar broadening parameters used in the log-Gaussian broadening for 2p NRG computations, as specified in Eqs.\,(17b) and (21) of Ref.\,\onlinecite{Lee2016}.

\section{Conversions between mpNRG and {quantum field theory} conventions}\label{app:conventions}

To convert the Keldysh vertex from the conventions of NRG, as, e.g., outlined in Ref.\,\onlinecite{Lihm2024}, to the conventions of the \textsc{KeldyshQFT} code \cite{Walter2022, Ge2024, Ritz2024}, the following steps must be taken:
\begin{enumerate}[label=(\roman*)]
    \item multiply the vertex by a global sign
    \item swap the middle Keldysh indices $(12|34) \leftrightarrow (13|24)$
    \item swap $K_{2,p} \leftrightarrow K_{2',p}$
    \item convert the frequency parametrization according to
\end{enumerate}

\begin{subequations}
\begin{align}
\left[\omega_t\right]^{\mathrm{NRG}} &= -\omega_t \\
\left[\nu_t\right]^{\mathrm{NRG}} &= \nu'_t + \tfrac{\omega_t}{2}\\
\left[\nu'_t\right]^{\mathrm{NRG}} &= \nu_t + \tfrac{\omega_t}{2}\, .
\end{align}
\end{subequations}
Using the conversions between the $t$-channel and the $a$- and $p$-channel parametrizations as given in App.\,A of Ref.\,\onlinecite{Walter2022}, we further have
\begin{subequations}
    \begin{align}
        \left[\omega_t\right]^{\mathrm{NRG}} &= \nu'_a - \nu_a = \nu'_p - \nu_p \\
        \left[\nu_t\right]^{\mathrm{NRG}} &= \nu_a - \tfrac{\omega_a}{2} = \nu_p + \tfrac{\omega_p}{2} \\
        \left[\nu'_t\right]^{\mathrm{NRG}} &= \nu_a + \tfrac{\omega_a}{2} = - \nu'_p + \tfrac{\omega_p}{2}\, .
    \end{align}
\end{subequations}

\section{Fulfillment of the BSE at large frequencies}\label{app:BSE_asymp}

\begin{figure}
    \centering
    \includegraphics[width=\linewidth]{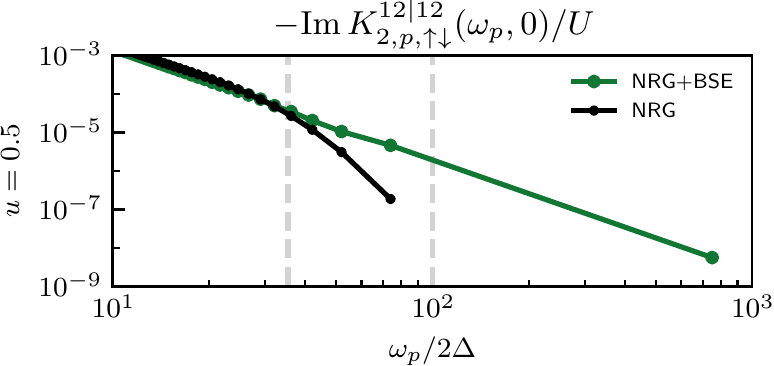}
    \caption{One-dimensional slice through one component of $K_2$, cf.~Fig.~\ref{fig:K2_nu0}, on logarithmic axes. The dots
    indicate the frequency grid points chosen in the \textsc{KeldyshQFT} code used to evaluate the BSE, onto which the NRG data was interpolated. The first vertical dashed line marks the maximal frequency at which \textsl{all} NRG vertex components required for evaluating the BSE were available. The second dashed line marks the maximal frequency for which the shown vertex component was computed by NRG. Starting at the first dashed line, we see very minor deviations below $0.1\%$ compared to the maximal value of the vertex component shown.
    }
    \label{fig:K2_p_asymp}
\end{figure}

In this section, we show that the finite extent of the frequency grid only minimally influences the fulfillment of the BSE. Figure \ref{fig:K2_p_asymp} shows a one-dimensional slice through one component of $K_2$, corresponding to one panel in Fig.~\ref{fig:K2_nu0}, focusing on the region at very large frequencies. The black line shows the vertex component as computed by NRG, already interpolated onto the frequency grid chosen in the \textsc{KeldyshQFT} code. The green line shows the same component after one evaluation of the BSE with the same code. Two vertical dashed lines highlight special points on the frequency axis: The one at $\frac{\omega_p}{2\Delta}=100$ marks the maximal frequency for which the shown vertex component had been computed by NRG. The other one at $\frac{\omega_p}{2\Delta}=\frac{100}{2\sqrt{2}} \approx 35$ marks the maximal frequency where all NRG vertex components needed for evaluating the BSE were available. It is smaller than the other frequency due to the $\omega/2$ shifts in the {quantum field theory} parametrizations and the rotations required when transforming between native channel parametrizations, see App.~\ref{app:conventions}. We see that, up to this point, the fulfillment of the BSE is close to perfect. Afterwards, small deviations appear, which is to be expected, since not all components required on the RHS of the BSE are available anymore. However, the deviations are smaller than $0.1\%$, compared to the height of the peak of the component shown, cf.~Fig.~\ref{fig:K2_nu0}, and hence numerically negligible. Beyond the second dashed line, no NRG data is available anymore. 

\section{Full frequency dependence of $K_2$}\label{app:K2}

\begin{figure*}
    \centering
    \includegraphics[height=.61\linewidth]{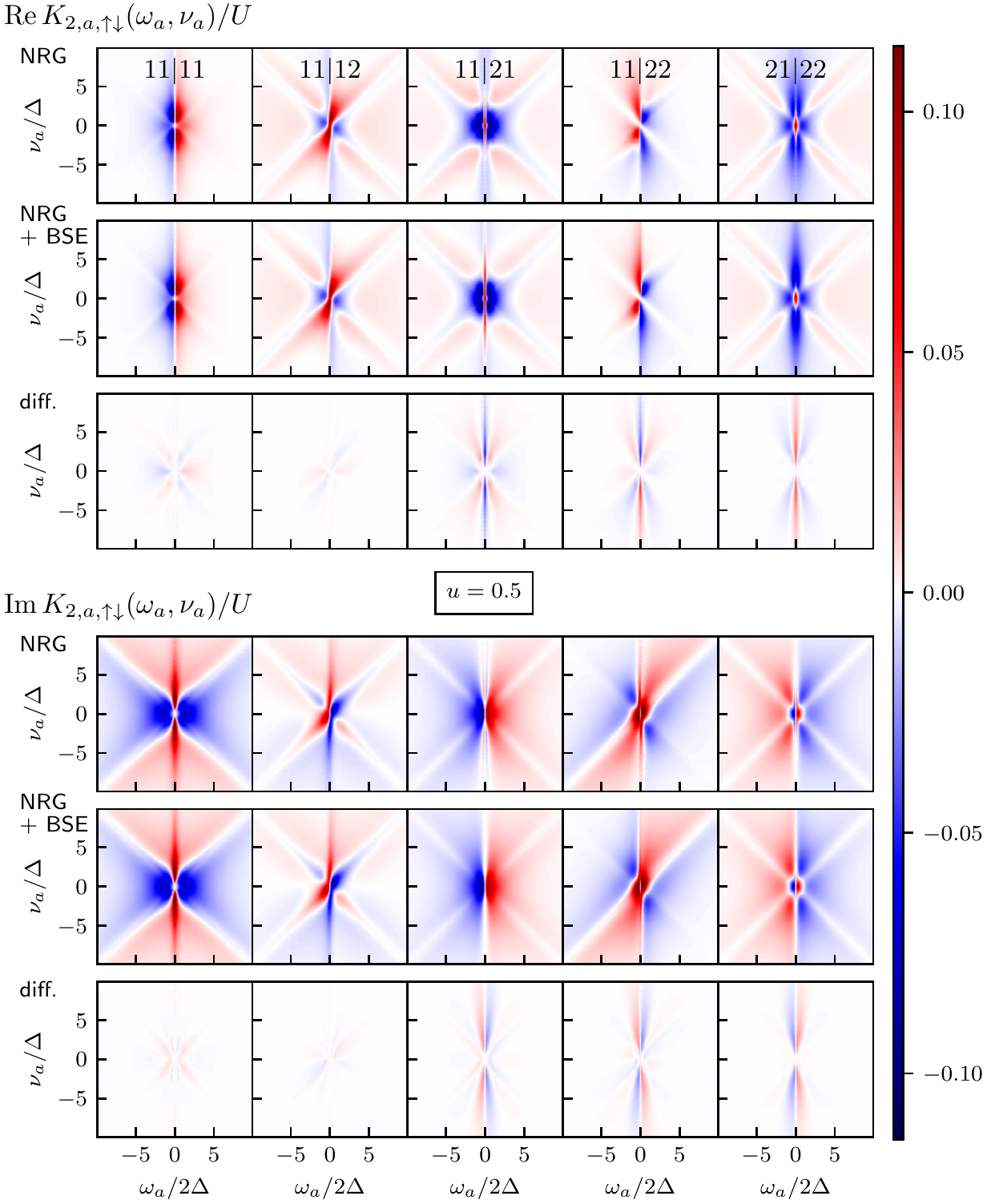}\vline
    \includegraphics[height=.61\linewidth]{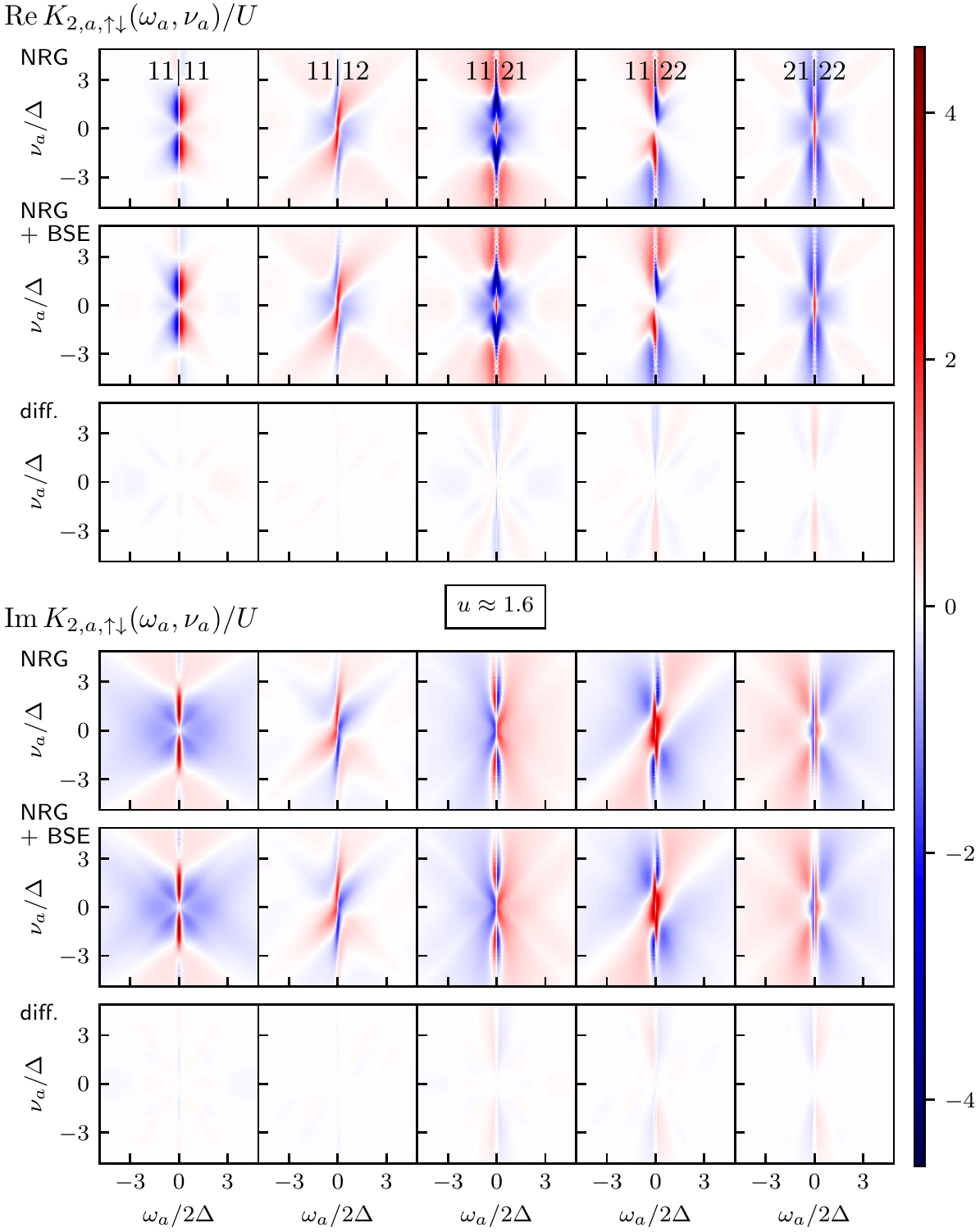}
    \caption{Fulfillment of the BSE for the full frequency dependence of both the real and imaginary parts of $K_2$ in the $a$ channel. The first rows show the vertex components as produced by NRG, the second the result after a single evaluation of the BSE~\eqref{eq:BSE_K2} and the third their absolute differences. Results for weak (strong) interaction are shown on the left (right). At weak interaction, we restrict the shown frequency intervals to $\pm 10\Delta$. At strong interaction, we zoom into a smaller region of $\pm 5 \Delta$ to highlight the increasingly sharp structures of the vertex. We observe good agreement of the BSE throughout.}
    \label{fig:K2_a}
\end{figure*}

\begin{figure*}
    \centering
    \includegraphics[height=.61\linewidth]{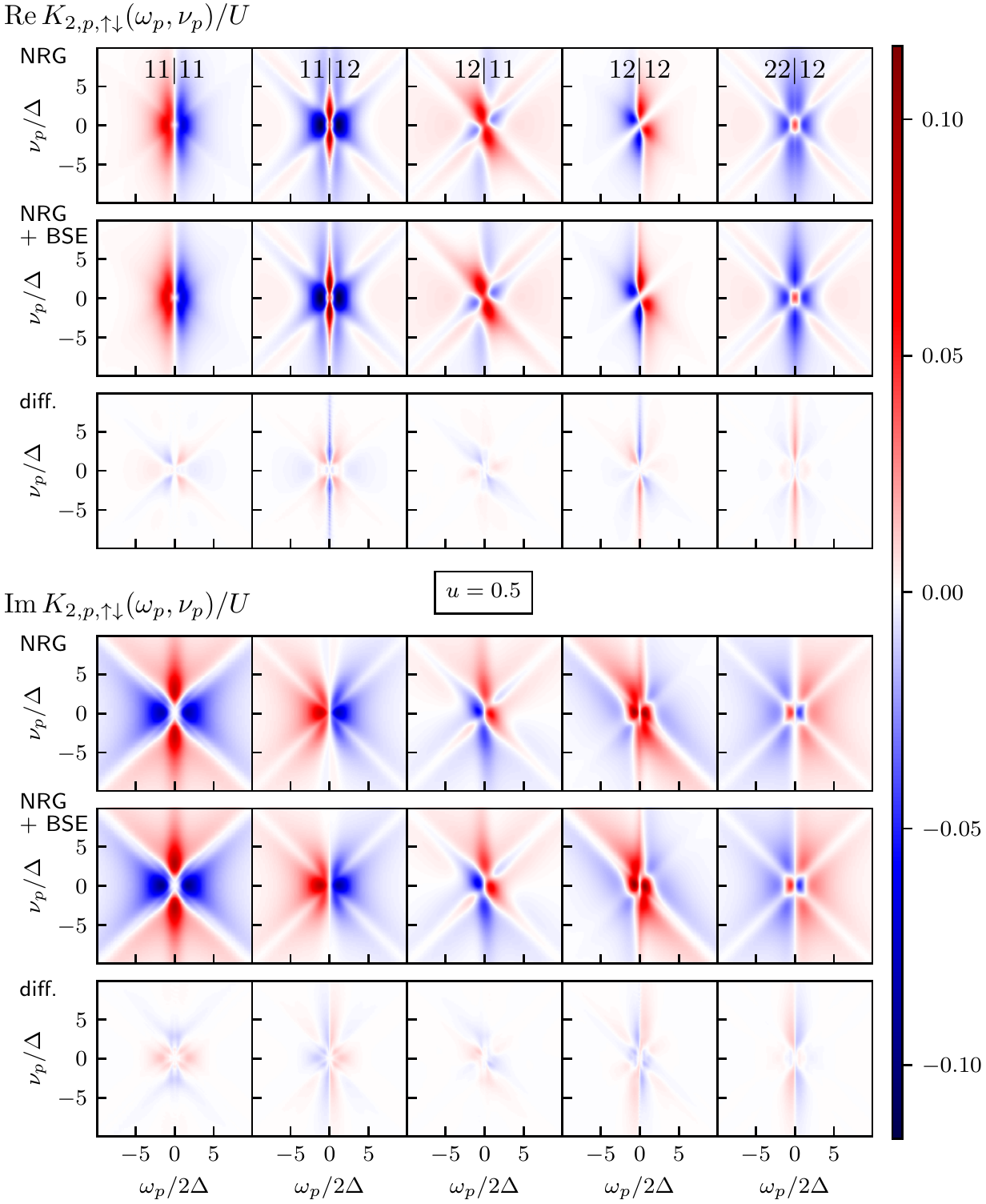}\vline
    \includegraphics[height=.61\linewidth]{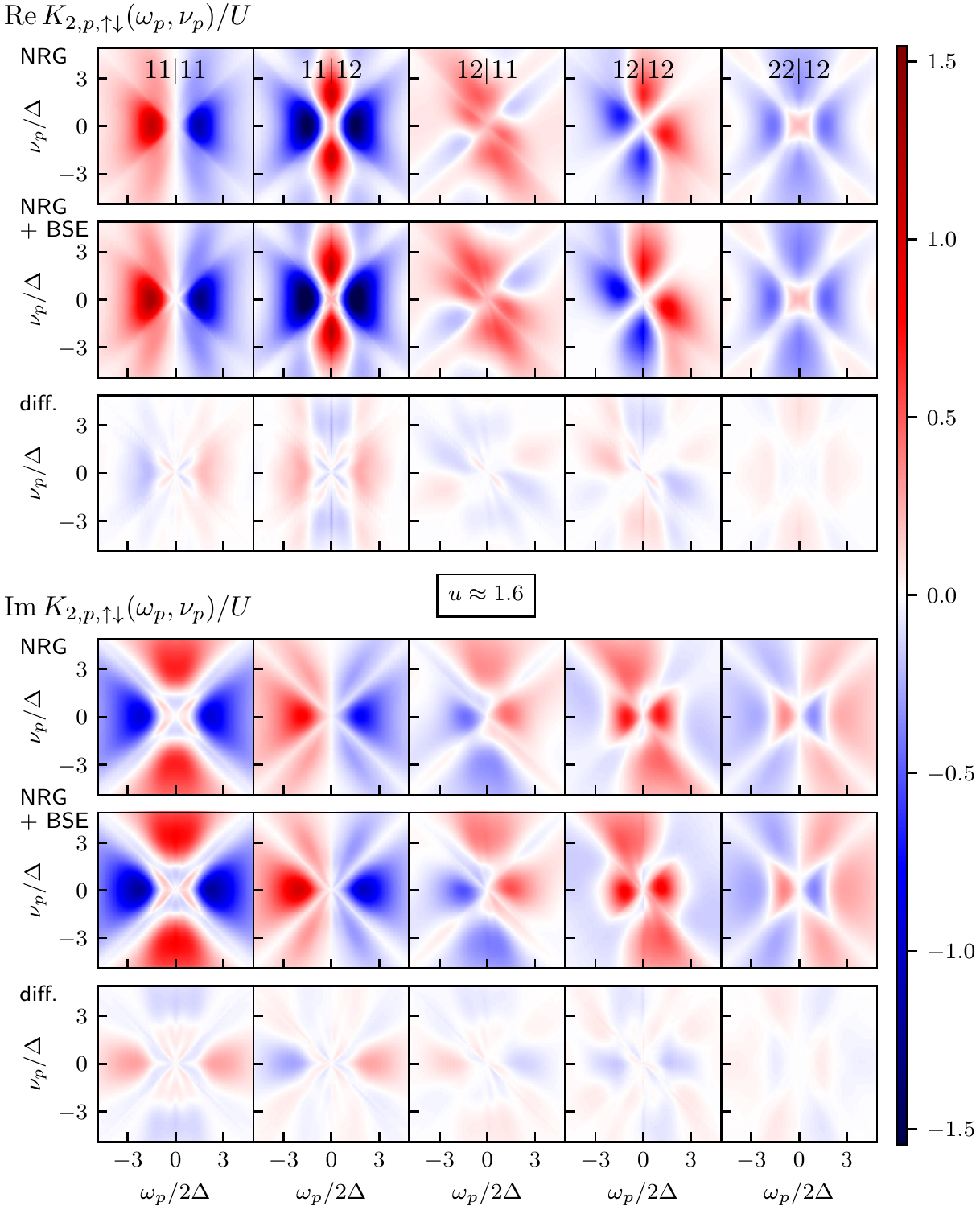}
    \caption{Same as in Fig.\,\ref{fig:K2_a} for the $p$ channel.}
    \label{fig:K2_p}
\end{figure*}

\begin{figure*}
    \centering
    \includegraphics[height=.61\linewidth]{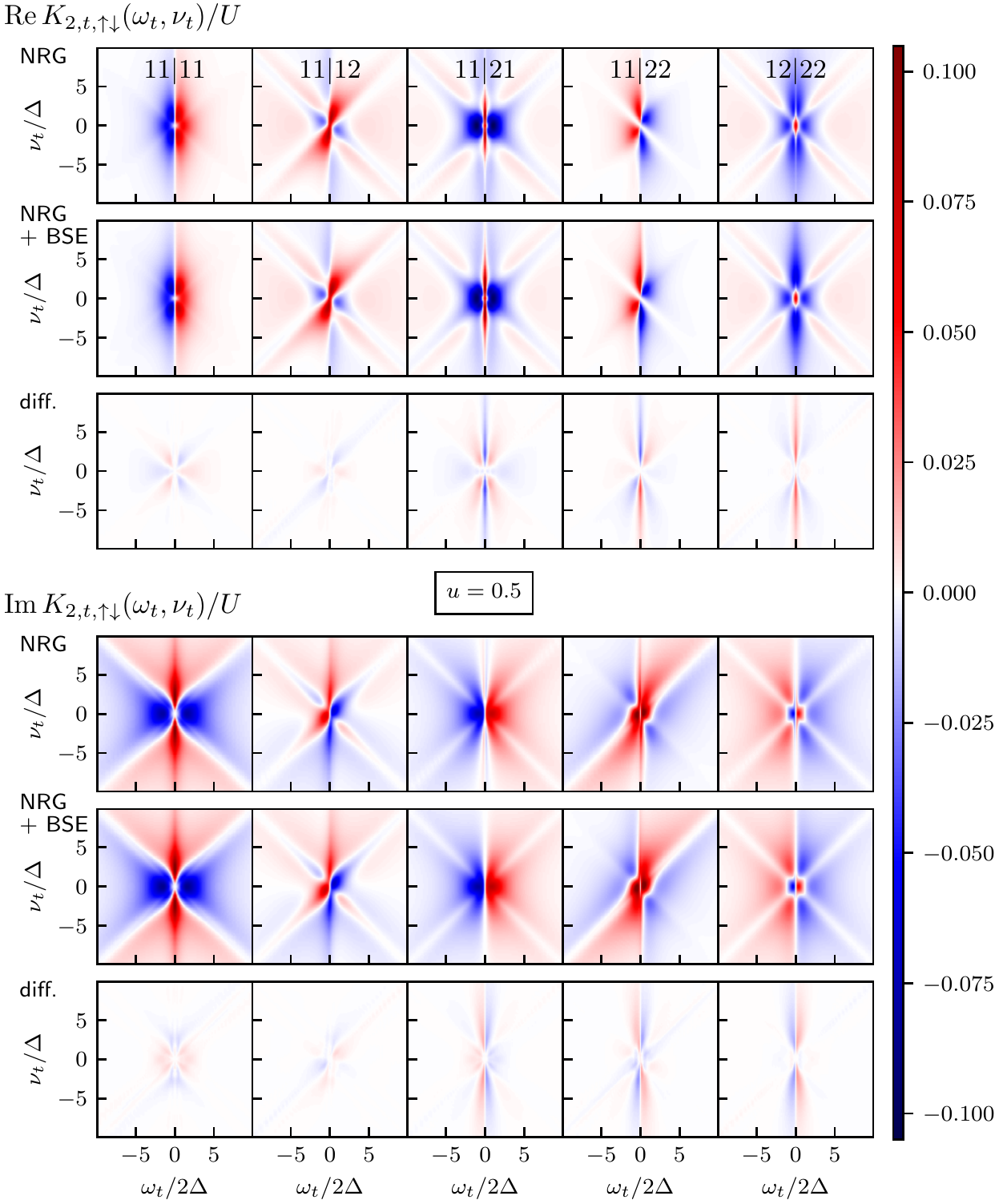}\vline
    \includegraphics[height=.61\linewidth]{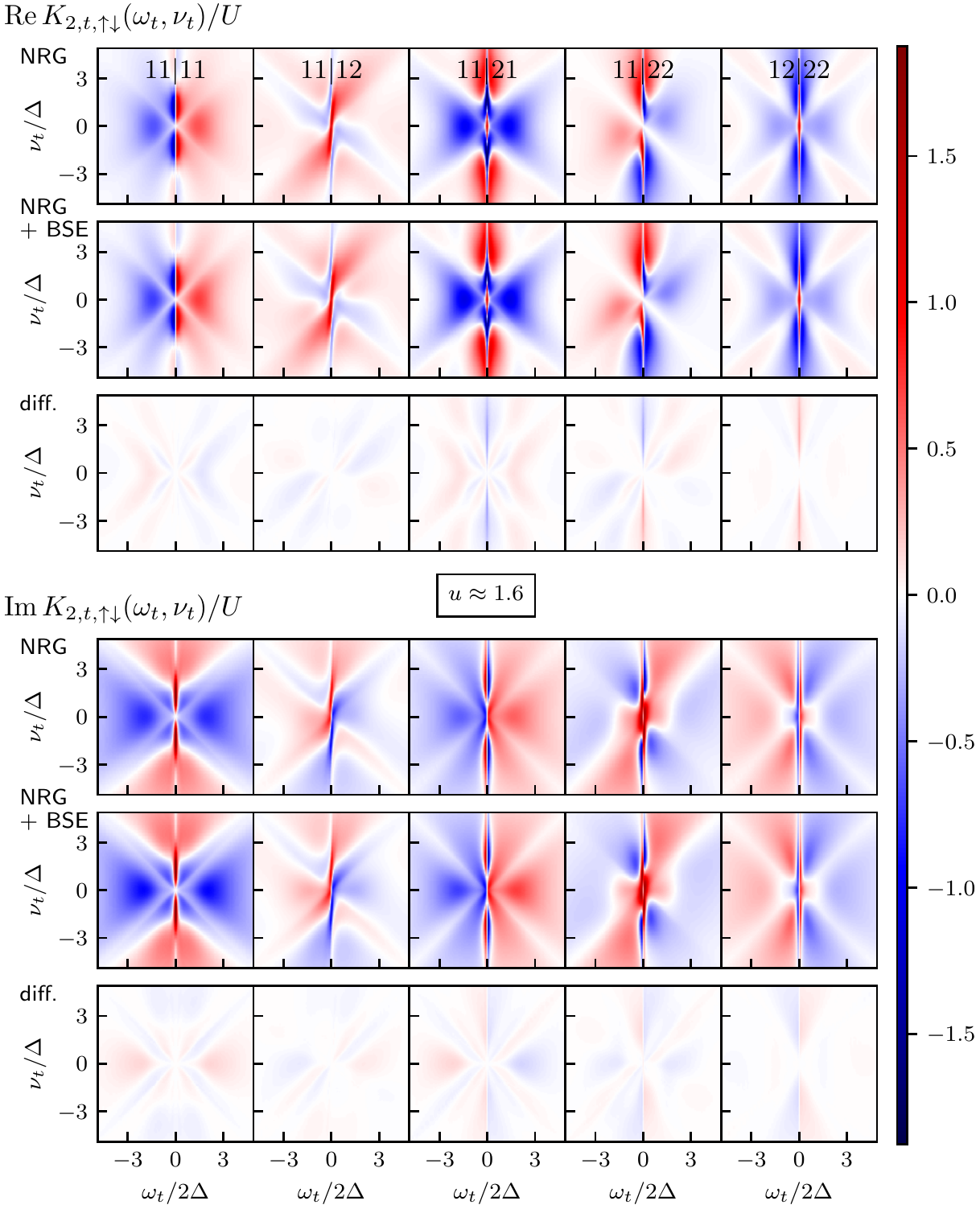}
    \caption{Same as in Figs.\,\ref{fig:K2_a} and \ref{fig:K2_p} for the $t$ channel.
    }
    \label{fig:K2_t}
\end{figure*}

In Sec.\,\ref{sec:BSE}, we restricted the discussion of the fulfillment of the BSEs for $K_2$ to two one-dimensional slices through a single Keldysh component. For completeness, we show the full two-dimensional frequency dependence of all five nonequivalent Keldysh components of $K_2$ in Figs.\,\ref{fig:K2_a}, \ref{fig:K2_p}, and \ref{fig:K2_t}. We plot the vertex components as produced by NRG in the first rows, the result after a single evaluation of the BSE~\eqref{eq:BSE_K2} in the second, and their absolute difference in the third. As always, we show data for both weak and strong interaction, whereas we restricted the frequency interval shown for strong interaction to smaller frequencies than for weak interaction to make the non-trivial structures of the vertex more clearly visible. In accordance with our discussion in Sec.\,\ref{sec:BSE}, we observe good agreement of the BSE throughout. Notably, many components show sharp structures around $\omega_r=0$, which are nevertheless extended along the $\nu_r$ direction, in particular in the $a$ and $t$ channels, already at weak interaction. Resolving these accurately poses a numerical challenge.

\section{Derivation of the generalized U(1) Ward identity in the KF}\label{app:WI}

The goal of this section is to provide a self-contained derivation of one of the main results of this work, namely the general two-dimensional form of the U(1) WI in the KF, Eq.\,\eqref{eq:WI_main}. We start from textbook definitions of the basic quantities involved and lay out all required calculations without omitting technical details.

\subsection{Setup and definitions}\label{app:setup}
Our starting point is the partition function expressed using a functional integral and the action, which contains a non-interacting as well as an interacting term,
\begin{align}
    \mathcal{Z} &= \int \mathcal{D}[\overline{d},d]\, e^{iS[\overline{d},d]} \\
    S[\overline{d},d] &= S_0[\overline{d},d] + S_{\mathrm{int}} [\overline{d},d] \nonumber \\ 
    &= \int_\mathcal{C}  dt  \Big\{ \int_\mathcal{C} dt' \overline{d}^{j_{1'}}_{\sigma_{1'}}(t') \left[G_0^{-1}\right]^{j_{1'}|j_{1}}_{\sigma_{1'}|\sigma_{1}}(t'|t) \, d^{j_{1}}_{\sigma_{1}}(t) \nonumber \\
    &\qquad + \tfrac{1}{4} \overline{d}^{j_{1'}}_{\sigma_{1'}}(t) \overline{d}^{j_{2'}}_{\sigma_{2'}}(t) \left[\Gamma_0\right]^{j_{1'}j_{2'}|j_{1}j_{2}}_{\sigma_{1'}\sigma_{2'}|\sigma_{1}\sigma_{2}} {d}^{j_{2}}_{\sigma_{2}}(t) {d}^{j_{1}}_{\sigma_{1}}(t)   \Big\}
\end{align}
\begin{align}
\left[G_0^{-1}\right]^{j_{1'}|j_{1}}_{\sigma_{1'}|\sigma_{1}}(t'|t)  &=  \delta_\mathcal{C}(t'-t) \delta_{j_{1'}, j_{1}} \delta_{\sigma_{1'},\sigma_{1}} i\partial_t - h^{j_{1'}|j_{1}}_{\sigma_{1'}|\sigma_{1}}(t'|t). \label{eq:G0}
\end{align}
Here, $G_0^{-1}$ is the inverse bare propagator and $h^{j_{1'}|j_{1}}_{\sigma_{1'}|\sigma_{1}}(t'|t)$ the single-particle Hamiltonian, which for the SIAM contains the level shift and the hybridization function. In this expression, repeated indices are meant to be summed over and time integrations are performed over the Keldysh contour $\mathcal{C}$, see, e.g., \cite{Walter2022} for details. In the context of this work, the single-particle term is diagonal in the spin indices, but we keep both indices for now, to make the discussion general enough to still apply to a model that, e.g., includes an external magnetic field.

To make the following computations more compact, we introduce a multi-index notation, writing
\begin{align}
    S[\overline{d},d] &= \int_{\bm{t}\bm{t'}} \overline{d}_{\bm{t'}} [G_0^{-1}]_{\bm{t'}|\bm{t}} d_{\bm{t}} \nonumber\\
    &+ \tfrac{1}{4} \sum_{1'2'12} \int_\mathcal{C} dt\,  \overline{d}_{{1'}}(t)  \overline{d}_{{2'}}(t) [\Gamma_0]_{{1'}{2'}|{1}{2}} d_{{2}}(t) d_{{1}}(t) \label{eq:action}
\end{align}
\begin{align}
    [G_0^{-1}]_{\bm{t'}|\bm{t}} = \delta(\bm{t'} - \bm{t}) i\partial_t - h_{\bm{t'}|\bm{t}} \label{eq:G0_inv} 
\end{align}
where non-bold indices ($1',2',1,2$ in Eq.\,\eqref{eq:action}) comprise Keldysh indices, spin indices and more general quantum numbers one might consider, and the bold indices combine the non-bold indices with time indices. 

Using this notation, correlation functions are defined as follows. The two-point (2p) and four-point (4p) functions read,
\begin{align}
    G_{\bm{1}|\bm{1'}} &= -i \langle d_{\bm{1}} \overline{d}_{\bm{1'}} \rangle \label{eq:G} \\
    G^{(4)}_{\bm{1}\bm{2}|\bm{1'} \bm{2'}} &= i \langle d_{\bm{1}} d_{\bm{2}} \overline{d}_{\bm{2'}} \overline{d}_{\bm{1'}} \rangle\, , \label{eq:G4}
\end{align}
where the bracket $\langle \ldots \rangle$ denotes the standard functional integral
\begin{align}
    \langle \ldots \rangle &= \frac{1}{\mathcal{Z}}\int \mathcal{D}[\overline{d},d]\, (\ldots) \, e^{iS[\overline{d},d]}\, ,
\end{align}
corresponding to expectation values of operators time-ordered on the Keldysh contour $\mathcal{C}$.
The self-energy $\Sigma$ is introduced via the Dyson equation,
\begin{align}
    G_{\bm{1}|\bm{1'}} &= [G_0]_{\bm{1}|\bm{1'}} + \int_{\bm{2'}\bm{2}} [G_0]_{\bm{1}|\bm{2'}} \Sigma_{\bm{2'}|\bm{2}} G_{\bm{2}|\bm{1'}} \\ \Leftrightarrow \quad G^{-1}_{\bm{1'}|\bm{1}} &= [G_0^{-1}]_{\bm{1'}|\bm{1}} - \Sigma_{\bm{1'}|\bm{1}}, \label{eq:Dyson}
\end{align}
and, after employing the tree expansion for the 4p function,
\begin{align}
    iG^{(4)}_{\bm{1}\bm{2}|\bm{1'}\bm{2'}} = G_{\bm{1}|\bm{1'}} G_{\bm{2}|\bm{2'}} - G_{\bm{1}|\bm{2'}} G_{\bm{2}|\bm{1'}} + iG^{(4)}_{c;\, \bm{1}\bm{2}|\bm{1'}\bm{2'}}, \label{eq:tree-expansion}
\end{align}
the 4p vertex $\Gamma$ is introduced via the connected part of the 4p function,
\begin{align}
    G^{(4)}_{c;\, \bm{1}\bm{2}|\bm{1'}\bm{2'}} = - \int_{\bm{3'}\bm{4'}\bm{3}\bm{4}} G_{\bm{1}|\bm{3'}} G_{\bm{2}|\bm{4'}} \Gamma_{\bm{3'}\bm{4'}|\bm{3}\bm{4}} G_{\bm{3}|\bm{1'}} G_{\bm{4}|\bm{2'}}. \label{eq:vertex}
\end{align}

\subsection{Equation of motion for the equal-time Green function}

We consider the infinitesimal U(1) gauge transformation
\begin{subequations}
\begin{align}
    d^j_\sigma(t) &\longrightarrow d^j_\sigma(t) \underbrace{+ i \varepsilon^j_\sigma(t) d^j_\sigma(t)}_{\delta d^j_\sigma(t)} \\
    \overline{d}^{j'}_{\sigma'}(t') &\longrightarrow \overline{d}^{j'}_{\sigma'}(t') \underbrace{- i \varepsilon^{j'}_{\sigma'}(t') \overline{d}^{j'}_{\sigma'}(t')}_{\delta \overline{d}^{j'}_{\sigma'}(t')},
\end{align}
\end{subequations}
or, written in multi-index notation,
\begin{subequations}
\begin{align}
    d_{\bm{t}} &\longrightarrow d_{\bm{t}} + i \varepsilon_{\bm{t}} d_{\bm{t}} \equiv  d_{\bm{t}} + \delta d_{\bm{t}} \\
    \overline{d}_{\bm{t'}} &\longrightarrow \overline{d}_{\bm{t'}} - i \varepsilon_{\bm{t'}} \overline{d}_{\bm{t'}} \equiv \overline{d}_{\bm{t'}} + \delta \overline{d}_{\bm{t'}}\, .
\end{align}
\end{subequations}

Here and from now on, repeated indices are not summed over, unless indicated explicitly. Since this transformation is supposed to be a symmetry of the theory to $\mathcal{O}(\varepsilon)$, we demand invariance of $\mathcal{Z}$ as well as all correlation functions under this transformation to $\mathcal{O}(\varepsilon)$. This generates an infinite set of consistency relations between correlation functions. 

As the U(1) transformation is non-anomalous, meaning that the path integral measure is invariant under this transformation, we therefore require
\begin{align}
    0 \overset{!}{=} \delta \mathcal{Z} = \int \mathcal{D}[\overline{d},d]\, \delta S[\overline{d},d] e^{iS[\overline{d},d]} . \label{eq:Z-inv}
\end{align}
{The notation $\overset{!}{=}$ is used to clarify that we demand the equation to hold.}
Since $S_\mathrm{int}$ is trivially invariant, the only contribution comes from the non-interacting part $S_0$. We have
\begin{align}
    \delta S_0[\overline{d},d] &= \int_{\bm{t}\bm{t'}} \left\{ \delta\overline{d}_{\bm{t'}} [G_0^{-1}]_{\bm{t'}|\bm{t}} d_{\bm{t}} + \overline{d}_{\bm{t'}} [G_0^{-1}]_{\bm{t'}|\bm{t}} \delta d_{\bm{t}} \right\} \nonumber \\
    &= \int_{\bm{t}\bm{t'}} \big\{  - i \varepsilon_{\bm{t'}} \overline{d}_{\bm{t'}} \left[ \delta(\bm{t'} - \bm{t}) i\partial_t - h_{\bm{t'}|\bm{t}} \right] d_{\bm{t}} \nonumber \\
    &\qquad \qquad + \overline{d}_{\bm{t'}} \left[ \delta(\bm{t'} - \bm{t}) i\partial_t - h_{\bm{t'}|\bm{t}} \right] i \varepsilon_{\bm{t}} d_{\bm{t}} \big\}  \nonumber \\
    &= - \int_{\bm{t}} \overline{d}_{\bm{t}} (\partial_t \varepsilon_{\bm{t}}) d_{\bm{t}} + \int_{\bm{t}\bm{t'}} \overline{d}_{\bm{t'}} (i\varepsilon_{\bm{t'}} - i\varepsilon_{\bm{t}}) h_{\bm{t'}|\bm{t}} d_{\bm{t}} \nonumber \\
    &= \int_{\bm{t}}  \varepsilon_{\bm{t}} \partial_t (\overline{d}_{\bm{t}} d_{\bm{t}}) + \int_{\bm{t}\bm{t'}} \overline{d}_{\bm{t'}} (i\varepsilon_{\bm{t'}} - i\varepsilon_{\bm{t}}) h_{\bm{t'}|\bm{t}} d_{\bm{t}}\, \nonumber \\
    &= \int_{\bm{t}} \varepsilon_{\bm{t}} \Big\{ \partial_{t} (\overline{d}_{\bm{t}} d_{\bm{t}}) 
    + i \int_{\bm{\tilde{t}}} \overline{d}_{\bm{t}} h_{\bm{t}|\bm{\tilde{t}}} d_{\bm{\tilde{t}}}
    - i \int_{\bm{\tilde{t}}} \overline{d}_{\bm{\tilde{t}}} h_{\bm{\tilde{t}}|\bm{t}} d_{\bm{t}}
    \Big\}  
    , \label{eq:delta_S0}
\end{align}
where we applied the product rule and integrated by parts in the second to last step. Since $\varepsilon_{\bm{t}}$ is an arbitrary function, using this result in Eq.\,\eqref{eq:Z-inv}, we get
\begin{align}
    0 \overset{!}{=} \partial_{t} \langle \overline{d}_{\bm{t}} {d}_{\bm{t}} \rangle + i \int_{\bm{\tilde{1}}} (h_{\bm{t|
    \bm{\tilde{1}}}} \langle \overline{d}_{\bm{t}} {d}_{\bm{\tilde{1}}} \rangle  -   h_{\bm{\tilde{1}} | \bm{t}} \langle \overline{d}_{\bm{\tilde{1}}} {d}_{\bm{t}} \rangle ),
\end{align}
where we performed a relabelling of all indices. Employing the definition of the 2p function in Eq.\,\eqref{eq:G}, together with the anticommutation property of the Grassmann variables, we write this result as
\begin{equation}
    i \partial_{t} G_{\bm{t}|\bm{t}} = \int_{\bm{\tilde{1}}} \left[h_{\bm{t}|\bm{\tilde{1}}} G_{\bm{\tilde{1}}|\bm{t}} - (\bm{t}\leftrightarrow\tilde{\bm{1}})\right] \ , \label{eq:eom-superindex}
\end{equation}
which is an equation of motion for the equal-time Green's function. This equation is trivially fulfilled if time-translation invariance is assumed. We state it here primarily for later use.

\subsection{First-order WI}

The first-order WI is derived by requiring that the 2p function remain invariant under the U(1) transformation. {Note that Ward identities are often derived from the invariance of the generating functional and subsequently taking functional derivatives w.r.t.\,the source fields, see, e.g., Ref.\,\onlinecite{Kopietz2010}. This approach yields an infinite hierarchy of identities, relating correlation functions of all orders, and is equivalent to demanding invariance of all correlation functions individually. Since we are here only interested in the first-order WI relating 2p and 4p functions, it is sufficient to require invariance of the 2p function, as outlined before, e.g., in Ref.\,\onlinecite{Heyder2017}. This strategy makes the derivation somewhat simpler, as it does not require any source fields.}
Using the definition, Eq.\,\eqref{eq:G}, we have
\begin{align}
    0 &\overset{!}{=} \delta G_{\bm{1}|\bm{1'}} \nonumber \\
    &= -i \int \mathcal{D}[\overline{d},d] \big\{ \delta d_{\bm{1}} \overline{d}_{\bm{1'}} + d_{\bm{1}} \delta \overline{d}_{\bm{1'}} \nonumber \\
    &\hspace{15ex} + i d_{\bm{1}} \overline{d}_{\bm{1'}} \delta S[\overline{d},d] \big\} e^{iS[\overline{d},d]} \nonumber \\
    &= -i \int \mathcal{D}[\overline{d},d] \big\{ (i\varepsilon_{\bm{1}} d_{\bm{1}}) \overline{d}_{\bm{1'}} +  d_{\bm{1}} (-i \varepsilon_{\bm{1'}}  \overline{d}_{\bm{1'}}) \nonumber \\
    &\hspace{15ex} + i d_{\bm{1}} \overline{d}_{\bm{1'}} \delta S_0[\overline{d},d] \big\} e^{iS[\overline{d},d]} \nonumber  \\
    &=  \int \mathcal{D}[\overline{d},d] \, d_{\bm{1}}  \overline{d}_{\bm{1'}} (\varepsilon_{\bm{1}} - \varepsilon_{\bm{1'}} + \delta S_0[\overline{d},d])  e^{iS[\overline{d},d]}\, .
\end{align}
Again, this must hold for arbitrary $\varepsilon_{\bm{\tilde{t}}}$, so that, using Eq.\,\eqref{eq:delta_S0},
\begin{align}
    0 &\overset{!}{=} \int \mathcal{D}[\overline{d},d] \, d_{\bm{1}}  \overline{d}_{\bm{1'}} \Big\{ \delta(\bm{1}-\bm{\tilde{t}}) -  \delta(\bm{1'}-\bm{\tilde{t}}) + \partial_{\tilde{t}} (\overline{d}_{\bm{\tilde{t}}} d_{\bm{\tilde{t}}}) \nonumber \\
    &\hspace{15ex} + i \int_{\bm{t}} (\overline{d}_{\bm{\tilde{t}}} h_{\bm{\tilde{t}}|\bm{t}} d_{\bm{t}} -  \overline{d}_{\bm{t}} h_{\bm{t}|\bm{\tilde{t}}} d_{\bm{\tilde{t}}} ) \Big\} e^{iS[\overline{d},d]} \nonumber \\
    &= \left[\delta(\bm{1}-\bm{\tilde{t}}) -  \delta(\bm{1'}-\bm{\tilde{t}})\right] \langle d_{\bm{1}}  \overline{d}_{\bm{1'}}\rangle + \partial_{\tilde{t}}\langle d_{\bm{1}}  \overline{d}_{\bm{1'}} \overline{d}_{\bm{\tilde{t}}} d_{\bm{\tilde{t}}} \rangle \nonumber\\
    &\quad + i \int_{\bm{t}} \left[ h_{\bm{\tilde{t}}|\bm{t}} \langle d_{\bm{1}}  \overline{d}_{\bm{1'}} \overline{d}_{\bm{\tilde{t}}}  d_{\bm{t}}  \rangle  - (\bm{\tilde{t}} \leftrightarrow \bm{t})\right] \, .
\end{align}
Using the definition of the 4p function, Eq.\,\eqref{eq:G4}, and relabeling indices, we obtain
\begin{align}
    0 &\overset{!}{=} \left[\delta(\bm{1} - \bm{t}) - \delta(\bm{1'} - \bm{t})\right] iG_{\bm{1}|\bm{1'}} \nonumber \\
    &\quad - i \partial_{t} G^{(4)}_{\bm{1} \bm{t} |\bm{t} \bm{1'}} + \int_{\bm{\tilde{1}}} \left[ h_{\bm{t}|\bm{\tilde{1}}}  G^{(4)}_{\bm{1} \bm{\tilde{1}} |\bm{t} \bm{1'}} -(\bm{t}\leftrightarrow\tilde{\bm{1}})\right] \ . \label{eq:WI_time}
\end{align}\\
This is the first-order U(1) WI, expressed through real-time arguments in the contour basis.

Next, we insert the tree expansion for the 4p function, Eq.\,\eqref{eq:tree-expansion}, into Eq.\,\eqref{eq:WI_time} and use the Dyson equation, Eq.\,\eqref{eq:Dyson}, as well as Eq.\,\eqref{eq:vertex} to express the U(1) WI in terms of the self-energy $\Sigma$ and the 4p vertex $\Gamma$. This gives 
\begin{align}
    &i \Sigma_{\bm{s'}|\bm{t}} \delta(\bm{t} - \bm{s}) - \delta(\bm{s'}-\bm{t}) i \Sigma_{\bm{t}|\bm{s}} \overset{!}{=}   \nonumber \\
    &\quad \int_{\bm{\tilde{1}}\bm{4'}\bm{3}}  \Big\{  \overset{\rightarrow}{\left[G_0^{-1}\right]}_{\bm{t}|\bm{\tilde{1}}} G_{\bm{\tilde{1}}|\bm{4'}}\Gamma_{\bm{s'}\bm{4'}|\bm{3}\bm{s}} G_{\bm{3}|\bm{t}} \nonumber \\
    &\hspace{10ex} -  G_{\bm{t}|\bm{4'}} \Gamma_{\bm{s'}\bm{4'}|\bm{3}\bm{s}} G_{\bm{3}|\bm{\tilde{1}}} \overset{\leftarrow}{\left[G_0^{-1}\right]}_{\bm{\tilde{1}}|\bm{t}} \Big\} \ . \label{eq:WI_sigma_gamma}
\end{align}
The derivation of this result can be found in App.\,\ref{app:rewrite_WI}.

Next, a Keldysh rotation is performed, and the open index $\bm{t}$ is contracted, as detailed in App.\,\ref{sec:Keldysh_rotation}. We furthermore assume time translation invariance and use a Fourier transform to frequency space, see App.\,\ref{app:fourier}. We also impose SU(2) spin symmetry, see App.\,\ref{app:spin}. With the short-hand notation $\nu_{\pm} = \nu \pm\frac{\omega}{2}$ (and, likewise, for $\tilde{\nu}$), the resulting equation then reads
\begin{align}
&\Sigma^{\alpha_{1'}|\overline{\alpha}_{1}}(\nu_-) -  \Sigma^{\overline{\alpha}_{1'}|\alpha_1}(\nu_+)  \nonumber\\
    &=\sum_{\alpha_{2'} \alpha_2  \alpha_{\tilde{2}} \alpha_{\tilde{1}}} \int_{\tilde{\nu}} \frac{d\tilde{\nu}}{2\pi i} \Big\{  G^{\alpha_{\tilde{2}}|\alpha_{2'}}(\tilde{\nu}_+)  \Gamma^{ \alpha_{2'} \alpha_{1'} | \alpha_2 \alpha_1}_{D}(\omega, \nu, \tilde{\nu}) \nonumber\\[-2.5ex]
    &\hspace{17ex}\times G^{\alpha_2|\alpha_{\tilde{1}}}(\tilde{\nu}_-) \left[G_0^{-1}\right]^{\alpha_{\tilde{1}}|\overline{\alpha}_{\tilde{2}}}(\tilde{\nu}_-) \nonumber \\
    &\hspace{13ex} -\left[G_0^{-1}\right]^{\overline{\alpha}_{\tilde{2}}|\alpha_{\tilde{1}}}(\tilde{\nu}_+) G^{\alpha_{\tilde{1}}|\alpha_{2'}}(\tilde{\nu}_+) \nonumber\\
    &\hspace{17ex}\times \Gamma^{ \alpha_{2'} \alpha_{1'}| \alpha_2 \alpha_1}_{D}(\omega, \nu, \tilde{\nu}) G^{\alpha_2|\alpha_{\tilde{2}}}(\tilde{\nu}_-)  \Big\}\, ,
\label{eq:WI_sigma_gamma_fullyparametrized}
\end{align}
where we applied crossing symmetry in the first two arguments of $\Gamma$ and performed a relabeling of the Keldysh indices compared to App.\,\ref{sec:Keldysh_rotation}. The U(1) WI for the self-energy has been derived in the context of lattice problems in the {Matsubara formalism} before, see, e.g., App.~A in Ref.\,\onlinecite{Krien2017} or Sec.\,E.1 in Ref.\,\onlinecite{Krien2018}. Equation \eqref{eq:WI_sigma_gamma_fullyparametrized} can be seen as a generalization of these results to the KF. The simpler form of the WI in those works, however, involves the 2PI vertex, which is at present not accessible with NRG in the KF. We therefore use the form of Eq.\,\eqref{eq:WI_sigma_gamma_fullyparametrized}, which involves only the full 4p vertex.
{Note that the WI in the KF could, in principle, also be derived from the WI in the Matsubara formalism by analytic continuation. However, such a derivation is not trivial, since it requires the analytic continuation of multipoint functions, which was only recently worked out in full detail in Ref.\,\onlinecite{Halbinger2024}. Correctly matching all Keldysh components turns out to be rather tedious, so we refrain from doing the explicit calculation here.}

Finally, using the explicit form of the inverse bare propagator for the single-impurity Anderson model without a magnetic field,
\begin{align}
    [G_0^{-1}]^{\alpha_{1'}|\alpha_1}(\nu) = \delta_{\alpha_{1'}, \overline{\alpha}_1} (\nu - \epsilon_d) - \Delta^{\alpha_{1'}|\alpha_1}(\nu)\, ,
\end{align}
see App.\,\ref{sec:G0-FT} for details, we obtain Eq.\,\eqref{eq:WI_main} from the main text.

\section{Explicit calculations}\label{app:calculations}

Most of the calculations below follow standard text-book strategies, which we formulate here in general notation, adapted to our conventions.

\subsection{Representation of Eq.\,\eqref{eq:WI_time} in terms of $\Sigma$ and $\Gamma$}\label{app:rewrite_WI}

Inserting the tree expansion for the 4p function, Eq.\,\eqref{eq:tree-expansion}, into Eq.\,\eqref{eq:WI_time} gives
\begin{align}
    0 &\overset{!}{=} \left[\delta(\bm{1} - \bm{t}) - \delta(\bm{1'} - \bm{t})\right] iG_{\bm{1}|\bm{1'}}  \nonumber \\
    &\quad -  \partial_{t}\left[ G_{\bm{1}|\bm{t}} G_{\bm{t}|\bm{1'}}\right] +  G_{\bm{1}|\bm{1'}} \partial_{t} G_{\bm{t}|\bm{t}} \nonumber \\
    &\quad -i \int_{\tilde{\bm{1}}} \left[ h_{\bm{t}|\tilde{\bm{1}}} \left(G_{\bm{1}|\bm{t}} G_{\tilde{\bm{1}}|\bm{1'}} - G_{\bm{1}|\bm{1'}} G_{\tilde{\bm{1}}|\bm{t}} \right) - (\bm{t} \leftrightarrow \tilde{\bm{1}}) \right] \nonumber \\
    &\quad -i \partial_t G^{(4)}_{c;\, \bm{1}\bm{t}|\bm{t}\bm{1'}} + \int_{\tilde{\bm{1}}} \left[h_{\bm{t}|\tilde{\bm{1}}} G^{(4)}_{c;\, \bm{1}\tilde{\bm{1}}|\bm{t}\bm{1'}}   - (\bm{t} \leftrightarrow \tilde{\bm{1}})\right] \nonumber \\
    &= \left[\delta(\bm{1} - \bm{t}) - \delta(\bm{1'} - \bm{t})\right] iG_{\bm{1}|\bm{1'}} \nonumber \\
    &\quad - \partial_{t}\left[ G_{\bm{1}|\bm{t}} G_{\bm{t}|\bm{1'}}\right] -i \int_{\tilde{\bm{1}}} \left[ h_{\bm{t}|\tilde{\bm{1}}} G_{\bm{1}|\bm{t}} G_{\tilde{\bm{1}}|\bm{1'}} - (\bm{t} \leftrightarrow \tilde{\bm{1}}) \right] \nonumber \\
    &\quad -i \partial_t G^{(4)}_{c;\, \bm{1}\bm{t}|\bm{t}\bm{1'}} + \int_{\tilde{\bm{1}}} \left[h_{\bm{t}|\tilde{\bm{1}}} G^{(4)}_{c;\, \bm{1}\tilde{\bm{1}}|\bm{t}\bm{1'}}   - (\bm{t} \leftrightarrow \tilde{\bm{1}})\right] \, . \label{eq:WI-connected}
\end{align}
In Eq.\,\eqref{eq:WI-connected}, we used Eq.\,\eqref{eq:eom-superindex} for $i \partial_{t} G_{\bm{t}|\bm{t}}$, leading to a cancellation of some terms. Now, we write the inverse bare Green's function, Eq.\,\eqref{eq:G0_inv}, as
\begin{align}
    \left[G_0^{-1}\right]_{\bm{\tilde{t}}|\bm{t}} &= \delta(\bm{\tilde{t}} - \bm{t}) i\partial_t - h_{\bm{\tilde{t}}|\bm{t}} = - \delta(\bm{\tilde{t}} - \bm{t}) i \overset{\leftarrow}{\partial_{\tilde{t}}} - h_{\bm{\tilde{t}}|\bm{t}} \nonumber\\
    \Rightarrow  \int_{\bm{\tilde{t}}}  \left[G_0^{-1}\right]_{\bm{\tilde{t}}|\bm{t}} &= i\partial_t - \int_{\bm{\tilde{t}}}  h_{\bm{\tilde{t}}|\bm{t}} = -i\overset{\leftarrow}{\partial_t} - \int_{\bm{\tilde{t}}}  h_{\bm{\tilde{t}}|\bm{t}}, \label{eq:G0_1} \\ \nonumber \\
    \text{and}\quad \left[G_0^{-1}\right]_{\bm{t}|\bm{\tilde{t}}} &= \delta(\bm{t} - \bm{\tilde{t}}) i\partial_{\tilde{t}} - h_{\bm{t}|\bm{\tilde{t}}} = -\delta(\bm{t} - \bm{\tilde{t}}) i\overset{\leftarrow}{\partial_{{t}}} - h_{\bm{t}|\bm{\tilde{t}}} \nonumber \\
    \Rightarrow  \int_{\bm{\tilde{t}}}  \left[G_0^{-1}\right]_{\bm{t}|\bm{\tilde{t}}} &= i\partial_t - \int_{\bm{\tilde{t}}}  h_{\bm{t}|\bm{\tilde{t}}} = -i\overset{\leftarrow}{\partial_t} - \int_{\bm{\tilde{t}}}  h_{\bm{t}|\bm{\tilde{t}}}. \label{eq:G0_2}
\end{align}
The second formulation arises from an integration by parts in the non-interacting action, letting the time derivative act on the barred Grassmann variable to the left of $G_0^{-1}$ in Eq.\,\eqref{eq:action}. The arising boundary term vanishes due to the closed time contour in the KF: As the time evolution returns to the same (in this case thermal) density matrix it started from at the initial time $t_0$, the Grassmann variables at the initial and final times can differ by at most a phase.
For the product $\overline{d} d$, the two phases cancel exactly. Therefore, the boundary term
\begin{align}
    \int_\mathcal{C}dt\, \partial_t [\overline{d}^j(t) d^j(t)] = \overline{d}^+(t_0) d^+(t_0) - \overline{d}^-(t_0) d^-(t_0)
\end{align}
vanishes.
We can thus rewrite the disconnected part (second line) of Eq.\,\eqref{eq:WI-connected} as
\begin{align}
    &- i \partial_{t}\left[ G_{\bm{1}|\bm{t}} G_{\bm{t}|\bm{1'}}\right] + \int_{\tilde{\bm{1}}} \left[ h_{\bm{t}|\tilde{\bm{1}}} G_{\bm{1}|\bm{t}} G_{\tilde{\bm{1}}|\bm{1'}} - (\bm{t} \leftrightarrow \tilde{\bm{1}}) \right] \nonumber\\
    &= \left[- i \partial_{t} G_{\bm{1}|\bm{t}}\right]G_{\bm{t}|\bm{1'}} -  \int_{\tilde{\bm{1}}}  h_{\tilde{\bm{1}}|\bm{t}} G_{\bm{1}|\bm{\tilde{1}}} G_{\bm{t}|\bm{1'}}  \nonumber \\
    &\quad + G_{\bm{1}|\bm{t}} \left[- i \partial_{t} G_{\bm{t}|\bm{1'}}\right] + \int_{\tilde{\bm{1}}} h_{\bm{t}|\tilde{\bm{1}}} G_{\bm{1}|\bm{t}} G_{\tilde{\bm{1}}|\bm{1'}} \nonumber \\
    &= \left[- i  G_{\bm{1}|\bm{t}} \overset{\leftarrow}{\partial_{t}} -  \int_{\tilde{\bm{1}}}   G_{\bm{1}|\bm{\tilde{1}}} h_{\tilde{\bm{1}}|\bm{t}} \right] G_{\bm{t}|\bm{1'}} \nonumber\\
    &\quad + G_{\bm{1}|\bm{t}} \left[- i \partial_{t} G_{\bm{t}|\bm{1'}} +   \int_{\tilde{\bm{1}}}  h_{\bm{t}|\tilde{\bm{1}}}G_{\tilde{\bm{1}}|\bm{1'}} \right] \nonumber\\
    &= \int_{\tilde{\bm{1}}} \left\{ G_{\bm{1}|\bm{\tilde{1}}} \overset{\leftarrow}{\left[G_0^{-1}\right]}_{\bm{\tilde{1}}|\bm{t}} G_{\bm{t}|\bm{1'}}  - G_{\bm{1}|\bm{t}} \overset{\rightarrow}{\left[G_0^{-1}\right]}_{\bm{t}|\bm{\tilde{1}}} G_{\tilde{\bm{1}}|\bm{1'}} \right\}.
\end{align}
Introducing the 4p vertex $\Gamma$ via Eq.\,\eqref{eq:vertex}, the 4p part (third line) of Eq.\,\eqref{eq:WI-connected} is written as
\begin{align}
    &-i \partial_t G^{(4)}_{c;\, \bm{1}\bm{t}|\bm{t}\bm{1'}} + \int_{\tilde{\bm{1}}} \left[h_{\bm{t}|\tilde{\bm{1}}} G^{(4)}_{c;\, \bm{1}\tilde{\bm{1}}|\bm{t}\bm{1'}}   - (\bm{t} \leftrightarrow \tilde{\bm{1}})\right] \nonumber \\
    &=\int_{\bm{3'}\bm{4'}\bm{3}\bm{4}} \Big\{ i \partial_t \left[  G_{\bm{1}|\bm{3'}} G_{\bm{t}|\bm{4'}} \Gamma_{\bm{3'}\bm{4'}|\bm{3}\bm{4}} G_{\bm{3}|\bm{t}} G_{\bm{4}|\bm{1'}} \right] \nonumber\\
    &\quad - \int_{\tilde{\bm{1}}} \big[h_{\bm{t}|\tilde{\bm{1}}} G_{\bm{1}|\bm{3'}} G_{\bm{\tilde{1}}|\bm{4'}} \Gamma_{\bm{3'}\bm{4'}|\bm{3}\bm{4}} G_{\bm{3}|\bm{t}} G_{\bm{4}|\bm{1'}} \nonumber\\
    &\qquad \qquad - h_{\tilde{\bm{1}}|\bm{t}} G_{\bm{1}|\bm{3'}} G_{\bm{t}|\bm{4'}} \Gamma_{\bm{3'}\bm{4'}|\bm{3}\bm{4}} G_{\bm{3}|\bm{\tilde{1}}} G_{\bm{4}|\bm{1'}}  \big] \Big\} \nonumber \\
    &=- \int_{\bm{3'}\bm{4'}\bm{3}\bm{4}} \Big\{ \nonumber\\
    &\quad G_{\bm{1}|\bm{3'}} \left[-i\partial_t G_{\bm{t}|\bm{4'}}  + \int_{\tilde{\bm{1}}}  h_{\bm{t}|\tilde{\bm{1}}} G_{\bm{\tilde{1}}|\bm{4'}} \right] \Gamma_{\bm{3'}\bm{4'}|\bm{3}\bm{4}} G_{\bm{3}|\bm{t}} G_{\bm{4}|\bm{1'}} \nonumber\\
    &\quad + G_{\bm{1}|\bm{3'}} G_{\bm{t}|\bm{4'}} \Gamma_{\bm{3'}\bm{4'}|\bm{3}\bm{4}} \left[-i G_{\bm{3}|\bm{t}} \overset{\leftarrow}{\partial_t} - \int_{\tilde{\bm{1}}} G_{\bm{3}|\bm{\tilde{1}}} h_{\tilde{\bm{1}}|\bm{t}} \right] G_{\bm{4}|\bm{1'}} \Big\} \nonumber \\
    &= \int_{\bm{3'}\bm{4'}\bm{3}\bm{4}} \int_{\tilde{\bm{1}}} \Big\{ G_{\bm{1}|\bm{3'}} \overset{\rightarrow}{\left[G_0^{-1}\right]}_{\bm{t}|\bm{\tilde{1}}} G_{\bm{\tilde{1}}|\bm{4'}} \Gamma_{\bm{3'}\bm{4'}|\bm{3}\bm{4}} G_{\bm{3}|\bm{t}} G_{\bm{4}|\bm{1'}} \nonumber \\
    &\qquad \qquad -  G_{\bm{1}|\bm{3'}} G_{\bm{t}|\bm{4'}} \Gamma_{\bm{3'}\bm{4'}|\bm{3}\bm{4}} G_{\bm{3}|\bm{\tilde{1}}} \overset{\leftarrow}{\left[G_0^{-1}\right]}_{\bm{\tilde{1}}|\bm{t}} G_{\bm{4}|\bm{1'}} \Big\}.
\end{align}
We thus obtain
\begin{align}
    0 &\overset{!}{=} \left[\delta(\bm{1} - \bm{t}) - \delta(\bm{1'} - \bm{t})\right] iG_{\bm{1}|\bm{1'}} \nonumber \\
    &\quad + \int_{\tilde{\bm{1}}} \Big\{ (-i)G_{\bm{1}|\bm{\tilde{1}}} \overset{\leftarrow}{\left[G_0^{-1}\right]}_{\bm{\tilde{1}}|\bm{t}} G_{\bm{t}|\bm{1'}}  +i G_{\bm{1}|\bm{t}} \overset{\rightarrow}{\left[G_0^{-1}\right]}_{\bm{t}|\bm{\tilde{1}}} G_{\tilde{\bm{1}}|\bm{1'}}  \nonumber\\
    &\quad + \int_{\bm{3'}\bm{4'}\bm{3}\bm{4}}  \Big(G_{\bm{1}|\bm{3'}} \overset{\rightarrow}{\left[G_0^{-1}\right]}_{\bm{t}|\bm{\tilde{1}}} G_{\bm{\tilde{1}}|\bm{4'}} \Gamma_{\bm{3'}\bm{4'}|\bm{3}\bm{4}} G_{\bm{3}|\bm{t}} G_{\bm{4}|\bm{1'}} \nonumber\\
    &\qquad - G_{\bm{1}|\bm{3'}} G_{\bm{t}|\bm{4'}} \Gamma_{\bm{3'}\bm{4'}|\bm{3}\bm{4}} G_{\bm{3}|\bm{\tilde{1}}} \overset{\leftarrow}{\left[G_0^{-1}\right]}_{\bm{\tilde{1}}|\bm{t}} G_{\bm{4}|\bm{1'}} \Big) \Big\}. \label{eq:WI_withG0}
\end{align}
Inserting the Dyson equation, Eq.\,\eqref{eq:Dyson}, into the second and third single-particle term and using that $\int_{\tilde{\bm{1}}} G^{-1}_{\bm{1}|\bm{\tilde{1}}} G_{\bm{\tilde{1}}|\bm{1'}} =  \int_{\tilde{\bm{1}}} G_{\bm{1}|\bm{\tilde{1}}} G^{-1}_{\bm{\tilde{1}}|\bm{1'}} = \delta(\bm{1} - \bm{1'}),
$
we get
\begin{align}
    &\int_{\tilde{\bm{1}}} \Big\{ G_{\bm{1}|\bm{\tilde{1}}} \overset{\leftarrow}{\left[G_0^{-1}\right]}_{\bm{\tilde{1}}|\bm{t}} G_{\bm{t}|\bm{1'}}  - G_{\bm{1}|\bm{t}} \overset{\rightarrow}{\left[G_0^{-1}\right]}_{\bm{t}|\bm{\tilde{1}}} G_{\tilde{\bm{1}}|\bm{1'}} \Big\} \nonumber\\ \nonumber \\
    &= \underbrace{\delta(\bm{1}-\bm{t}) G_{\bm{t}|\bm{1'}}}_{\delta(\bm{1}-\bm{t}) G_{\bm{1}|\bm{1'}}} - \underbrace{G_{\bm{1}|\bm{t}}\delta(\bm{t}-\bm{1'})}_{G_{\bm{1}|\bm{1'}}\delta(\bm{t}-\bm{1'})} \nonumber \\
    &\qquad + \int_{\tilde{\bm{1}}} \Big\{ G_{\bm{1}|\bm{\tilde{1}}} \Sigma_{\bm{\tilde{1}}|\bm{t}} G_{\bm{t}|\bm{1'}}  - G_{\bm{1}|\bm{t}} \Sigma_{\bm{t}|\bm{\tilde{1}}} G_{\tilde{\bm{1}}|\bm{1'}} \Big\},
\end{align}
and hence, using the cancellation with the first term of Eq.\,\eqref{eq:WI_withG0},
\begin{align}
    0 &\overset{!}{=} \int_{\tilde{\bm{1}}} \Big\{ i G_{\bm{1}|\bm{t}} \Sigma_{\bm{t}|\bm{\tilde{1}}} G_{\tilde{\bm{1}}|\bm{1'}} -i G_{\bm{1}|\bm{\tilde{1}}} \Sigma_{\bm{\tilde{1}}|\bm{t}} G_{\bm{t}|\bm{1'}}  \nonumber\\
    &\quad + \int_{\bm{3'}\bm{4'}\bm{3}\bm{4}}  \Big(G_{\bm{1}|\bm{3'}} \overset{\rightarrow}{\left[G_0^{-1}\right]}_{\bm{t}|\bm{\tilde{1}}} G_{\bm{\tilde{1}}|\bm{4'}} \Gamma_{\bm{3'}\bm{4'}|\bm{3}\bm{4}} G_{\bm{3}|\bm{t}} G_{\bm{4}|\bm{1'}} \nonumber \\
    &\qquad -  G_{\bm{1}|\bm{3'}} G_{\bm{t}|\bm{4'}} \Gamma_{\bm{3'}\bm{4'}|\bm{3}\bm{4}} G_{\bm{3}|\bm{\tilde{1}}} \overset{\leftarrow}{\left[G_0^{-1}\right]}_{\bm{\tilde{1}}|\bm{t}} G_{\bm{4}|\bm{1'}} \Big) \Big\}.
\end{align}
Multiplying with $\int_{\bm{1}} G^{-1}_{\bm{s'}|\bm{1}}$ from the left and with $\int_{\bm{1'}} G^{-1}_{\bm{1'}|\bm{s}}$ from the right, we obtain
\begin{align}
    0 &\overset{!}{=} \int_{\tilde{\bm{1}}} \Big\{ i \delta(\bm{s'}-\bm{t}) \Sigma_{\bm{t}|\bm{\tilde{1}}} \delta(\bm{\tilde{1}}-\bm{s}) -i \delta(\bm{s'}-\bm{\tilde{1}}) \Sigma_{\bm{\tilde{1}}|\bm{t}} \delta(\bm{t} - \bm{s})  \nonumber\\
    &\quad + \int_{\bm{3'}\bm{4'}\bm{3}\bm{4}} \Big(\delta(\bm{s'}-\bm{3'}) \overset{\rightarrow}{\left[G_0^{-1}\right]}_{\bm{t}|\bm{\tilde{1}}} G_{\bm{\tilde{1}}|\bm{4'}} \Gamma_{\bm{3'}\bm{4'}|\bm{3}\bm{4}} G_{\bm{3}|\bm{t}} \delta(\bm{4}-\bm{s}) \nonumber \\
    &\qquad -  \delta(\bm{s'}-\bm{3'}) G_{\bm{t}|\bm{4'}} \Gamma_{\bm{3'}\bm{4'}|\bm{3}\bm{4}} G_{\bm{3}|\bm{\tilde{1}}} \overset{\leftarrow}{\left[G_0^{-1}\right]}_{\bm{\tilde{1}}|\bm{t}} \delta(\bm{4}-\bm{s}) \Big) \Big\} \nonumber \\
    &= \delta(\bm{s'}-\bm{t}) i \Sigma_{\bm{t}|\bm{s}} -  \delta(\bm{t} - \bm{s}) i \Sigma_{\bm{s'}|\bm{t}} \nonumber \\
    &\quad + \int_{\bm{\tilde{1}}\bm{4'}\bm{3}}  \Big\{  \overset{\rightarrow}{\left[G_0^{-1}\right]}_{\bm{t}|\bm{\tilde{1}}} G_{\bm{\tilde{1}}|\bm{4'}}\Gamma_{\bm{s'}\bm{4'}|\bm{3}\bm{s}} G_{\bm{3}|\bm{t}} \nonumber \\
    &\qquad \qquad \quad -  G_{\bm{t}|\bm{4'}} \Gamma_{\bm{s'}\bm{4'}|\bm{3}\bm{s}} G_{\bm{3}|\bm{\tilde{1}}} \overset{\leftarrow}{\left[G_0^{-1}\right]}_{\bm{\tilde{1}}|\bm{t}} \Big\}\, ,
\end{align}
which is Eq.\,\eqref{eq:WI_sigma_gamma}.

\subsection{Keldysh rotation of Eq.\,\eqref{eq:WI_sigma_gamma}}\label{sec:Keldysh_rotation}

The Green's functions in the Keldysh and contour bases are related by the Keldysh rotation $G^{\alpha|\alpha'} = D^{\alpha|j} G^{j|j'} (D^{-1})^{j'|\alpha'}$, with the matrices
\begin{align}
    D &= \frac{1}{\sqrt{2}}
    \begin{pmatrix}
        1 & -1 \\
        1 & 1
    \end{pmatrix}\, ; &
    D^{-1} &= \frac{1}{\sqrt{2}}
    \begin{pmatrix}
         1 & 1 \\
        -1 & 1
    \end{pmatrix}\\ \nonumber \\
    (G^{\alpha|\alpha'}) &=
    \begin{pmatrix}
        0   & G^A \\
        G^R & G^K
    \end{pmatrix}\, ; &
    (G^{j|j'}) &= 
    \begin{pmatrix}
        G^{-|-} & G^{-|+} \\
        G^{+|-} & G^{+|+}
    \end{pmatrix}.
\end{align}
The inverse transformation is $G^{j|j'} = (D^{-1})^{j|\alpha} G^{\alpha|\alpha'} D^{\alpha'|j'}$ (summation convention implied). The same transformation applies to the self-energy, whose Keldysh structure reads
\begin{align}
    &(\Sigma^{\alpha'|\alpha}) = 
    \begin{pmatrix}
        \Sigma^{1|1} & \Sigma^{1|2} \\
        \Sigma^{2|1} & \Sigma^{2|2}  
    \end{pmatrix} =
    \begin{pmatrix}
        \Sigma^K   & \Sigma^R \\
        \Sigma^A & 0
    \end{pmatrix}
    \, . \label{eq:Sigma_Keldysh_structure}
\end{align}
Likewise, for the vertex one has
\begin{align}
    \Gamma^{j_{1'}j_{2'}|j_1 j_2} = (D^{-1})^{j_{1'}|\alpha_{1'}} (D^{-1})^{j_{2'}|\alpha_{2'}} \Gamma^{\alpha_{1'}\alpha_{2'}|\alpha_1 \alpha_2} D^{\alpha_1|j_1} D^{\alpha_2|j_2}\, .
\end{align}

To perform the Keldysh rotation of Eq.\,\eqref{eq:WI_sigma_gamma}, we proceed as follows. First, to avoid a trivially vanishing result after contracting the open multi-index $\bm{t}$, we multiply the whole equation with the contour index $-j_t$. We then contract $\bm{t}$, leaving out the integration over time for now, as that will follow later when doing the Fourier transformation into frequency space. Only focusing on the Keldysh index structure, this gives
\begin{align}
    (-j_s)\Sigma^{j_{s'}|j_s} -  (-j_{s'})\Sigma^{j_{s'}|j_s}
\end{align}
for the LHS of Eq.\,\eqref{eq:WI_sigma_gamma}. The Keldysh rotation is now performed by multiplying with $D$ from the left and with $D^{-1}$ from the right. To compute the Keldysh rotation of $-j_{s'}\Sigma^{s'|s}$, we write it as a matrix product, $-j_{s'}\Sigma^{s'|s} =  \sum_{\tilde{s}} \sigma_z^{s'|\tilde{s}}\Sigma^{\tilde{s}|s} =  (\sigma_z \Sigma^c)^{s'|s}$, where $\sigma_z$ is the third Pauli matrix and the superscript $c$ of $\Sigma$ in the last expression indicates that it is given in the contour basis. For the Keldysh basis, we use the superscript $k$. Applying the Keldysh rotation and inserting an identity gives
\begin{align}
     (D\sigma_z \Sigma^c D^{-1})^{\alpha'|\alpha} &=  (D\sigma_z D^{-1} D \Sigma^c D^{-1})^{\alpha'|\alpha} = (\sigma_x \Sigma^k) ^{\alpha'|\alpha} \nonumber \\
     &= \Sigma^{\overline{\alpha}'|\alpha}.
\end{align}
Here, we used $D\sigma_z D^{-1}=\sigma_x$. This first Pauli matrix flips the corresponding Keldysh index, which is what the bar over the first Keldysh index denotes in the final expression. Concretely, $\bar{1}=2;\, \bar{2}=1$.
The other term, $-j_{s}\Sigma^{s'|s} =  (\Sigma^c \sigma_z)^{s'|s}$, is transformed analogously and gives $\Sigma^{\alpha'|\overline{\alpha}}$. After the Keldysh rotation, the LHS of Eq.\,\eqref{eq:WI_sigma_gamma} thus reads $ \Sigma^{\alpha_{s'}|\overline{\alpha}_s} - \Sigma^{\overline{\alpha}_{s'}|\alpha_s}$.

The right-hand side of the WI, Eq.\,\eqref{eq:WI_sigma_gamma}, is transformed analogously. Again focusing only on the Keldysh index structure, after contracting $\bm{t}$, the first term can be written as
\begin{align}
    &\sum_{j_t, j_{\tilde{1}}, j_{4'}, j_3} (-j_t) [G_0^{-1}]^{j_t|j_{\tilde{1}}} G^{j_{\tilde{1}}|j_{4'}} \Gamma^{j_{s'}j_{4'}|j_3 j_{s}} G^{j_3|j_t} \nonumber \\
    &= \mathrm{Tr}\left\{\sigma_z [G_0^{-1}]^c G^c \Gamma^{c; j_{s'}|j_s} G^c\right\} \nonumber\\
    &= \mathrm{Tr}\left\{D\sigma_z D^{-1} D [G_0^{-1}]^c D^{-1} D G^c D^{-1} D \Gamma^{c; j_{s'}|j_s} D^{-1} D G^c D^{-1}\right\}\, ,
\end{align}
where we inserted identities and used the cyclicity of the trace. Again using $D\sigma_z D^{-1}=\sigma_x$, which flips the corresponding Keldysh index, and performing the Keldysh rotation for the two remaining open indices in $\Gamma$, the first term of the RHS of Eq.\,\eqref{eq:WI_sigma_gamma} reads
\begin{align}
    \sum_{\alpha_t, \alpha_{\tilde{1}}, \alpha_{4'}, \alpha_3} [G_0^{-1}]^{\overline{\alpha}_t|\alpha_{\tilde{1}}} G^{\alpha_{\tilde{1}}|\alpha_{4'}} \Gamma^{\alpha_{s'}\alpha_{4'}|\alpha_3 \alpha_{s}} G^{\alpha_3|\alpha_t}\, .
\end{align}
The second term is transformed analogously, such that the Keldysh structure of the full WI, Eq.\,\eqref{eq:WI_sigma_gamma}, reads
\begin{align}
    &\Sigma^{\alpha_{s'}|\overline{\alpha}_s} - \Sigma^{\overline{\alpha}_{s'}|\alpha_s} \nonumber \\
    &= \sum_{\alpha_t, \alpha_{\tilde{1}}, \alpha_{4'}, \alpha_3} \Big\{ [G_0^{-1}]^{\overline{\alpha}_t|\alpha_{\tilde{1}}} G^{\alpha_{\tilde{1}}|\alpha_{4'}} \Gamma^{\alpha_{s'}\alpha_{4'}|\alpha_3 \alpha_{s}} G^{\alpha_3|\alpha_t} \nonumber \\
    &\hspace{12 ex} - G^{\alpha_{t}|\alpha_{4'}} \Gamma^{\alpha_{s'}\alpha_{4'}|\alpha_3 \alpha_{s}} G^{\alpha_3|\alpha_{\tilde{1}}} [G_0^{-1}]^{{\alpha}_{\tilde{1}}|\overline{\alpha}_{t}} \Big\}
\end{align}
after Keldysh rotation. In a final step, we apply crossing symmetry to the first two arguments of $\Gamma$ for a favorable frequency parametrization later on. This yields an additional minus sign and swaps the first two Keldysh indices of the vertices, such that the Keldysh structure of Eq.\,\eqref{eq:WI_sigma_gamma} can be written as
\begin{align}
    &\Sigma^{\alpha_{s'}|\overline{\alpha}_s} - \Sigma^{\overline{\alpha}_{s'}|\alpha_s} \nonumber \\
    &= \sum_{\alpha_t, \alpha_{\tilde{1}}, \alpha_{4'}, \alpha_3} \Big\{  G^{\alpha_{t}|\alpha_{4'}} \Gamma^{\alpha_{4'}\alpha_{s'}|\alpha_3 \alpha_{s}} G^{\alpha_3|\alpha_{\tilde{1}}} [G_0^{-1}]^{{\alpha}_{\tilde{1}}|\overline{\alpha}_{t}}  \nonumber \\
    &\hspace{12 ex} - [G_0^{-1}]^{\overline{\alpha}_t|\alpha_{\tilde{1}}} G^{\alpha_{\tilde{1}}|\alpha_{4'}} \Gamma^{\alpha_{4'}\alpha_{s'}|\alpha_3 \alpha_{s}} G^{\alpha_3|\alpha_t}\Big\}
\end{align}

\subsection{Fourier transform of Eq.\,\eqref{eq:WI_sigma_gamma}}\label{app:fourier}

We insert the Fourier transforms of all functions, which read $G(t_1|t_{1'}) = \int_{\nu_1 \nu_{1'}} e^{i\nu_1 t_1}  G(\nu_1| \nu_{1'}) e^{-i\nu_{1'} t_{1'}}$ for all 2p functions $G$, $G_0^{-1}$, and $\Sigma$ and $\Gamma(t_{1'}t_{2'}|t_1 t_2) =  \int_{\nu_{1'}\nu_{2'}\nu_1\nu_2} e^{i\nu_{1'} t_{1'}} e^{i\nu_{2'} t_{2'}}  \Gamma(\nu_{1'}\nu_{2'}|\nu_1 \nu_2) e^{-i\nu_1 t_1}  e^{-i\nu_2 t_2}$ for the 4p vertex. Here and from now on, we use the compact notation $\int\frac{d\nu}{2\pi i} = \int_\nu$ for frequency integrals. In this section, we temporarily drop the Keldysh and spin indices of all functions and purely focus on their time- and frequency-dependence.

We transform the whole Eq.\,\eqref{eq:WI_sigma_gamma} with respect to $t_s$ and $t_{s'}$ by applying $\int_{t_{s'} t_s} e^{-i\nu' t_{s'}} e^{i\nu t_s}$ on both sides. We furthermore divide the whole equation by $i$. For the LHS, we get
\begin{align}
    &\int_{t_{s'} t_s} e^{-i\nu' t_{s'}} e^{i\nu t_s} \Big\{ \delta(t_t-t_s) \int_{\nu_{s'}\nu_t} e^{i\nu_{s'}t_{s'}} e^{-i\nu_t t_t} \Sigma(\nu_{s'}|\nu_t) \nonumber \\
    &\qquad - \delta(t_{s'}-t_t) \int_{\nu_t \nu_s} e^{i\nu_t t_t} e^{-i\nu_s t_s}  \Sigma(\nu_t|\nu_s) \Big\} \nonumber \\
    &=   e^{i\nu t_t} \int_{\nu_{s'}\nu_t} e^{-i\nu_t t_t} \Sigma(\nu_{s'}|\nu_t) \int_{t_{s'}} e^{i(\nu_{s'}-\nu')t_{s'}} \nonumber \\
    &\quad -  e^{-i\nu' t_t} \int_{\nu_t \nu_s} e^{i\nu_t t_t}  \Sigma(\nu_t|\nu_s) \int_{t_s} e^{i(\nu-\nu_s)t_s} \nonumber \\
    &=   e^{i\nu t_t} \int_{\nu_t} e^{-i\nu_t t_t} \Sigma(\nu'|\nu_t) -  e^{-i\nu' t_t} \int_{\nu_t } e^{i\nu_t t_t}  \Sigma(\nu_t|\nu) \nonumber\\
    &= e^{i(\nu - \nu') t_t}\left[ \Sigma(\nu') -   \Sigma(\nu) \right]\, .
\end{align}
In the last step, we imposed time-translation invariance, which entails frequency conservation, $\Sigma(\nu'|\nu) \equiv \Sigma(\nu) \delta(\nu' - \nu)$.

The transformation of the RHS is more tedious, but straightforward, as proceeds analogously. It gives
\begin{widetext}
\begin{align}
    e^{i(\nu - \nu') t_t} \int_{\nu_t} \Big\{ \left[G_0^{-1}\right](\nu_t) G(\nu_t) \Gamma(\nu', \nu_t|\nu'-\nu+\nu_t, \nu) G(\nu'-\nu+\nu_t) - G(\nu_t + \nu - \nu') \Gamma(\nu', \nu_t+\nu-\nu'|\nu_t, \nu) G({\nu}_t) \left[G_0^{-1}\right](\nu_t) \Big\}\, .
\end{align}
\end{widetext}
Here, we used energy conservation both for the 2p functions and for the 4p vertex, for which we have $\Gamma(\nu_{1'}\nu_{2'}|\nu_1\nu_2)\equiv \Gamma(\nu_{1'},\nu_{2'}|\nu_1,\nu_2) \delta(\nu_{1'}+\nu_{2'}-\nu_1-\nu_2) $. We now perform a final Fourier transform with respect to $t_t$, applying $\int_{t_t} e^{i\omega t_t}$ to the full equation with the transfer frequency $\omega$. This yields the delta function $\delta(\nu - \nu' +\omega)$, which allows us to replace $\nu'=\nu+\omega$ by formally integrating over $\nu'$. The full WI in frequency space thus reads
\begin{align}
    &\Sigma(\nu + \omega) -  \Sigma(\nu)\nonumber\\
    &= \int_{\nu_t} \Big\{ \left[G_0^{-1}\right](\nu_t) G(\nu_t) \Gamma(\nu + \omega, \nu_t|\nu_t + \omega, \nu) G(\nu_t + \omega) \nonumber \\
    &\qquad - G(\nu_t - \omega) \Gamma(\nu + \omega, \nu_t - \omega|\nu_t, \nu) G({\nu}_t) \left[G_0^{-1}\right](\nu_t) \Big\}\, . \\ \nonumber
\end{align}
To make the frequency parametrizations of the vertices of both terms on the RHS match, we now shift $\nu_t \rightarrow \nu_t + \omega$ in the second term and subsequently rename $\nu_t\rightarrow\tilde{\nu}$, which gives
\begin{align}
    &\Sigma(\nu + \omega) -  \Sigma(\nu)\nonumber\\
    &= \int_{\tilde{\nu}} \Big\{ \left[G_0^{-1}\right](\tilde{\nu}) G(\tilde{\nu}) \Gamma(\nu + \omega, \tilde{\nu}|\tilde{\nu} + \omega, \nu) G(\tilde{\nu} + \omega) \nonumber \\
    &\qquad - G(\tilde{\nu}) \Gamma(\nu + \omega, \tilde{\nu}|\tilde{\nu} + \omega, \nu) G(\tilde{\nu} + \omega) \left[G_0^{-1}\right](\tilde{\nu} + \omega) \Big\}\, .
\end{align}
Finally, we shift the external fermionic frequency $\nu \rightarrow \nu - \omega/2$ and the integration frequency $\tilde{\nu} \rightarrow \tilde{\nu} - \omega/2$ and subsequently flip $\omega \rightarrow - \omega$ to symmetrize the equation. Using the short-hand notation $\nu_{\pm} = \nu \pm\frac{\omega}{2}$ (and, likewise, for $\tilde{\nu}$) again, we arrive at
\begin{align}
    &\Sigma(\nu_-) -  \Sigma(\nu_+) \nonumber\\
    &= \int_{\tilde{\nu}} \Big\{ \left[G_0^{-1}\right](\tilde{\nu}_+) G(\tilde{\nu}_+) \Gamma(\nu_-, \tilde{\nu}_+|\tilde{\nu}_-, \nu_+) G(\tilde{\nu}_-) \nonumber \\
    &\qquad \quad - G(\tilde{\nu}_+) \Gamma(\nu_-, \tilde{\nu}_+|\tilde{\nu}_-, \nu_+) G(\tilde{\nu}_-) \left[G_0^{-1}\right](\tilde{\nu}_-) \Big\}\, .
\end{align}
This way, the vertex is parametrized in the $a$ channel convention as defined in App.\,A of Ref.\,\onlinecite{Walter2022}. In a final step, we apply crossing symmetry in the first two arguments of $\Gamma$:
\begin{align}
    &\Sigma(\nu_-) -  \Sigma(\nu_+) \nonumber \\
    &= \int_{\tilde{\nu}} \Big\{ G(\tilde{\nu}_+) \Gamma(\tilde{\nu}_+, \nu_-|\tilde{\nu}_-, \nu_+) G(\tilde{\nu}_-) \left[G_0^{-1}\right](\tilde{\nu}_-) \nonumber \\
    &\qquad \quad  - \left[G_0^{-1}\right](\tilde{\nu}_+) G(\tilde{\nu}_+) \Gamma(\tilde{\nu}_+, \nu_-|\tilde{\nu}_-, \nu_+) G(\tilde{\nu}_-) \Big\}\, .
\end{align}
At the expense of a minus sign, the vertex is then parametrized in the $t$ channel parametrization and we will susequently write $\Gamma(\tilde{\nu}_+, \nu_-|\tilde{\nu}_-, \nu_+) = \Gamma_t(\omega, \nu, \tilde{\nu})$.

\subsection{Spin structure of Eq.\,\eqref{eq:WI_sigma_gamma} in the case of SU(2) symmetry}\label{app:spin}
After contracting the open index $\sigma_t$, the spin structure of Eq.\,\eqref{eq:WI_sigma_gamma} reads 
\begin{align}
    &\mathrm{LHS}_{\sigma_s'|\sigma_s} = \sum_{\sigma_t,\sigma_{\tilde{1}},\sigma_{4'},\sigma_3} \Big([G_0^{-1}]_{\sigma_t|\sigma_{\tilde{1}}} G_{\sigma_{\tilde{1}}|\sigma_{4'}} \Gamma_{\sigma_{s'}\sigma_{4'}|\sigma_{3}\sigma_s} G_{\sigma_3|\sigma_t} \nonumber \\
    &\qquad\qquad\quad - G_{\sigma_t|\sigma_{4'}} \Gamma_{\sigma_{s'}\sigma_{4'}|\sigma_{3}\sigma_s} G_{\sigma_3|\sigma_{\tilde{1}}} [G_0^{-1}]_{\sigma_{\tilde{1}}|\sigma_t} \Big)\, ,
\end{align}
where we abbreviated the left-hand side as 
\begin{align}
    \Sigma^{\alpha_{1'}|\overline{\alpha}_{1}}_{\sigma_{s'}|\sigma_s}(\nu - \tfrac{\omega}{2}) -  \Sigma^{\overline{\alpha}_{1'}|\alpha_1}_{\sigma_{s'}|\sigma_s}(\nu + \tfrac{\omega}{2}) \equiv \mathrm{LHS}_{\sigma_s'|\sigma_s}\, .
\end{align}
We now consider the case in which SU(2) symmetry holds. This implies that all 2p functions are diagonal in their spin arguments, e.g. $\Sigma_{\sigma_{1'}|\sigma_1} \sim \delta_{\sigma_{1'},\sigma_1}$. For the 4p vertex, we have $\Gamma_{\sigma_{1'} \sigma_{2'}|\sigma_1 \sigma_2} \sim \delta_{\sigma_{1'}+\sigma_{2'},\sigma_1+\sigma_2}$. Restricting ourselves to $\sigma_{s'}=\sigma_s=\uparrow$, we have
\begin{align}
    &\mathrm{LHS}_{\uparrow|\uparrow} = \sum_{\sigma} \Big([G_0^{-1}]_{\sigma|\sigma} G_{\sigma|\sigma} \Gamma_{\uparrow\sigma|\sigma\uparrow} G_{\sigma|\sigma} \nonumber \\
    &\qquad \qquad \qquad - G_{\sigma|\sigma} \Gamma_{\uparrow\sigma|\sigma\uparrow} G_{\sigma|\sigma} [G_0^{-1}]_{\sigma|\sigma} \Big)\, . \label{eq:spin-structure}
\end{align}
Using $G_{\uparrow|\uparrow} = G_{\downarrow|\downarrow}$, we can suppress the spin-indices for the 2p functions and write
\begin{align}
    \mathrm{LHS} = \Big([G_0^{-1}]\, G\, \Gamma_{\uparrow\uparrow + \overline{\uparrow\downarrow}}\, G - G\, \Gamma_{\uparrow\uparrow + \overline{\uparrow\downarrow}}\, G\, [G_0^{-1}] \Big)\, . \label{eq:spin-structure2}
\end{align}
where we used the notation introduced in App.\,\ref{app:BSE-violation},
$\Gamma_{\uparrow\downarrow|\uparrow\downarrow} \equiv \Gamma_{\uparrow\downarrow}$, $\Gamma_{\uparrow\downarrow|\downarrow\uparrow} \equiv \Gamma_{\overline{\uparrow\downarrow}}$, $\Gamma_{\uparrow\uparrow|\uparrow\uparrow} \equiv \Gamma_{\uparrow\uparrow}$ and $ \Gamma_{\overline{\uparrow\downarrow} + \uparrow\uparrow} =\Gamma_{\overline{\uparrow\downarrow}} +  \Gamma_{\uparrow\uparrow}$. Again applying crossing symmetry in the first two arguments of $\Gamma$ yields the $\uparrow\uparrow + \uparrow\downarrow = D$ spin component, so we write
\begin{align}
    \mathrm{LHS} = \Big(G\, \Gamma_{D}\, G\, [G_0^{-1}]  - [G_0^{-1}]\, G\, \Gamma_{D}\, G \Big)\, .
\end{align}
As mentioned in Sec.\,\ref{sec:WI}, in addition to U(1) symmetry, the SU(2) symmetry of the action can be exploited as well to derive another, almost identical, WI. Its derivation works in almost the same way, the only difference being that the generators of SU(2) transformations, i.e.\,the Pauli matrices, modify the spin structure of the equation. As explained in Ref.\,\cite{Kopietz2010}, the result is given by a slight modification of Eq.\,\eqref{eq:spin-structure},
\begin{align}
    &\mathrm{LHS}_{\uparrow|\uparrow} = \sum_{\sigma} \sigma \Big([G_0^{-1}]_{\sigma|\sigma} G_{\sigma|\sigma} \Gamma_{\uparrow\sigma|\sigma\uparrow} G_{\sigma|\sigma}  \nonumber \\
    &\qquad \qquad\qquad \qquad   - G_{\sigma|\sigma} \Gamma_{\uparrow\sigma|\sigma\uparrow} G_{\sigma|\sigma} [G_0^{-1}]_{\sigma|\sigma} \Big)\, ,
\end{align}
where $\sigma=\, \uparrow \ \rightarrow +1$ and $\sigma=\, \downarrow \ \rightarrow -1$. Compared to Eq.\,\eqref{eq:spin-structure}, this only changes the sign with which the $\overline{\uparrow\downarrow}$ component enters in Eq.\,\eqref{eq:spin-structure2}. Once again applying crossing symmetry to parametrize the vertex in the $t$ channel yields the $\uparrow\uparrow - \uparrow\downarrow = M$. The rest of the WI is unchanged. In this work, we do not discuss the SU(2) WI further.

\subsection{Fourier transform, Keldysh rotation, and explicit form of $G_0^{-1}$ for the single-impurity Anderson model}\label{sec:G0-FT}

In this section, we compute the Fourier transform of the inverse bare Green's function $G_0^{-1}$ and its Keldysh rotation explicitly. As seen in Eqs.\,\eqref{eq:G0_1} and \eqref{eq:G0_2}, we can write $G_0^{-1}$ using derivatives acting either to the left or to the right. Both versions must yield the same result for the Fourier transform, which we will now show. Starting with the derivative acting to the right, we compute
\begin{align}
    &[G_0^{-1}]_{1'|1}(\nu_{1'}|\nu_1) = \int_{t_{1'}t_1} e^{-i\nu_{1'} t_{1'}}  [\overset{\rightarrow}{G_0^{-1}}]_{1'|1}(t_{1'}|t_1)\, e^{i\nu_{1} t_{1}} \nonumber\\
    &= \int_{t_{1'}t_1} e^{-i\nu_{1'} t_{1'}}  \left[\delta_{1',1} \delta_\mathcal{C}(t_{1'}-t_1) i \overset{\rightarrow}{\partial}_{t_1} - h_{1'|1}(t_{1'}|t_1)\right] e^{i\nu_{1} t_{1}} \nonumber \\
    &= \int_{t_{1'}t_1} e^{-i\nu_{1'} t_{1'}}  \left[\delta_{1',1} \delta_\mathcal{C}(t_{1'}-t_1) (-\nu_1) - h_{1'|1}(t_{1'}|t_1)\right] e^{i\nu_{1} t_{1}} \nonumber \\
    &= j_{1'} \delta_{1',1} \nu_1 \int_{t_1} e^{i(\nu_1-\nu_{1'})t_1} - \int_{t_{1'}t_1} e^{-i\nu_{1'} t_{1'}}  h_{1'|1}(t_{1'}|t_1) e^{i\nu_{1} t_{1}} \nonumber \\
    &= j_{1'} \delta_{1'|,1} \nu_1 \delta(\nu_1 - \nu_{1'}) - h_{1'|1}(\nu_{1'}|\nu_1).
\end{align}
Likewise, using the derivative acting to the left, we obtain
\begin{align}
    &\int_{t_{1'}t_1} e^{-i\nu_{1'} t_{1'}}  \left[\delta_{1',1} \delta_\mathcal{C}(t_{1'}-t_1) (-\nu_{1'}) - h_{1'|1}(t_{1'}|t_1)\right] e^{i\nu_{1} t_{1}} \nonumber \\
    &= j_{1} \delta_{1',1} \nu_1 \delta(\nu_1 - \nu_{1'}) - h_{1'|1}(\nu_{1'}|\nu_1),
\end{align}
which is the same result. Writing the first term in matrix form 
    $([G_0^{-1}]^{j'|j}_{\nu-\mathrm{part}}) = {-\nu \; 0 \choose 0 \ \nu }$,
we perform a Keldysh rotation as in Sec.\,\ref{sec:Keldysh_rotation}, multiplying with $D^{-1}$ from the left and with $D$ from the right to obtain
     $([G_0^{-1}]^{\alpha'|\alpha}_{\nu-\mathrm{part}}) = {0 \; \nu \choose \nu \; 0}$,
which is the expected result. Using energy conservation, writing $[G_0^{-1}]_{1'|1}(\nu_{1'}|\nu_1) = [G_0^{-1}]_{1'|1}(\nu_1) \delta(\nu_{1'} - \nu_1)$, we therefore have 
\begin{align}
    [G_0^{-1}]_{\sigma_{1'}|\sigma_1}^{\alpha_{1'}|\alpha_1}(\nu) = \delta_{\alpha_{1'}, \overline{\alpha}_1} \delta_{\sigma_{1'},\sigma_1}\, \nu - h_{\sigma_{1'}|\sigma_1}^{\alpha_{1'}|\alpha_1}(\nu)\, .
\end{align}
For the single-impurity Anderson model without a magnetic field, the single-particle Hamiltonian is given by the shift of the impurity level plus the hybridization function, 
\begin{align}
    h_{\sigma_{1'}|\sigma_1}^{\alpha_{1'}|\alpha_1}(\nu) = \delta_{\alpha_{1'}, \overline{\alpha}_1} \delta_{\sigma_{1'},\sigma_1}\, \epsilon_d + \Delta_{\sigma_{1'}|\sigma_1}^{\alpha_{1'}|\alpha_1}(\nu)\, .
\end{align}

\subsection{Derivation of Heyder's result for the special case $\omega=0$}\label{app:Heyder}
We obtain the special case of the WI already studied in the literature \cite{Heyder2017, Walter2022} by taking
$\alpha_{1'} = \alpha_1 = 2$ and setting $\omega\equiv 0$ in Eq.\,\eqref{eq:WI_main}. The LHS of Eq.\,\eqref{eq:WI_main} then becomes
\begin{align*}
    \Sigma^{2|1}(\nu) - \Sigma^{1|2}(\nu) &=  \Sigma^{A}(\nu) -  \Sigma^{R}(\nu) = -2i\, \mathrm{Im} \Sigma^{R}(\nu)\, .
\end{align*}
Using that the frequency arguments of all 2p functions are identical in this case, we focus only on the Keldysh structure of $-$RHS of Eq.\,\eqref{eq:WI_main}, which we write as the trace over matrix products in Keldysh space,
\begin{align}
    &\mathrm{Tr} \Bigg\{ 
    \begin{pmatrix}
        \Delta^A & 0 \\ \Delta^K & \Delta^R
    \end{pmatrix}\! 
    \begin{pmatrix}
        0 & G^A \\ G^R & G^K
    \end{pmatrix}\!
    \begin{pmatrix}
        \Gamma^{1|1} & \Gamma^{1|2} \\ \Gamma^{2|1} & 0
    \end{pmatrix}\!
    \begin{pmatrix}
        0 & G^A \\ G^R & G^K
    \end{pmatrix}\! \nonumber \\
    &-
    \begin{pmatrix}
        0 & G^A \\ G^R & G^K
    \end{pmatrix}
    \begin{pmatrix}
        \Gamma^{1|1} & \Gamma^{1|2} \\ \Gamma^{2|1} & 0
    \end{pmatrix}
    \begin{pmatrix}
        0 & G^A \\ G^R & G^K
    \end{pmatrix}
    \begin{pmatrix}
        \Delta^R & \Delta^K \\ 0 &  \Delta^A
    \end{pmatrix}
    \Bigg\} \, .
\end{align}
The first term on the RHS of Eq.\,\eqref{eq:WI_main}, being $\sim \omega$, obviously vanishes. Here, we have already flipped the Keldysh index $\alpha_{\tilde{2}}$ of the hybridization functions and fixed the Keldysh indices $\alpha_{1'}$ and $\alpha_1$ of $\Gamma$ to $2$, using that $\Gamma^{22|22}=0$ by causality. Evaluating the matrix product and computing the trace gives
\begin{align}
    &\Delta^R G^R (\Gamma^{1|1} G^A + \Gamma^{1|2} G^K) + (\Delta^K G^A + \Delta^R G^K) \Gamma^{2|1} G^A \nonumber \\
    &- (G^R \Gamma^{1|1} + G^K \Gamma^{2|1}) G^A \Delta^A  - G^R \Gamma^{1|2} (G^R\Delta^K + G^K \Delta^A)\, .
\end{align}
Reshuffling terms and using the FDR $G^K = \mathrm{th} (G^R - G^A)$ and likewise for $\Delta^K$, where ``$\mathrm{th}$'' is a short-hand notation for $\tanh(\tfrac{\nu}{2T}) = 1 - 2 n_\mathrm{F}(\nu)$, several terms cancel and we obtain
\begin{align}
    &G^R G^A (\Delta^R - \Delta^A) \Gamma^{1|1} + \mathrm{th} \big[ \Gamma^{1|2} (G^R \Delta^A G^A - G^R \Delta^R G^A) \nonumber \\
    &\qquad \qquad\qquad \qquad \qquad + \Gamma^{2|1}(\Delta^R G^A G^R - G^A \Delta^A G^R) \big] \nonumber \\
    &= G^R G^A (\Delta^R - \Delta^A) \left[\Gamma^{1|1} - \mathrm{th} (\Gamma^{1|2} - \Gamma^{2|1}) \right]\, .
\end{align}
In the wide-band limit, where $\Delta^R - \Delta^A = -2i\Delta$, the whole equation becomes
\begin{align}
    &2\, \mathrm{Im} \Sigma^{R}(\nu) = 2i \Delta \int_{\tilde{\nu}} G^R(\tilde{\nu}) G^A(\tilde{\nu}) \Big\{ \Gamma^{21|12}(\nu,\tilde{\nu}|\tilde{\nu},\nu) \nonumber \\
    &\quad - [1 - 2 n_\mathrm{F}(\tilde{\nu})] \left[ \Gamma^{21|22}(\nu,\tilde{\nu}|\tilde{\nu},\nu) - \Gamma^{22|12}(\nu,\tilde{\nu}|\tilde{\nu},\nu)\right] \Big\}\, ,
\end{align}
where we reinstated the frequency arguments. Multiplying the whole equation with $(-1)$ and using crossing symmetry for the vertices twice, this becomes precisely Eq.\,(8.13) in Ref.\,\onlinecite{Walter2022}.

\section{Diagrammatic representation of the U(1) WI}

In this section, we provide a compact diagrammatic representation of the U(1) WI. This representation is useful to motivate the result of the Keldysh rotation and of the Fourier transform carried out explicitly in in App.~\ref{sec:Keldysh_rotation} and App.~\ref{app:fourier}.

Introducing the bare 3p ``Hedin'' vertex as
\begin{align}
    \delta_{\bm{1'}\bm{t}\bm{1}} &= 
    \delta(\bm{1'}-\bm{t}) \delta(\bm{t}-\bm{1})
    =
    \tikzm{Hedin}{
    \threepointvertexleftarrows{$\delta$}{0}{0}{1}
    \boson{0.6}{0}{1.2}{0}
    \node at (-0.5, 0.75) {$\bm{1}$};
    \node at (-0.5, -0.75) {$\bm{1'}$};
     \node at (1.4, 0) {$\bm{t}$};
    }
    \, ,
\label{eq:def_bareHedinVertex}
\end{align}
where $\bm{t}$ labels a ``bosonic'' multi-index that is contracted, Eq.~\eqref{eq:WI_sigma_gamma} can be written as
\begin{align}
    &i 
    \Sigma_{\bm{s'}|\bm{1}} \delta_{\bm{1}\bm{t} \bm{s}} - i  \delta_{\bm{s'}\bm{t}\bm{1}}  \Sigma_{\bm{1}|\bm{s}} =   \nonumber 
    \delta_{\bm{2}\bm{t}\bm{2'}} \Gamma_{\bm{s'}\bm{4'}|\bm{3}\bm{s}} 
    \times 
    \\&\quad
    \Big\{  
    \left[G_0^{-1}\right]_{\bm{2}|\bm{\tilde{1}}} G_{\bm{\tilde{1}}|\bm{4'}} G_{\bm{3}|\bm{2'}} 
    -  G_{\bm{2}|\bm{4'}} G_{\bm{3}|\bm{\tilde{1}}} \left[G_0^{-1}\right]_{\bm{\tilde{1}}|\bm{2'}} \Big\} \ . \label{eq:WI_sigma_gamma_diagrammatically}
\end{align}
For ease of notation, repeated multi-indices are meant to be contracted.

Introducing a diagrammatic notation for $G_0^{-1}$,
\begin{align*}
    [G_0^{-1}]_{\bm{1'}|\bm{1}} = 
    \tikzm{Gbareinv}{
        \Gbareinvwithlegs{$G_0^{-1}$}{0}{0}{1};
        \node at (-0.8, -0.3) {$\bm{1'}$};
        \node at (0.8, -0.3) {$\bm{1}$};
    }\, ,
\end{align*}
we can depict Eq.\,\eqref{eq:WI_sigma_gamma_diagrammatically} diagrammatically as
\begin{align}
    &\tikzm{WI_LHS1}{
    \threepointvertexleft{$\delta$}{0}{0}{1};
    \draw[lineWithArrowCenterEnd] (-0.3, 0.7) -- (0, 0.4);
    \boson{0.6}{0}{1.2}{0};
    \node at (-0.5, 0.75) {$\bm{s}$};
    \node at (1.4, 0) {$\bm{t}$};
    \selfenergy{$\Sigma$}{-0.23}{-0.63}{1};
    \draw[lineWithArrowCenterEnd] (-0.43, -0.83) -- (-0.73, -1.13);
    \node at (-0.9, -1.3) {$\bm{s'}$};
    }
    -
    \tikzm{WI_LHS2}{
    \threepointvertexleftarrows{$\delta$}{0}{0}{1};
    \boson{0.6}{0}{1.2}{0};
    \node at (-0.5, -0.75) {$\bm{s'}$};
    \node at (1.4, 0) {$\bm{t}$};
    \selfenergy{$\Sigma$}{-0.23}{0.63}{1};
    \draw[lineWithArrowCenterEnd] (-0.73, 1.13) -- (-0.43, 0.83);
    \node at (-0.9, 1.3) {$\bm{s}$};
    } \nonumber \\
    &= 
    \tikzm{WI_RHS1}{
    \def\rightshift{0.5};
    \arrowslefthalffull{0}{0}{1};
    \fullvertex{$\Gamma$}{0}{0}{1};
    \threepointvertexleft{$\delta$}{\rightshift+1.5}{0}{1};
    \boson{\rightshift+2.1}{0}{\rightshift+2.7}{0};
    \node at (\rightshift+2.9, 0) {$\bm{t}$};
    \draw[lineWithArrowCenter] (\rightshift+1.5,-0.4) to [out=225, in=315] (0.3,-0.3) ;
    \Gbareinv{$G_0^{-1}$}{\rightshift+1.22}{0.68}{1};
    \draw[lineWithArrowCenter] (0.3,0.3) to [out=45, in=135] (\rightshift+0.94,0.97) ;
    \node at (-0.8, -0.8) {$\bm{s'}$};
    \node at (-0.8, 0.8) {$\bm{s}$};
    } \nonumber \\
    &\qquad \qquad - \tikzm{WI_RHS12}{
    \def\rightshift{0.5};
    \arrowslefthalffull{0}{0}{1};
    \fullvertex{$\Gamma$}{0}{0}{1};
    \threepointvertexleft{$\delta$}{\rightshift+1.5}{0}{1};
    \boson{\rightshift+2.1}{0}{\rightshift+2.7}{0};
    \node at (\rightshift+2.9, 0) {$\bm{t}$};
    \draw[lineWithArrowCenter] (\rightshift+0.94,-0.97) to [out=225, in=315] (0.3,-0.3) ;
    \Gbareinv{$G_0^{-1}$}{\rightshift+1.22}{-0.68}{1};
    \draw[lineWithArrowCenter] (0.3,0.3) to [out=45, in=135] (\rightshift+1.5,0.4) ;
    \node at (-0.8, -0.8) {$\bm{s'}$};
    \node at (-0.8, 0.8) {$\bm{s}$};
    } \, ,
\end{align}
where touching diagram components mean a direct contraction between the two, without a connecting propagator.

In these expressions, the Keldysh rotation and the Fourier transform are mere basis transformations to be carried out consistently.
After choosing a frequency convention and accordingly labeling the legs, the frequency arguments can be read off from the diagrams.
Hence, we merely need the Keldysh and frequency structure of the bare Hedin vertex $\delta$, which turns out to be very analogous to that of the bare interaction $\Gamma_0$.
First, in \eqref{eq:def_bareHedinVertex} the delta functions that enforce equal times simply become a delta function that frequency conservation.
Second, the Keldysh structure of $\delta$ is given by
\begin{subequations}
\begin{align}
    \delta^{\alpha_{1'}1\alpha_{1}}
    &=
    \begin{pmatrix}
    0 & 1 \\ 1 & 0
    \end{pmatrix}
    =\sigma^{\alpha_{1'}\alpha_{1}}_x\, ,
    \\
    \delta^{\alpha_{1'}2\alpha_{1}}
    &=
    \begin{pmatrix}
    1 & 0 \\ 0 & 1
    \end{pmatrix}
    = \delta_{\alpha_{1'},\alpha_{1}}
\end{align}
\end{subequations}
for $\alpha_t=1$ and $\alpha_t=2$, respectively. Equation \eqref{eq:WI_main} is obtained for the choice $\alpha_t=1$.

\bibliography{bibliography}

\end{document}